%% file: paper.tex
\documentstyle[11pt,epsf]{article}

\include{mymacros}

\newtheorem{CSTHEOREM}{Theorem}[section]
\newenvironment{cstheorem}{\begin{CSTHEOREM} \hspace{-.85em} {\bf :} }%
                        {\end{CSTHEOREM}}
\newtheorem{CSCOROLLARY}[CSTHEOREM]{Corollary}
\newenvironment{cscorollary}{\begin{CSCOROLLARY} \hspace{-.85em} {\bf :} }%
                          {\end{CSCOROLLARY}}

\newtheorem{CSLEMMA}[CSTHEOREM]{Lemma}
\newenvironment{cslemma}{\begin{CSLEMMA} \hspace{-.85em} {\bf :} }%
                      {\end{CSLEMMA}}
\newtheorem{CSEXAMPLE}[CSTHEOREM]{Example}
\newenvironment{csexample}{\begin{CSEXAMPLE} \hspace{-.85em} {\bf :} \rm}%
                            {\end{CSEXAMPLE}}
\newcommand{\csxam}{\begin{csexample}}
\newcommand{\csexam}{\csbbox\end{csexample}}
\newcommand{\csbbox}{\vrule height7pt width4pt depth1pt}

\newcommand{\csdfn}{\begin{csdefinition}}
\newcommand{\csedfn}{\csbbox\end{csdefinition}}
\newcommand{\csthm}{\begin{cstheorem}}
\newcommand{\cscor}{\begin{cscorollary}}
\newcommand{\csecor}{\end{cscorollary}}
\newcommand{\csethm}{\csbbox\end{cstheorem}}
\newcommand{\cslem}{\begin{cslemma}}
\newcommand{\cselem}{\csbbox\end{cslemma}}

\newtheorem{defn}{Definition}[section]
\newtheorem{thm}{Theorem}[section]
\newtheorem{lem}{Lemma}[section]
\newtheorem{prop}{Proposition}[section]

\newcounter{examplectr}

\font\boldsl=cmbxsl10 at 12pt
\font\ssl=cmsl8
\newcommand\FELLINI{{\boldmath$\cal F$\kern-.09em{\boldsl ellini\ }}}
\newcommand\FEllini{{$\cal F$\kern-.09em{\sl ellini\ }}}
\newcommand\Fellini{{$\cal F$\kern-.09em{\ssl ellini\ }}}

\newcommand{\eat}[1]{}

\newcommand{\np}{$\cal NP$}

\newcommand{\vvec}{\overline{v}}
\newcommand{\wvec}{\overline{w}}
\newcommand{\Vvec}{\overline{V}}
\newcommand{\Wvec}{\overline{W}}
\newcommand\OpSched{{\sc OpSched}}
\newcommand\PipeSched{{\sc PipeSched}}
\newcommand\LevelSched{{\sc LevelSched}}

\newcommand\TreeSched{{\sc TreeSched}}

\newcommand\tsr{{\bf{\sc ts}}}
\newcommand\ssr{{\bf{\sc ss}}}
\newcommand\op{{\tt op}}
\newcommand\ophjoin{{\tt hash-join}}
\newcommand\opsmjoin{{\tt sortmerge-join}}
\newcommand\opmergeruns{{\tt mergeruns}}
\newcommand\opformruns{{\tt formruns}}
\newcommand\opscan{{\tt scan}}
\newcommand\opbuild{{\tt build}}
\newcommand\opBuild{{\tt Build}}
\newcommand\opprobe{{\tt probe}}
\newcommand\opselect{{\tt select}}
\newcommand\opstore{{\tt store}}
\newcommand\mop{\mbox{\tt op}}
\newcommand\mopt{{\mathrm OPT}}
\newcommand\msched{{\mathrm SCHED}}
\newcommand\cgf{$\mbox{CG}_{f}$}
\newcommand\procarea{{\cal A}_p}
\newcommand\commarea{{\cal A}_c}
\newcommand\maxmem{{\cal V}}

\newcommand\gsched{{\tt GlobalScheduler}}
\newcommand\qsched{{\tt QueryScheduler}}
\newcommand\zsched{{\sc Zeta}}
\newcommand\hier{{\sc Hier}}

\newcommand\fdef{0.002}
\newcommand\ldef{0.2}

\title{
{\bf Multi-Resource Parallel Query Scheduling\\
	    and Optimization}
      }

\author{
    {\bf Minos Garofalakis}\thanks{Part of this work was done while the first 
	author was a graduate student at the University of 
	Wisconsin--Madison.}\\
             {\small School of Electronic and Computer Engineering}\\
             {\small Technical University of Crete}\\
             {\small Hellas (Greece)}\\
             {\small\em minos@softnet.tuc.gr}\\
     \and
    {\bf Yannis E. Ioannidis}\\
             {\small Department of Informatics}\\
             {\small University of Athens}\\ 
             {\small Hellas (Greece)}\\
             {\small\em yannis@di.uoa.gr}\\
}

\date{}

\begin{document}
\maketitle
\thispagestyle{empty}

\begin{DoubleSpace}

\input{abstract}

\newpage

\input{intro}

\input{problemform}

\input{partitioned}

\input{sched}

\input{zsim}
\input{pqopt}

\input{related}

\input{concl}


\begin{SingleSpace}

{\small

\newcommand{\etalchar}[1]{$^{#1}$}

}

\end{SingleSpace}


\appendix
\input{appendix}


\end{DoubleSpace}
\end{document}

%% file: mymacros.tex
\newcommand\ms{\:}
\newcommand\ts{\;}

\newcommand{\vthm}{\vspace{5ex}}

\newcommand\vscaption{\vspace{-3ex}}
\newcommand\vsfigbot{\vspace{0ex}}

\newcommand{\myhrule}{\rule[.5pt]{\hsize}{.5pt}}
\newcommand\fig[3]{
 {
 \begin{figure}[htb]
 \myhrule\vskip 1pt
 \begin{center}
  \vbox{\hfil \vbox{#1} \hfil}
 \end{center}
  \vscaption
  \caption{#2}
 \vspace{-1ex}
 \myhrule
 \vsfigbot
 \label{#3}
 \end{figure}
 }
}

\newcommand\figp[3]{
 {
 \begin{figure}[htbp]
 \myhrule\vskip 5pt
  \vbox{\hfil \vbox{#1} \hfil}
  \vscaption
  \caption{#2}
  \label{#3}
 \vspace{-1ex}
 \myhrule
 \vsfigbot
 \end{figure}
 }
}

\newcommand\epsfiglow[2]{\hfil\epsfxsize=#1\hsize \epsfbox{#2}\hfil}

\newcommand\epsfig[4]{
	\fig{\epsfxsize=#1\hsize \epsfbox{#2}}{#3}{#4}
}

\newcommand\depsfig[6] {
\fig{
\hbox{
\begin{minipage}[htb]{0.5\hsize}
\hfil
\epsfiglow{#1}{#2}
\hfil
\end{minipage} \ 
\begin{minipage}[htb]{0.5\hsize}
\hfil
\epsfiglow{#3}{#4}
\hfil
\end{minipage}
}
}{#5}{#6}
}

\newcommand\quadepsfig[7] {
\fig{
\hbox{
\begin{minipage}[htb]{0.5\hsize}
\hfil
\epsfiglow{#5}{#1}
\hfil
\end{minipage} \
\begin{minipage}[htb]{0.5\hsize}
\hfil
\epsfiglow{#5}{#2}
\hfil
\end{minipage}
}
\hbox{
\begin{minipage}[htb]{0.5\hsize}
\hfil
\epsfiglow{#5}{#3}
\hfil
\end{minipage} \
\begin{minipage}[htb]{0.5\hsize}
\hfil
\epsfiglow{#5}{#4}
\hfil
\end{minipage}
}
}{#6}{#7}
}

\newcommand\sixepsfig[9] {
\figp{
\hbox{
\begin{minipage}[htb]{0.5\hsize}
\hfil
\epsfiglow{#7}{#1}
\hfil
\end{minipage} \
\begin{minipage}[htb]{0.5\hsize}
\hfil
\epsfiglow{#7}{#2}
\hfil
\end{minipage}
}
\hbox{
\begin{minipage}[htb]{0.5\hsize}
\hfil
\epsfiglow{#7}{#3}
\hfil
\end{minipage} \
\begin{minipage}[htb]{0.5\hsize}
\hfil
\epsfiglow{#7}{#4}
\hfil
\end{minipage}
}
\hbox{
\begin{minipage}[htb]{0.5\hsize}
\hfil
\epsfiglow{#7}{#5}
\hfil
\end{minipage} \
\begin{minipage}[htb]{0.5\hsize}
\hfil
\epsfiglow{#7}{#6}
\hfil
\end{minipage}
}
}{#8}{#9}
}

\def\titlenote #1 raised #2 {
\vspace{-#2}
{\bf\normalsize
 \begin{flushright}
 \begin{tabular}{|p{4.5in}|}\hline
  {#1}\\ \hline
 \end{tabular}
 \end{flushright}
}
\vspace{#2}
}

\def\papernumber #1 raised #2 {
\vspace{-#2}
\vbox to 0pt{\hfill\framebox{\bf Paper Number #1}}
\vspace{#2}
}

\newcommand{\alg}[4]{
\begin{figure}[htb]
\begin{center}
\begin{tabular}{|l|}\hline
\begin{minipage}{#2\hsize}
#1 
\end{minipage}\\ \hline
\end{tabular}
\end{center}
\hfil
\caption{#3}\label{#4}
\end{figure}}

%% file: abstract.tex
%

\begin{AbstSpace}

\begin{abstract}

\noindent
Scheduling query execution plans is a particularly complex problem in shared-nothing  parallel
systems, where each site consists of a collection of local time-shared (e.g., CPU(s) or disk(s))
and space-shared (e.g., memory) resources and communicates with remote sites by message-passing.
Earlier work on parallel query scheduling  employs either (a) one-dimensional models of 
parallel task scheduling, effectively ignoring the potential benefits of resource sharing, or
(b) models of globally accessible resource units, which are appropriate only for shared-memory
architectures, since they cannot capture the affinity of system resources to sites.  
In this paper, we develop a general approach capturing the full complexity of scheduling distributed,
multi-dimensional resource units for all forms of parallelism within and across queries and operators.
We present a level-based list scheduling heuristic algorithm for independent query tasks (i.e., physical
operator pipelines) that is provably near-optimal for given degrees of partitioned parallelism (with a
worst-case performance ratio that depends on the number of time-shared and space-shared
resources per site and the granularity of the clones). 
We also propose 
extensions to handle blocking constraints in logical operator (e.g., hash-join)
pipelines and bushy query plans as well as on-line task arrivals (e.g., in a dynamic or multi-query
execution environment).
%
%
Experiments with our scheduling algorithms implemented on top of a detailed simulation 
model verify their effectiveness compared to existing 
approaches in a realistic setting.
Based on our analytical and experimental results, we revisit
the open problem of designing efficient cost models for parallel query
optimization and propose a solution that captures all the important
parameters of parallel execution.

\end{abstract}

\end{AbstSpace}

\eat{
\vspace{3em}
\begin{SingleSpace}

{\small
\noindent
\underline{\bf Note to referees:}\hspace{1em}
Parts of this paper have appeared in the {\em Proceedings of the 1996 ACM 
SIGMOD International Conference on Management of Data (SIGMOD'96)\/}~\cite{gi:sigmod96}
and in the {\em Proceedings of the 23rd International Conference on Very Large Data Bases
(VLDB'97)\/}~\cite{gi:vldb97}.
This paper extends these earlier conference publications with the
following new material:
(a) more extensive descriptions and examples of the underlying models
and algorithms;
(b) experimental results from an implementation of our scheduling
algorithms as well as other algorithms on top of a detailed simulation
model of a parallel DBMS (Section~\ref{sec.zsim}); and
(c) proofs of all theoretical results (Appendix~\ref{sec.proofs}).
}

\end{SingleSpace}
}


%% file: intro.tex
\section{Introduction}
\label{sec.intro}
Parallelism has been recognized as a powerful and cost-effective means of handling the 
projected increases in  data size and query complexity in future database applications. 
Among all proposals, the shared-nothing~\cite{ston:debull86,dgsb:tkde90,dg:cacm92} and, 
recently, the more general
hierarchical (or, hybrid)~\cite{graefe:surveys93,bfv:vldb96,nzt:sigmodrec96} multiprocessor
architectures have emerged as the most scalable to support very large database management.
In these systems, each {\em site\/} consists of its own set of local resources (such as,
memory, CPU(s), and disk(s)) and communicates with other sites only by message-passing
over a high-speed interconnect.
Despite the popularity of these architectures, the development of effective and efficient
query processing and optimization techniques to exploit their full potential still remains
an issue of concern~\cite{ghk:sigmod92,valduriez:dpdbs93,hfv:sigmodrec96}.
%
%
%

Perhaps the major difference between parallel query optimization and its well-understood 
centralized counterpart lies in the choice of {\em response time\/} as a more appropriate
optimization metric for parallel database systems.
Ideally, this means that a scheduling algorithm should be run on every
candidate query execution plan so that its response time is estimated
while taking parallelism into account.
To avoid the significant increase in optimization cost that this
implies, a {\em two-phase optimization\/} is often advocated:
the optimal sequential plan is identified in a first phase using
the traditional, work (i.e., total resource consumption), metric for
plan comparison; a scheduling algorithm is run on that plan in a second
phase to identify its optimal parallelization.
Although prior work has demonstrated that this often leads to plans
that are inherently sequential and, consequently, unable to exploit the
available parallelism~\cite{jps:apads93,bfgh:ibmsj95,lvz:vldb93},
the technique is quite attractive due to its speed.
Whether in a one-phase or a two-phase optimizer, however, it is clear
that effective scheduling algorithms are still needed to intelligently
{\em parallelize\/} query execution plans~\cite{hm:vldb94,chm:pods95,hasan:phdthesis95}.

The above observation has lead to significant research activity on the complex problem of 
{\em parallelizing\/} a query execution plan; that is, scheduling the plan's operators
to execute on the resources of a parallel system with the objective of minimizing the
response time of the 
query~\cite{chm:pods95,gw:rutgers93,hm:vldb94,hasan:phdthesis95,hong:sigmod92,hcy:sigmod94,lcry:sigmod93,schneider:phdthesis90}.
%
%
Most of these earlier efforts, however, are based on simplifying assumptions 
(e.g., restricted forms of intra-query parallelism) that tend to limit their applicability.
%

One of the main sources of complexity of query plan scheduling is the 
{\em multi-dimensionality\/} of the resource needs of database queries; that is, during
their execution queries typically require multiple different resources, such as memory 
buffers, and CPU and disk bandwidth.
It is therefore important to employ an appropriate multi-dimensional cost model that can
capture a query's demand for individual system resources as separate components (i.e.,
dimensions).
The reason, of course, is that such a model introduces a range of possibilities for effectively
sharing system resources among concurrent query operators, which can substantially increase the
utilization of these resources and reduce the response time of the query.
Moreover, system resources can be categorized into two radically different classes with respect
to their mode of usage by query plan  operators:
\begin{itemize}
\item
\underline{Time-Shared (\tsr) (or, preemptable) resources} 
(e.g., CPUs, disks, network interfaces), that can be sliced across concurrent operators at
very low overhead~\cite{ghk:sigmod92,gi:sigmod96,gi:vldb97}. 
For a \tsr\  resource, operators specify an amount of {\em work\/} (i.e., the effective time for
which the resource is used) that can be appropriately ``stretched'' over the operator's execution
time (which, of course, depends on the level of contention at the \tsr\  resource).
\item
\underline{Space-Shared (\ssr) resources} (e.g., memory buffers), whose time-sharing among 
concurrent operators introduces prohibitively high overheads~\cite{ghk:sigmod92}.
For a \ssr\  resource, operators typically specify rigid capacity requirements that must
be satisfied throughout their execution\footnote{Our assumption of fixed memory requirements for query
operators is sufficiently realistic (especially for optimization-time query 
scheduling~\cite{chm:pods95,hasan:phdthesis95}) and has been used in several earlier studies 
of  resource scheduling issues in database systems (see, for example, 
\cite{bfv:vldb96,bfmv:icde00,lcry:sigmod93}).
Note, however, that {\em memory-adaptive operators\/} (like hybrid-hash joins~\cite{shapiro:tods86}) 
can actually operate under a range of possible memory allotments, with each allotment implying
different operator demands for \tsr\  resources.
Extending our work to deal with such {\em malleable\/} operators is a very challenging direction
for future research.}.
Of course, this means that the total \ssr\ requirements of concurrent operators cannot exceed 
the available \ssr\ resource capacity.
\end{itemize}

\csxam
Consider a parallel database system, with each site comprising one CPU, one disk, and $M$ pages 
of main memory (i.e., three dimensions).
Also, consider a relation $R_1$  whose tuples are stored on disk at some site of the system
and a selection operator \opselect$(R_1.A > 10)$, where $A$ is a numeric attribute of $R_1$.
Using DBMS catalog information (e.g., existence of an index on $R_1.A$) and 
simple cost-model equations, we can estimate the {\em work\/} (e.g., in msec) that 
\opselect$(R_1.A > 10)$ imposes on the CPU (${\tt work}_{\tt sel}(CPU)$) and 
disk (${\tt work}_{\tt sel}(disk)$)  resources of the site.
Further, assuming that \opselect\ operators use simple double buffering, 
the memory requirement of \opselect$(R_1.A > 10)$ from the site (in pages) 
is simply ${\tt mem}_{\tt sel} = 2$.

Additionally, consider a \opbuild\ operator running on the same site  concurrently with 
\opselect$(R_1.A > 10)$ to build an in-memory hash table for a different relation $R_2$.
The two concurrent operators will obviously time-share the CPU and (possibly) the disk 
resource at the site, so the execution times of both operators increase; that is, their
{\tt work} requirements on each \tsr\  resource are ``stretched'' over a longer period 
of time.
On the other hand, the physical memory allotted to each operator is fixed throughout its 
execution and memory buffers should not be shared among the two clones (in order to avoid
expensive paging overheads).
Thus, a requirement for the two operators to run in parallel is that their total memory 
requirement cannot exceed $M$ pages (the memory capacity of the site).
\label{eg.resmodel}
\csexam

Most previous work on parallel query scheduling has typically ignored
the multi-dimensional nature of database queries.
It has simplified the allocation of \tsr\  resources to a mere allocation of processors, hiding 
the multi-dimensionality of query operators under a scalar ``time'' 
cost metric~\cite{chm:pods95,gw:rutgers93,hm:vldb94,hcy:sigmod94,lcry:sigmod93}.
This one-dimensional model of scheduling is inadequate for database operations 
that impose a significant load on multiple system resources.
With respect to \ssr\ resources, all previous work has concentrated
on simplified models, assuming that the capacity of all such resources is
infinite or that they are all {\em globally accessible\/}
to all tasks~\cite{gg:siamjc75,st:ipl99,nshl:dimacs95,cm:spaa96}.
Clearly, such models do not account for the physical distribution of resource
units or the possibilities of \ssr\ resource fragmentation.
This limits the usefulness of these models to a {\em shared-memory\/} context, 
where all processors have equal access to memory and disks~\cite{dg:cacm92}.

\vspace*{.5em}\noindent
{\bf Our Contributions.}
In this paper, we address the problem of effective
{\em optimization-time parallelization\/} of complex query execution plans 
over shared-nothing architectures.
This problem is, of course, of critical importance to both  the 
{\em one-phase\/}~\cite{lvz:vldb93}  and the 
{\em two-phase\/}~\cite{hs:pdis91,hong:sigmod92,hs:dpdbs93}
approach to parallel query optimization.
We propose a general framework for multi-dimensional \tsr\ and \ssr\ resource 
scheduling in  parallel database systems and employ it to address the  query plan scheduling
problem in its most general form, assuming the full variety of bushy plans and schedules that
incorporate {\em independent\/} and {\em pipelined\/} forms of inter-operation parallelism as
well as intra-operation (i.e., {\em partitioned}) parallelism.
Our multi-dimensional framework  represents query operator costs as pairs of {\em work\/} 
and {\em demand\/} vectors  with one dimension per \tsr\ and \ssr\ resource, respectively.
The basic idea is to accurately capture the operator's requirements of individual \tsr (\ssr)
resources in the system in the corresponding work (resp., demand) vector components.
Referring back to Example~\ref{eg.resmodel}, the work vector for our example
\opselect$(R_1.A > 10)$ operator is  $[ {\tt work}_{\tt sel}(CPU), {\tt work}_{\tt sel}(disk) ]$
and its (one-dimensional) demand vector is simply $[{\tt mem}_{\tt sel}]$.
We observe that the inclusion of the \ssr\ resource dimension(s) gives rise to certain
interesting tradeoffs with respect to the degree of partitioned parallelism.
Smaller degrees result in reduced communication overhead and, therefore, increased 
total work (i.e., \tsr\ resource requirements) for the operator execution 
(i.e., {\em coarse grain\/} parallel executions~\cite{gw:rutgers93,ggs:pods96}).
On the other hand, larger degrees of parallelism in general imply smaller \ssr\ 
requirements for each operator clone, thus allowing for better load-balancing 
opportunities and tighter schedulability conditions. 
The importance of such tradeoffs for parallel query processing and optimization has
been stressed earlier~\cite{hfv:sigmodrec96} and is addressed in this work.

Based on our framework, we develop a fast resource scheduling algorithm for physical operator 
pipelines called \PipeSched\ that belongs to the class of 
{\em list scheduling algorithms\/}~\cite{graham:belltj66}.
The {\em response time\/} (or, {\em makespan\/}) of the parallel schedule produced by \PipeSched\ 
is analytically shown to be within $d(1 + \frac{s}{1-\lambda})$ of the optimal schedule length for
given degrees of partitioned parallelism, where $d$ and $s$ are the dimensionalities of the \tsr\ 
and \ssr\ resource vectors respectively, and $\lambda$ is an upper bound on the (normalized) \ssr\
demands of any clone in  the pipeline. 
We then extend our approach to multiple independent pipelines, using a level-based scheduling 
algorithm~\cite{cgjt:siamjc80,twpy:sigmetrics92} that treats \PipeSched\ as a subroutine within 
each level.
The resulting algorithm, termed \LevelSched, is analytically shown to be near-optimal for given
degrees of operator parallelism as well.
Furthermore, we propose heuristic extensions to \LevelSched\  that allow
the resulting algorithm, \TreeSched\, to handle 
{\em blocking constraints\/} that can arise in logical operator pipelines and
arbitrary query plans, and demonstrate how the algorithm can handle 
{\em on-line\/} task arrivals (e.g., in a dynamic or multi-query execution
environment).

We also present the results of a performance study comparing our parallel query scheduling 
strategies and other existing algorithms implemented on top of a detailed simulation model~\cite{brown:prpl94}
for a shared-nothing parallel database system based on the 
Gamma parallel database machine~\cite{dgsb:tkde90}.

Finally, we revisit the open problem of designing effective cost models
that would capture the essense of parallel execution and allow traditional
query optimization approaches, without explicit scheduling methods, to
be effectively used for parallel query optimization.
These would be applicable to both one-phase optimizers that avoid
scheduling or to the first phase of two-phase optimizers.
We start at the work of Ganguly et al.~\cite{ggs:pods96},
which identified two important ``bulk parameters'' of
a parallel query execution plan, namely average work and critical path length, 
that are important to characterizing its expected response time.
Based on our analytical and experimental
results, we identify the importance of a third
parameter, the {\em average volume\/} (i.e., the resource-time product) for \ssr\
resources, which captures the constraints on query execution that derive from such
resources.
We believe that 
a plan cost model that captures these three ``bulk parameters'' is sufficient
for efficient and effective parallel query optimization.

\vspace*{.5em}\noindent
{\bf Roadmap.}
The remainder of this paper is organized as follows.
Section~\ref{sec.problemform}, presents a formulation of the query scheduling
problem and an overview of our approach.
Our multi-dimensional model of \tsr\ and \ssr\ resource usage and quantification
of partitioned operator parallelism are developed in Section~\ref{sec.partitioned}.
In Section~\ref{sec.sched},  we develop the \PipeSched\ and \LevelSched\ algorithms
and prove constant bounds on their worst-case performance ratios.
Following that, we propose heuristic extensions to deal with blocking constraints
(e.g., in logical operator pipelines or bushy query plans) and discuss how our
approach can handle dynamic task arrivals.
Section~\ref{sec.zsim} presents our experimental findings from
the implementation of our  query scheduling algorithms in
a detailed simulation model of a parallel database system.
In Section~\ref{sec.pqopt} we discuss the implications of our results 
on the design of cost models for parallel query optimizers.
Section~\ref{sec.related} reviews related work in the
areas of parallel query processing and deterministic
scheduling.
Finally, Section~\ref{sec.concl} concludes the paper and
identifies directions for future research.
Proofs of theoretical results presented in this paper can be found in 
Appendix~\ref{sec.proofs}.

%% file: problemform.tex
\section{Problem Formulation}
\label{sec.problemform}
In this section, we provide a more detailed description of our target parallel execution
architecture and discuss our model and terminology for parallel query execution plans.
We then give a concise statement of the general query scheduling problem addressed in 
this paper, and provide a high-level overview of our proposed solution.
Finally, we discuss the main assumptions that underlie our methodology.

\subsection{System and Query Plan Model}
\label{subsec.definitions}
\noindent
{\bf System Architecture.}
%
We consider
{\em shared-nothing parallel database architectures}~\cite{ston:debull86,dgsb:tkde90,dg:cacm92},
comprising  several {\em identical\/} multi-resource sites.
Each system site consists of (potentially) several CPUs, several disks, and a
memory shared by all CPUs.
Inter-site  communication is done via message-passing over a high-speed interconnect, while
CPUs within a site can communicate much more effectively via shared-memory.
(The presentation in this paper focuses on the shared-nothing model but, since our methodology
can capture multiple resources at each system site, it is equally applicable to hierarchical
architectures.)

\begin{SingleSpace}

\epsfig{0.8}{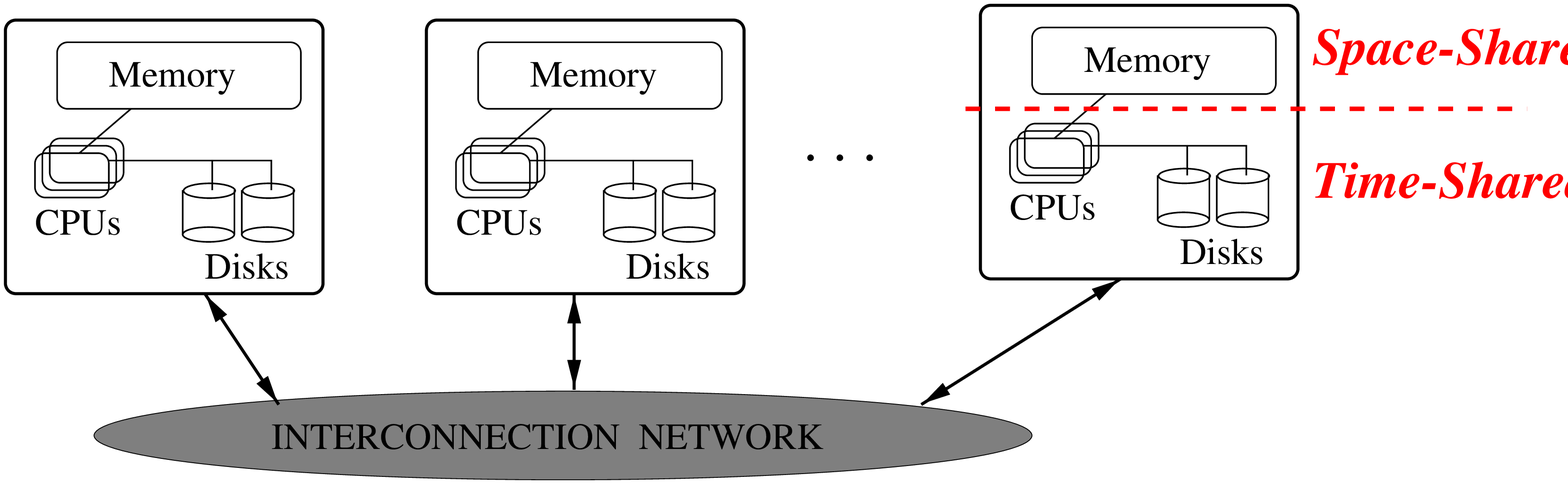}
{A shared-nothing parallel database system.}
{fig.sysmodel}

\end{SingleSpace}

For our more abstract scheduling problem formulation, we view each site as a collection of $d$
\tsr\ resources (capturing the site's  CPUs, disks, and, possibly, other resources  that are
typically time-shared like network interfaces or communication processors) 
and $s$ \ssr\ resources (capturing the site's memory capacity).  
Although memory is probably the only \ssr\ resource that comes to mind when discussing traditional
query processing operators, often the distinction between time-  and space-sharing depends on the 
needs of a particular application.
As an example, the playback of a digitized video stream from disk is an operation that requires
a specific fraction of disk bandwidth throughout its execution~\cite{ors:icmcs96,ors:pods97,gio:vldb98}.
Clearly, such an operation views the disk as an \ssr\ resource with a fixed bandwidth capacity, 
whereas traditional query processing operators do not have similar ``hard real-time''
requirements for data delivery and view it as a \tsr\ resource.
For this reason, we have decided to address the scheduling problems for general $s$ rather than 
restricting our discussion to the special case of $s=1$ (i.e., a single dimension for memory capacity).
An obvious advantage of this general formulation is that it allows us the flexibility to ``draw the line''
between time-  and space-sharing at any boundary, depending on factors such as application
requirements or user view of resources.

\vspace*{.5em}\noindent
{\bf Query Execution Plan Model.}
A {\em query execution plan\/} comprises a tree of {\em logical operator} nodes, 
like \ophjoin\ and \opsmjoin\ (Figure~\ref{fig.opertree}(a)).
The shape of the plan tree can be left-deep, right-deep, or bushy.
Bushy trees capture the most general space of query execution plans and are the 
most appealing for parallel query execution, since they offer the best opportunities
for minimizing the size of intermediate results~\cite{syt:vldb93} and for exploiting
all forms of parallelism~\cite{valduriez:dpdbs93}.
Thus, in this paper, we consider the full space of bushy query execution plans.

\begin{SingleSpace}

\epsfig{1.0}{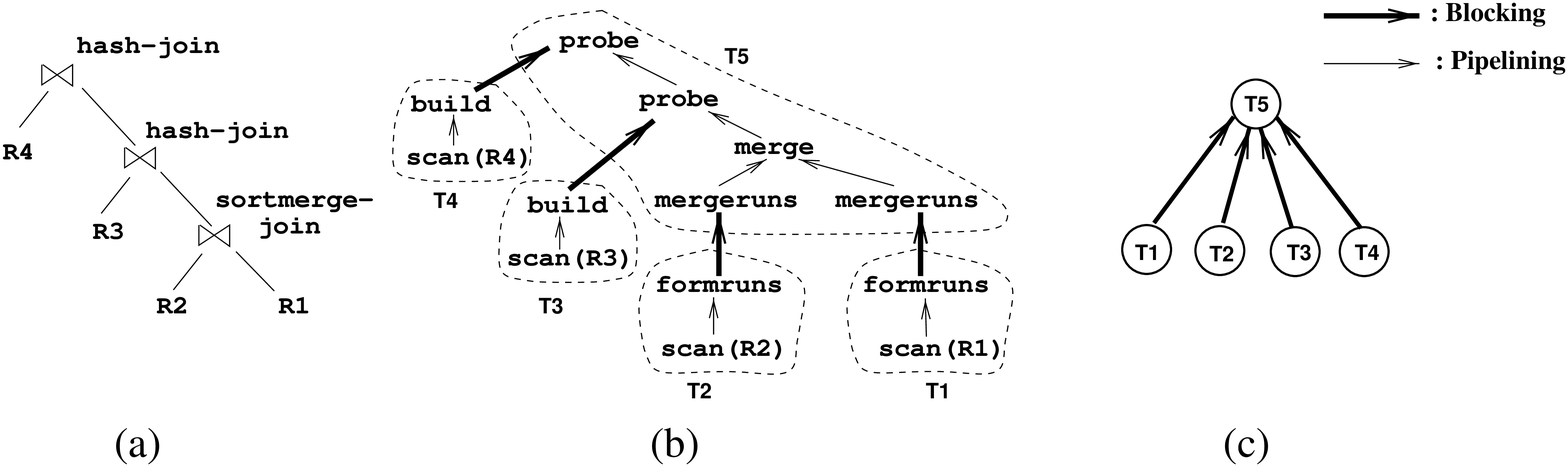} 
{(a) A right-deep query execution plan. (b) The corresponding
operator tree. (c) The corresponding query task tree. The thick edges
in (b) indicate blocking constraints.}
{fig.opertree}

\end{SingleSpace}

An {\em operator tree}~\cite{ghk:sigmod92,hong:sigmod92,schneider:phdthesis90} is created
as a ``macro-expansion'' of a query execution plan by refining each plan node into a subtree
of {\em physical operator\/} nodes, such as \opscan, \opprobe, and \opbuild\ 
(Figure~\ref{fig.opertree}(b)).
The operator tree representation exposes the available parallelism in the plan as well as
the specific  data flow  and  timing constraints between physical operators.
In our example operator tree of Figure~\ref{fig.opertree}(b), edges  represent the flow of
data as well as the two forms of timing constraints between physical operators: 
{\em pipelining\/} (thin edges) and {\em blocking\/} (thick edges).
Pipelining edges connect physical operators that work in parallel in ``producer/consumer''
mode, whereas blocking edges indicate that a parent operator cannot start execution 
until all its children via blocking edges have produced their full result.
Furthermore, blocking constraints often imply two forms of stronger
{\em execution dependencies\/} between physical operators.
\begin{enumerate}
\item
{\em Disk-Materialization Dependencies.\/}
If the child (tail) operator of a blocking edge materializes its result  on disk, then 
the parent (head) operator (e.g., \opscan) has to be executed at the same set of sites as the
child, since it needs to access the same set of disks.
For example, this is the case with the \opmergeruns\ and \opformruns\ operators of
Figure~\ref{fig.opertree}(b): each \opmergeruns\  needs to access the sorted runs
stored on disk by its child \opformruns.

\item
{\em Memory-Materialization Dependencies.\/}
If child operators of blocking edges materialize their result  in memory,
then not only do their parent operators have to be executed at the same set of sites,
but it also often needs to be guaranteed that
(a) all children are executed in parallel, and (b) their parents are executed  
immediately afterwards as well.
This is, for instance, the case with the \opbuild\ operators of Figure~\ref{fig.opertree}(b),
which must build their hash tables in memory in parallel, so that the corresponding \opprobe\ 
parent operators, being executed immediately after them, find those tables in memory.
\end{enumerate}
%

A {\em query task\/} is a maximal subgraph of the operator tree containing
only pipelining edges.
Intuitively, a query task is a fully-pipelined collection of physical operators.
A {\em query task tree\/} is created from an operator tree by representing
query tasks as single nodes (Figure~\ref{fig.opertree}(c)).
Note that a single logical operator pipeline, like the right-deep execution
plan of Figure~\ref{fig.opertree}(a), can give rise to multiple query tasks 
(i.e., physical operator pipelines) with various blocking constraints and
execution dependencies between them.

The above tree representations clarify the definitions of the three forms
of intra-query parallelism:
\begin{itemize}
\item
{\em Partitioned parallelism:} A single node of the operator tree (i.e.,
physical operator) is executed on a set of sites by appropriately partitioning
its input data set(s) across independent {\em operator clones}.
\item
{\em Pipelined parallelism:} The operators of a single node of the task tree
(i.e., physical operator pipeline) are executed on a set of sites in a pipelined
manner.
\item
{\em Independent parallelism:} Nodes of the task tree with no path between
them are executed in parallel on a set of sites independent of each other
(as mentioned above, they {\em must} be executed in parallel in certain cases).
For example, in Figure~\ref{fig.opertree}, tasks T1-T4 may all be executed
in parallel, with T3-T4 having no other choice but to be executed in parallel.
Task T5 must await the completion of T1-T4 before it is executed, which should
be immediately after T3-T4 complete.
\end{itemize}

The {\em home\/} of an operator is the set of sites allotted to its execution.
Each operator is either {\em rooted\/}, if its home is fixed by 
execution dependencies (e.g., scanning the materialized result of a child 
operator from disk or {\tt prob}ing the memory-resident result of a child operator), or
{\em floating\/}, if the resource scheduler is free to choose its home.

\subsection{Problem Statement and Overview of Approach}
\label{subsec.overview}
Consider a parallel query optimizer and a query plan tree that the optimizer
has generated during the course of its exploration for the optimal plan.
A {\em parallel execution schedule\/} for the query plan consists
of an allocation of system resources to physical operators in the
corresponding operator tree and a mapping
of these operators onto system sites, such that  
(1) the resource needs of all operators are satisfied without violating system capacity 
constraints, and  (2) all timing constraints and execution dependencies in the plan
are respected.
%
%
The general optimization-time query scheduling problem addressed 
in this work can be stated as follows.

\vspace*{-.6em}
\begin{center}
\begin{tabular}{|l|}\hline
\begin{minipage}{.9\hsize}
\begin{description}
\vspace*{-.0em}
\item[Given:]
A query execution plan to be parallelized over
a shared-nothing  parallel database system,
where each site comprises $d$  \tsr\ and $s$ \ssr\  resources.

\vspace*{-.5em}
\item[Find:]
A parallel execution schedule  that minimizes the response time for the
input query plan.
\end{description} 
\end{minipage}\vspace*{1pt}\\ 
\hline
\end{tabular}
\end{center}

%
 
We have devised an algorithm for parallelizing bushy query execution plans that
consists of the following general steps\footnote{The description here is 
necessarily abstract and does not directly address certain issues, such as the
\opbuild/\opprobe\ execution dependencies described earlier in this section.
The details will be filled in later in the paper.}:
\begin{enumerate}
\item
Construct the corresponding operator and task trees, and for each operator, determine
its individual resource requirements using hardware parameters, DBMS statistics, and
conventional optimizer cost models \cite{saclp:sigmod79,hcy:sigmod94}.
(Part of this information would already be available from earlier 
actions of the optimizer.)
\item
For each (floating) logical operator, determine the degree of parallelism based on 
the \tsr\ vs.\ \ssr\ resource tradeoffs mentioned earlier and
discussed in detail in Section~\ref{sec.partitioned}, taking into
account any execution dependencies involving the operator
(partitioned parallelism).
\item
Place the tasks corresponding to the leaf nodes of the task tree
in the {\em ready list\/} $L$ of the scheduler. 
While $L$ is not empty, perform the following steps:
\begin{enumerate}
\item[3.1.]
Determine a batch of tasks from $L$ that can be executed concurrently 
(physical operator pipelines)
and schedule them using a {\em provably near-optimal\/} multi-dimensional 
list scheduling heuristic (pipelined and independent parallelism).
\item[3.2.]
If there are tasks in the tree whose execution is enabled after Step 3.1,
place them in the ready list $L$.
\end{enumerate}
\end{enumerate}

Our scheduling strategy 
can be readily used to handle {\em on-line\/} task arrivals (e.g., in a dynamic or 
multi-query execution environment).

\eat{
Finally, we propose a technique for selecting an operator parallelization
that allows us to relax the coarse granularity restriction (Step~2).
Combining this technique with our list scheduling rule for independent
operators results in an algorithm that is provably near-optimal in
the space of {\em all possible\/} parallel executions.
}

\subsection{Assumptions}
\label{subsec.assumption}
Our resource scheduling model and optimality results are based on certain simplifying 
assumptions about the execution system.
We believe that most of these assumptions are reasonable for optimization-time
query scheduling, where decisions are made using a cost model that
abstracts away many aspects of run-time query execution.
We outline our assumptions along with a short discussion behind their rationale.

\vspace{.4em}\noindent {\bf A.1}
\underline{\bf No Time-Sharing Overhead for \tsr\ Resources.}
Following Ganguly et al.~\cite{ghk:sigmod92}, we assume that slicing a preemptable resource 
among multiple operators introduces no additional resource costs.
(Note that, a similar assumption is made in the cost model of the query optimizer for IBM's 
DB2 Parallel Edition~\cite{jmp:debull97}.) 
This is certainly realistic for \tsr\ resources with reasonably small context-switching
overheads, like CPUs or communication processors.
On the other hand, time-slicing  a disk resource across multiple requests can lead to
significant seek and latency overheads.
Nevertheless, most modern disk architectures employ sequential data prefetching along multiple
cache contexts, which means that they are typically able to ``hide'' such overheads and use the
disk effectively as long as the number of concurrent requests stays below the number of cache
contexts (typically $16$ or higher in modern SCSI disks~\cite{cheetah:spec00}).
Thus, even for disk resources, we believe that our assumption is justified for reasonably
high levels of concurrency.
(When these levels are exceeded, a multiplicative {\em penalty factor\/} could be used 
to model the effects of resource contention~\cite{ghk:sigmod92}.)

\vspace{.4em}\noindent {\bf A.2}
\underline{\bf Uniform \tsr\ Resource Usage.}
Following Ganguly et al.~\cite{ghk:sigmod92} again, we assume that the usage of a \tsr\
resource by an operator is uniformly spread over the execution of the operator.
Essentially, this implies that our scheduling  model does not capture operator 
``hot spots'' in CPU or disk usage  -- a reasonable assumption, we believe, for 
optimization-time query scheduling.
Note that this assumption is clearly true {\em on the average\/} over the lifetime
of the operator.

\vspace{.4em}\noindent {\bf A.3}
\underline{\bf Constant \ssr\ Resource Demand.}
The total \ssr\ requirements of an operator are assumed to be constant and independent 
of its degree of parallelism; that is, distributing the execution of an operator over
multiple clones does not increase its overall memory requirement.
As an example, the total amount of memory required by all the clones of a \opbuild\ operator
is constant and equal to the size of a hash table on the entire build relation.
Further, increasing the degree of parallelism does not increase the maximum \ssr\ demands of
individual clones.
These tend to be realistic assumptions for most query processing operators and all
reasonable data partitioning strategies.


\vspace{.4em}\noindent {\bf A.4}
\underline{\bf Dynamically Repartitioned Pipelined Inputs for Floating Operators.}
To compute the communication costs for floating operators and their children in an operator
pipeline, we assume that the output of the children is always dynamically redistributed to 
serve as input to their parents.
Dynamic data repartitioning is quite common, e.g., it is often required in join pipelines,
when the join attributes of pipelined joins are different, the degrees of
partitioned parallelism differ, or different declustering schemes must be used 
for load balancing.
%
On the other hand, repartitioning is not required for 
parallel join  strategies that try to benefit from 
existing data placement decisions, such as  {\em collocated joins\/} (where both join inputs
are already partitioned on the join attribute over the same set of sites so that no
repartitioning is needed)  or 
{\em directed joins\/} (where one of the join  inputs is already partitioned on the join
attribute and we only need to redistribute the tuples of the second input over the home of the
first) \cite{bfgh:ibmsj95}.
Note, however, that for such strategies the join  operators are obviously {\em rooted\/}
by data placement; for example, a collocated join  is always executed at
the home of its operands.
Thus, if the decisions of using collocated or directed joins  have already been made by 
earlier stages of the optimizer, our scheduling framework can readily incorporate 
them as rooted operation costs and our assumption remains valid.
If, however,  these decisions are to be made during the plan scheduling step, then 
we believe  that our scheduling framework can be extended to handle them effectively.
The key observation here is that only a small, constant number of alternatives need to be
explored for certain joins~\cite{bfgh:ibmsj95}.
Consequently, at the cost of an increase in complexity, our scheduler can investigate 
these alternatives and select the best one.

%
%


%% file: partitioned.tex
\section{Quantifying Partitioned Parallelism}
\label{sec.partitioned}
In this section, we offer a  formal description of our multi-dimensional
model of resource usage that captures the coexistence of \tsr\ and \ssr\ resources
at each system site.
Accounting for both \tsr\ and \ssr\ resource dimensions raises some
interesting tradeoffs with respect to the degree of partitioned parallelism
for operators.
We propose a quantification of these tradeoffs 
that allows us to derive the degree of partitioned parallelism
for operators based on system parameters and given bounds on the
granularity of parallel execution.

\subsection{A Multi-Dimensional Model for Time- and Space-Shared Resources}
\label{sec.usage}
For \tsr\ resources, our treatment 
builds on the model of preemptable resources proposed by Ganguly et al.~\cite{ghk:sigmod92},
which we briefly describe here.
The usage of a single \tsr\  resource by an operator is modeled by two parameters, 
$T$ and $W$, where $T$ is the elapsed time after which the resource is freed (i.e.,
the response time of the operator) and $W$ is the work measured as the effective 
time for which the resource is used by the operator.
Intuitively, during the execution of the operator, the \tsr\ resource is kept busy 
only for a fraction of the time equal to $W/T$.
Further, by Assumption (A.2), the utilization of the resource is ``stretched'' uniformly
over $T$.
For example,  a \opselect\ with a total CPU time requirement of $100$msec that runs over 
a period of $1$sec  only uses $10$\% of the CPU resource over that period --- the 
remaining $90$\% of the CPU can be used to accommodate other concurrent operations at
no overhead (Assumption (A.1)).
Thus, in conjunction with our first two assumptions, this model leads to straightforward 
quantification of the effects of  \tsr\ resource sharing.
%

%
In this paper, we propose a model that captures the coexistence of \tsr\ and \ssr\ resources 
at each system site and quantifies the effects of sharing such multi-resource sites among 
query operators.
The basic idea is to describe an isolated operator's usage of a site
comprising $d$ \tsr\ resources and $s$ \ssr\ resources by the triple 
$(T^{seq}, \overline{W}, \overline{V})$, where:
\begin{itemize}
\item
$T^{seq}$ is the {\em stand-alone}, sequential execution time
		of the operator at the site, i.e., its execution 
		time assuming no concurrent operators;
\item
$\overline{W}$ is a $d$-dimensional {\em work vector\/}
               whose components denote the work done on individual 
	       \tsr\ resources, i.e., the effective time~\cite{ghk:sigmod92,gi:sigmod96,gi:vldb97}
               for which each resource is used by the operator; and,
\item
$\overline{V}$ is an $s$-dimensional {\em demand vector\/}
               whose components denote the \ssr\ resource requirements 
	       of the operator throughout its execution.
	       For notational convenience we assume that the dimensions
	       of $\overline{V}$ are {\em normalized\/} using the
	       corresponding \ssr\ resource capacities of a single site.
\end{itemize}

\csxam
Continuing from Example~\ref{eg.resmodel}, Figure~\ref{fig.bin}(a) depicts our 
generalized view of a system site with $d=2$ \tsr\ resources (CPU and disk) 
and $s=1$ \ssr\ resource (memory). 
(We assume a fixed numbering of system resources for all sites; for example, dimensions
$1$ and $2$ of $\overline{W}$ correspond to CPU and disk, respectively.)
Figures~\ref{fig.bin}(b,c) depict the (stand-alone) usage of  a system site 
by our example \opselect\ and \opbuild\  operator clones.
Note that, since our \opbuild\ operator clone can fit its entire hash table in memory 
(Assumption (A.3)), it only makes use of the CPU resource and, therefore, its stand-alone
execution time is equal to its CPU work. 

The work vectors of these operators are calculated based on standard
cost models of the actual process of the operand.  For example,
assuming that there is no index on $R_1.A$ and using $\{ R_1\}$ ($|R_1|$) to
denote the number of tuples (resp., disk blocks) of relation $R_1$, we can estimate the
work vector components (in msec) of our example \opselect\ operator using
the following realistic cost model:
\begin{eqnarray*}
\small
W[CPU] &=& \left( I(init\_select) + I(init\_IO)\cdot |R_1| +
                   [I(read\_tuple) + I(apply\_pred)]\cdot \{ R_1\} + \right. \\
                 & & \left. I(copy\_tuple)\cdot sel_{A>10}(R_1)\cdot \{ R_1\} +
                 I(terminate\_select) \right) / (1000\cdot MIPS(CPU)) \\
W[disk] &=& avg\_block\_xfer\_time(disk)\cdot |R_1| +
                      min\_seek\_time(disk)\cdot \left\lceil\frac{|R_1|}{blocks\_per\_cylinder(disk)}\right\rceil,\\
\end{eqnarray*}
where $I(op)$ denotes the number of instructions for performing operation $op$, $sel_p$ is the
selectivity of predicate $p$, and hardware parameters (like $MIPS$ and $avg\_block\_xfer\_time$)
have the obvious interpretation.
(Note that our formula for $W[disk]$ assumes that $R_1$ is stored sequentially on disk.)
Similar formulas can also be derived for the \opbuild\ operator.
\csexam
%

\begin{SingleSpace}

\epsfig{1.0}{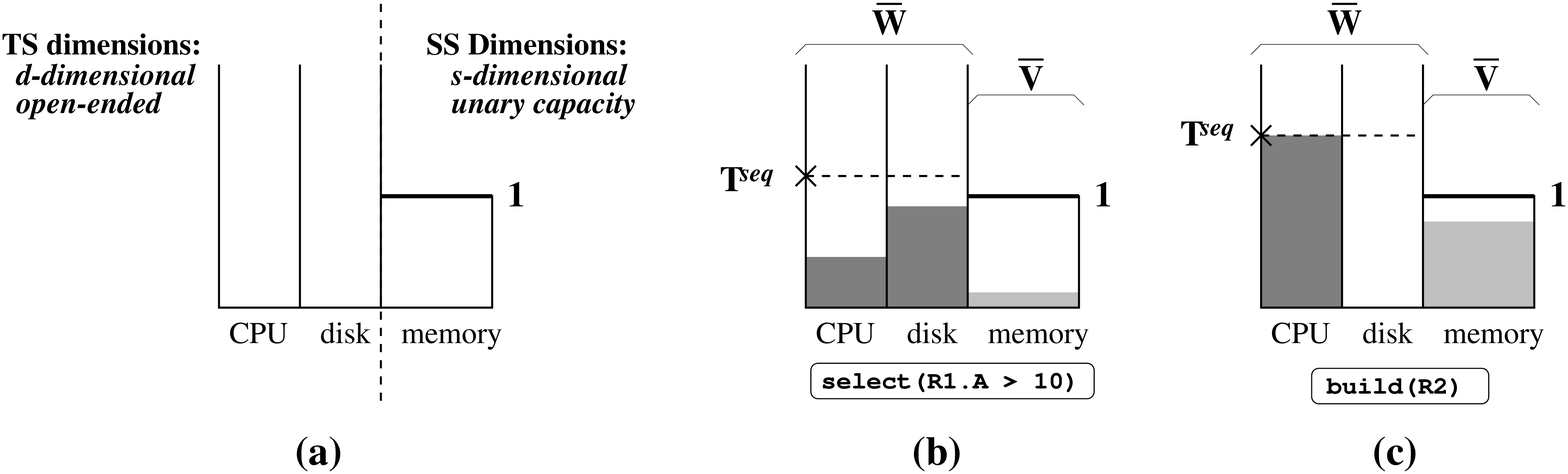}
{(a) A system site with $d=2$ \tsr\ resources and $s=1$ \ssr\ resource.
Stand-alone usage of the site from  (b) a \opselect\ and (c) a \opbuild\ operator.}
{fig.bin}

\end{SingleSpace}

The execution time $T^{seq}$ of an operator is actually a function of the operator's individual 
\tsr\  resource requirements, i.e., its work vector $\overline{W}$ (sometimes emphasized
by using $T^{seq}(\overline{W})$ instead of $T^{seq}$), and the amount of {\em overlap\/} 
that can be achieved between processing at different resources~\cite{gi:sigmod96,gi:vldb97}.
This overlap is a system parameter that depends on the hardware and software architecture
of the resource sites (e.g., buffering architecture for disk I/O) as well as  the algorithm
implementing the operator.
The operator's \ssr\ resource requirements ($\overline{V}$) depend primarily on the size of
its inputs and the algorithm used to implement the operator. 
On the other hand, the operator's work requirements ($\overline{W}$) depend on 
both of these parameters as well as  its \ssr\ resource allotment $\overline{V}$.

We should once again stress that our assumption of fixed \ssr\ resource demands
(i.e., $\overline{V}$ component values) for operators is sufficiently realistic for 
optimization-time query scheduling~\cite{chm:pods95,hasan:phdthesis95} and has been 
used in several earlier studies of  resource scheduling issues in database systems 
(e.g. , \cite{bfv:vldb96,bfmv:icde00,lcry:sigmod93}).
However, most real-life query execution engines support memory-adaptive operators,
whose  memory requirements are typically specified as a {\em range\/} of possible
memory allotments~\cite{yc:vldbj93}.
%
%
This flexibility adds an extra level of difficulty to our scheduling problem.
It means that the scheduler also has to select, for each operator, a specific 
memory allotment (i.e., \ssr\ demand vector $\overline{V}$)  such that the 
query response time is minimized over all possible $(\overline{W},\overline{V})$
combinations. 
This is a very challenging generalization of our resource scheduling problem
that we plan to pursue in future work.
%

\subsection{Quantifying the Granularity of Parallel Execution}
\label{sec.quant}
As is well known, increasing the parallelism of an operator reduces its execution time 
until a saturation point is reached, beyond which additional parallelism causes a speed-down,
due to excessive communication startup and coordination overhead over too many 
sites~\cite{dgsb:tkde90}.
To avoid operating beyond that point when parallelizing an operator, we want to ensure that
the granules of the parallel execution are sufficiently coarse~\cite{ggs:pods96,gi:sigmod96,gi:vldb97}.
However, \ssr\  resource requirements and capacity constraints introduce additional complexity
to the operator parallelization problem.
More specifically, in the presence of \ssr\ resources, it is not possible to
concurrently execute a set of clones at a site if their total \ssr\ requirements exceed the
site's capacity (in any of the $s$ dimensions).
Clearly, coarse operator clones imply that each clone has \ssr\ resource 
requirements that are relatively large.
This means that, when restricted to coarse grain operator executions, a scheduling 
method can be severely limited in its ability to balance the total work across
sites.
Furthermore, coarse \ssr\ requests can cause fragmentation that may
lead to underutilization of system resources.
Thus, taking both \tsr\ and \ssr\ resources into account gives rise to
interesting tradeoffs with respect to the granularity of operator clones.
Our analytical results in Section~\ref{sec.sched}
further demonstrate this effect.

\csxam
Consider the \opbuild\  operator depicted in Figure~\ref{fig.bin}(c).
As it stands, this \opbuild\  can only run on sites with  at least $F\cdot |R_2|$
free pages of memory, where $F$ is the standard ``fudge factor'' accounting for 
hash table overheads~\cite{shapiro:tods86}.
(To simplify the notation, we assume that memory pages and disk blocks are of equal
size.)
Assume that we decide to partition the \opbuild\ across three clones by partitioning $R_2$, 
with the first clone getting $1/2$ of $R_2$'s tuples and the remaining two each getting
$1/4$ of $R_2$'s tuples.
(The partition sizes can be estimated using statistical information on the partitioning
attribute and the form of the specific hash function used.)
Then, the first \opbuild\ clone can be placed at any site with 
$\frac{F}{2}\cdot |R_2|$ free memory pages, and the other two clones each require only
$\frac{F}{4}\cdot |R_2|$ pages for their execution.
This partitioning obviously gives the scheduler much more freedom to distribute the \opbuild\ 
work onto the system and reduce memory fragmentation at sites.
On the other hand, of course, it also implies the startup and communication overheads associated 
with the parallel \opbuild\  execution.
\csexam

We now introduce some formal machinery for the purpose of quantifying
the granularity of parallel operator execution  and  allowing us to
resolve the \tsr\ vs. \ssr\ resource tradeoffs discussed above in a 
methodical fashion.
Consider the execution of a parallel operator \op\ that has been
partitioned among $N$ clones in some specific fashion.
We view the granularity of that execution of \op\ as a parameter that depends  on
the ratio $\frac{\procarea(\mop)}{\commarea(\mop,N)}$ and $\maxmem(\mop,N)$, 
where:
\begin{itemize}
\item
$\procarea(\mop)$ denotes the total amount of work performed during the execution of 
\op\ on a single site, when all its operands are locally resident (i.e., zero communication
cost); it corresponds to the {\em processing area\/}~\cite{gw:rutgers93} of \op\ and is constant
for all possible executions of \op;
\item
$\commarea(\mop,N)$ denotes the total communication overhead incurred during the execution of 
\op\ partitioned in that way; it corresponds to the {\em communication area\/} of 
the partitioned execution of \op\ and is a non-decreasing function of $N$;
and,
\item
$\maxmem(\mop,N)$ denotes the maximum (normalized) \ssr\ resource requirement of any clone 
during the execution of \op\ partitioned in that way; it corresponds to the 
{\em \ssr\ grain size\/} of the partitioned execution of \op\ and is a non-increasing
function of $N$.
\end{itemize}

Note that the execution of \op\ with degree of partitioned parallelism equal to $N$ is
feasible only if $\maxmem(\mop,N)\leq 1$; that is, the partitioning of \op\ must be 
sufficiently fine grain for each clone to be able  to maintain its \ssr\ working set 
at a site.
The following two definitions use the notions defined above to quantify the 
granularity of parallel operator execution.

\begin{defn}
\rm
A parallel execution of an operator \op\ with degree of partitioned parallelism
equal to $N$ is {\em $\lambda$-granular\/} if $\maxmem(\mop,N)\leq \lambda$,
where $\lambda\leq 1$. \csbbox
\label{defn.lgranul}
\end{defn}

\begin{defn}
\rm
A parallel execution of an operator \op\ with degree of partitioned parallelism
equal to $N$ is {\em coarse grain with parameter f\/} (referred to as a
\cgf\ execution) if the communication area of the execution is no more than $f$
times the processing area of \op; that is,
$\commarea(\mop,N)\leq f\cdot \procarea(\mop)$. \csbbox
\label{def.cgf}
\end{defn}

The granularity parameters $\lambda$  and $f$ are system-wide parameters that control
the parallelization of query operators in the system.
The \ssr\  granularity parameter $\lambda$ essentially tries to restrict the maximum
memory requirements of any operator clone.
Smaller values for $\lambda$ imply higher degrees of partitioned parallelism, smaller
memory requirements for clones and, therefore, more scheduling freedom to ``pack'' these
requirements onto system sites and better expected utilization of  memory resources.
On the other hand, the communication granularity parameter $f$ tries to place an upper
bound on the overhead of operator parallelization.
Smaller values for $f$ imply more conservative parallelizations, with fewer clones per
query operator.

The  $\lambda$ and $f$  parameters can obviously place conflicting requirements
on the degree of operator parallelism.
In our model, we  give precedence to \ssr\ granularity requirements since they, in a sense, 
represent ``harder'' constraints that can determine the schedulability of query operators.
(This will become even more apparent in Section~\ref{sec.sched}.)
As an example, a \opbuild\ operator with a hash table larger than the size of a site's 
memory simply {\em cannot\/}  be scheduled on a single site, even if that is what 
the communication granularity requirement  dictates.
This is more formally expressed in the following definition.

\begin{defn}
\rm
A parallel execution of an operator \op\ with degree of partitioned parallelism
equal to $N$ is $\lambda$-granular \cgf, if the communication area of the
execution is no more than $f'$ times the processing area of \op, i.e.,
$\commarea(\mop,N)\leq f'\cdot \procarea(\mop)$, where $f'$ is the minimum 
value greater than or equal to $f$ such that $\maxmem(\mop,N)\leq\lambda$. \csbbox
\label{def.lcgf}
\end{defn}

Again, the intuition behind Definition~\ref{def.lcgf} is that, because of its importance
for operator schedulability, the $\lambda$-granularity restriction takes precedence
over communication granularity.
Thus, we may sometimes have to compromise our restrictions on communication overhead
to ensure that the parallelization is in the $\lambda$-granular region.
This is graphically demonstrated in Figure~\ref{fig.cgf}.


\epsfig{0.6}{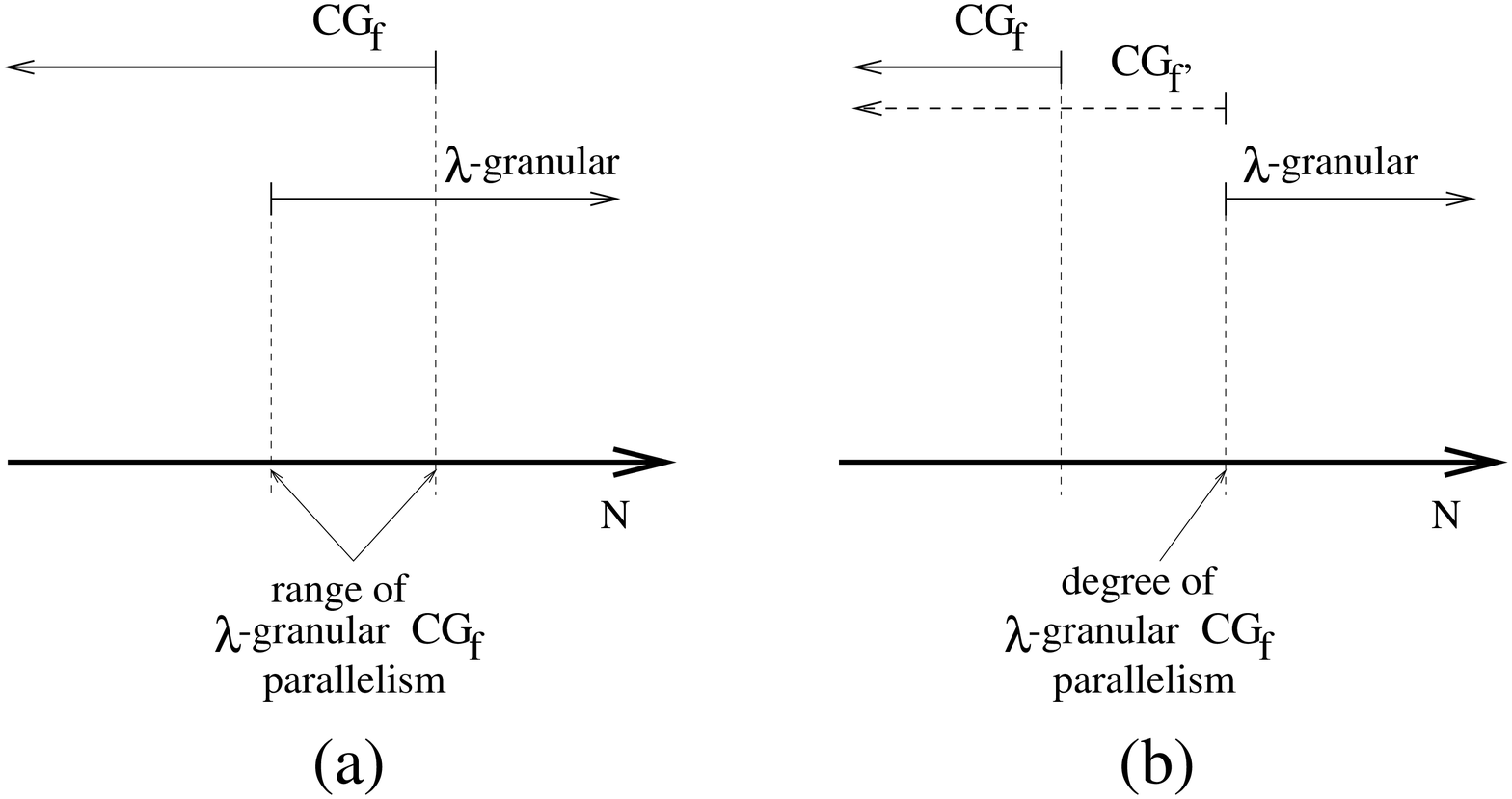}{$\lambda$-granular \cgf\ execution with (a) $f = f'$, and (b) $f < f'$.}
{fig.cgf}


\subsection{Determining the Degree of Partitioned Parallelism}
\label{sec.degree}
We now demonstrate how our multi-dimensional resource model and our quantification
of operator granularity can be employed to derive the degree of partitioned parallelism
for query operators.
Assuming zero communication costs, the \tsr\ and \ssr\ resource requirements of an
operator \op\  are  described by a $d$-dimensional work vector $\overline{W}$ and an 
$s$-dimensional demand vector $\overline{V}$ whose components can be derived from
system parameters and traditional optimizer cost models~\cite{saclp:sigmod79}.
By definition, the processing area of the operator $\procarea(\mop)$
is simply the sum of $\overline{W}$'s components, i.e.,
$\procarea(\mop) = \sum_{i=1}^{d}W[i]$.
Similarly, the \ssr\ grain size  $\maxmem(\mop,N)$ can be estimated using traditional 
optimizer cost models and statistics kept in the database catalogs on the distribution
of attribute values~\cite{ioan:vldb93,pi:vldb96}.
Finally, we can estimate the communication area $\commarea(\mop,N)$ using again
some standard cost models.
As such models are not ubiquitous in the database literature, we present one of them
for completeness, a simple
linear model of communication costs that has been adopted in previous 
studies of shared-nothing architectures~\cite{gmsy:orsaj93,wfa:vldb92,gi:sigmod96,gi:vldb97}
and validated on the Gamma research prototype~\cite{dgsb:tkde90}.
Specifically, if $D$ is the total size of the operator's input and output 
transferred over the interconnection network, then 
$\commarea(\mop,N) = \alpha\cdot N + \beta\cdot D$, where $\alpha$ and $\beta$ are 
architecture-specific parameters specified as follows:
(1) $\alpha$ is the {\em startup cost\/} for each operator clone, and 
(2) $\beta$ is the average time spent at the CPU and/or the network interface per
unit of data transferred (e.g., for packaging data tuples into network messages).
An experimental methodology for determining these communication cost parameters
is presented by Englert et al.~\cite{egh:sigmodrec95} in the context of NonStop
SQL/MP, a commercial shared-nothing database system from Tandem computers.
Their measurements on a $4$-site system show a cost of $\alpha = 0.5$ sec per
clone process that needs to be created and a cost of $\beta = 1.23$ sec/MByte 
for data repartitioning.
(They also note that $\alpha$ can be significantly reduced by re-using clone
processes.)
The following proposition is an immediate consequence of Definition~\ref{def.lcgf}.
\begin{prop}
\rm
Assuming a linear model for clone communication costs, the maximum allowable
degree of partitioned parallelism for a $\lambda$-granular \cgf\  
execution of operator \op\
is denoted by $N_{max}(\mop, f, \lambda)$ and is determined by the formula
\[
N_{max}(\mop, f, \lambda) = \max \{ \ts 1 \ts, \ts
                           \left\lfloor
                           \frac{f\cdot \procarea(\mop) -\beta\cdot D}{\alpha}
                           \right\rfloor \ts,\ts
                           \min \{\ms
                                  N: \maxmem(\op, N)\leq\lambda
                                \ms\}
                         \ts\} . \hspace*{1.5em}\csbbox
\]
\label{prop.degree}
\end{prop}

Proposition~\ref{prop.degree} provides a concise formula for determining
the degree of partitioned parallelism for all floating query operators,
based on our quantification of operator granularity and a concrete model
for communication costs.
It is important to note, however, that all our scheduling algorithms 
and analytical results (developed in Section~\ref{sec.sched}) are 
{\em independent\/} of the specifics of both the processing and the
communication models as well as of the
quantification of communication granularity being used.
In other words, our results are always valid for the  given degrees of 
partitioned parallelism for floating operators, regardless of the exact 
method used to derive these degrees.

%% file: sched.tex
\section{The Query Scheduling Algorithms}
\label{sec.sched}
In this section, we present our multi-resource scheduling algorithms for
parallel query execution plans. 
Our algorithms and analytical results assume that the 
degree of parallelism for query operators has been   fixed
based on system-wide granularity constraints (Section~\ref{sec.quant}).
We present the scheduling problems for parallel query plans in increasing order
of complexity\footnote{Note that all the scheduling problems formulated in this section are \np-hard.}:
\begin{enumerate}
\item
We demonstrate how our multi-dimensional resource model can account for 
concurrent execution and resource sharing among independent and pipelined operator
clones.
\item
We present a lower bound on the optimal parallel response time for a 
collection of {\em independent\/}  operators  and propose a fast algorithm that 
guarantees near-optimal schedules for such operators.
\item
We consider the operator scheduling problem in the presence of 
{\em pipelining\/} constraints  and propose a novel, provably near-optimal 
scheduling algorithm (termed \LevelSched) for independent physical 
operator pipelines.
\item
We propose heuristic extensions that allow \LevelSched\ to handle
{\em blocking\/} constraints and execution dependencies in logical operator 
pipelines and bushy query plans, as well as on-line task arrivals.
\end{enumerate}
We start by outlining some definitions and notational conventions that will be
used in the presentation.

\subsection{Notation and Definitions}
\label{sec.notation}
Table~\ref{tab.notation} summarizes the notation used in the development of our
query scheduling algorithms with a brief description of its semantics.
Detailed definitions of some of these parameters are given in the text.
Additional notation will be introduced when necessary.
In the rest of the paper, unless otherwise noted, all references to the
term ``operator'' imply a physical operator.

\begin{SingleSpace}

\begin{table}[tbh]
{\small
\begin{center}
\begin{tabular}{||c|l||}\hline\hline
{\bf Symbol} & {\bf Semantics} \\ \hline \hline
$P$            &  Number of system sites \\ \hline
$d$            &  Number of \tsr\ resources per site\\ \hline
$s$            &  Number of \ssr\ resources per site\\ \hline
$B_j$          &  System site (i.e., ``bin'') $j$ ($j=1,\ldots,P$)\\ \hline
$B_j^W$        &  Set of \tsr\ work vectors scheduled at site $B_j$\\ \hline
$B_j^V$        &  Set of \ssr\ demand vectors scheduled at site $B_j$\\ \hline
$T^{site}(B_j)$   &  Execution time for all operator clones at site $B_j$\\\hline 
$T^{par}(\msched, P)$   &  Response time for a query execution schedule SCHED on $P$ sites\\\hline \hline
$M$            &  Number of operators to be scheduled\\ \hline
$\mop_{i}$     &  Operator, e.g., \opscan, \opbuild\ ($i=1,\ldots,M$)\\ \hline
$N_{i}$        &  Degree of partitioned parallelism (number of clones) for $\mop_{i}$\\ \hline
$\overline{W}_{\mop_{i}}$ & Work vector for $\mop_{i}$ (including communication
			    costs for $N_{i}$ clones)\\ \hline
$\overline{V}_{\mop_{i}}$  & Demand vector for $\mop_{i}$\\ \hline
$T^{max}(\mop_{i}, N_{i})$ & Maximum execution time among the $N_i$ clones of $\mop_{i}$
                             while alone in system \\ \hline
$S$	       & Set of (floating) operator clones to be scheduled\\ \hline
$S_{C_i}$      & Set of (floating) clones in a physical operator pipeline $C_i$\\ \hline
$S^W$ ($S^V$)  & Set of  work (resp., demand) vectors for all clones in $S$\\ \hline
$S^{TV}$       & Set of {\em volume\/} (time $\times$ demand) vectors for all clones 
		 in $S$\\ \hline
$T^{max}(S)$   & Maximum stand-alone execution time among all operator clones in $S$\\ \hline
$l(\overline{v}), l(S^v)$  & Length of a vector $\overline{v}$ or set of vectors $S^v$\\\hline\hline
\end{tabular}
\end{center}
}
\caption{Notation}
\label{tab.notation}
\end{table}

\end{SingleSpace}

The collection of physical query operators to be scheduled is denoted by 
$\{ \mop_1,\ldots, \mop_M \}$ and $N_i$ denotes the fixed degree of partitioned
parallelism (i.e., number of clones) for operator $\mop_i$ (as determined by
communication and memory granularity constraints).
The work and demand vector for each operator $\mop_i$ are defined as earlier,
except that communication costs for the given degree of parallelism $N_i$
have already been factored in the operator's work requirements.
More specifically:
\begin{itemize}
\item
The demand vector $\overline{V}_{\mop_{i}}$  describes the total (normalized)
\ssr\ resource requirements of $\mop_i$. 
The components of $\overline{V}_{\mop_{i}}$ are computed  using architectural
parameters and database statistics, and are
independent of the degree of partitioned parallelism $N_i$.

\item
The work vector $\overline{W}_{\mop_{i}}$ describes the total
(i.e., processing and communication) \tsr\ resource requirements
of $\mop_{i}$, {\em given its degree of parallelism\/} $N_{i}$.
Using the notions of communication and processing area defined
in Section \ref{sec.partitioned}, the above is expressed as
$
\sum_{k=1}^{d}{W}_{\mop_{i}}[k] = \procarea(\mop_{i}) + \commarea(\mop_{i}, N_{i}) .
$
The individual components of $\overline{W}_{\mop_{i}}$ are computed using
architectural parameters and database statistics, as well as the
\ssr\ allotment for $\mop_i$ and a model for communication costs\footnote{The 
actual distribution of work among the vector's components is immaterial as far
as our model is concerned.}.
\end{itemize}

Each operator $\mop_i$ distributes its total \tsr\ and \ssr\ resource requirements 
$(\overline{W}_{\mop_{i}}, \overline{V}_{\mop_{i}})$  across $N_i$ (work vector, demand vector) pairs
corresponding to its clones.
The \tsr\  and \ssr\  requirements for each such clone are determined based on the
exact form of data repartitioning (e.g., range or hash) and statistical information
on the partitioning attribute typically kept in the DBMS catalogs~\cite{pi:vldb96}.
Note that, for the results presented in this paper, {\em no uniformity assumptions\/}
are made about the distribution of data and work/demand vectors among clones; this 
distribution can be arbitrary, possibly highly skewed, as long as the memory
granularity constraint $\lambda$ is observed.

Given  an operator clone with resource requirements $(\overline{W}, \overline{V})$
and a (stand-alone) execution time of $T = T^{seq}(\overline{W})$, we define the 
{\em volume vector\/} of the clone as the product $T\cdot\overline{V}$,
i.e., the resource-time product\footnote{The {\em volume\/} of an 
operator is defined as the product of the amount of resource(s) 
that the operator reserves during its execution
and its execution time.}
for the clone's execution~\cite{cm:spaa96,yc:vldbj93}.
We use $S^W$, $S^V$, and $S^{TV}$ to denote the set of work, demand, and volume
vectors (respectively) for a given set $S$ of floating operator clones.
The $W$, $V$, and $TV$ superscripts are used in this manner throughout the
paper.
For example, $B_j^W$ and $B_j^V$ denote the sets of work and demand vectors 
for all clones mapped to site $B_j$ by our scheduler.

The {\em length of an $n$-dimensional vector $\overline{v}$\/} is its
maximum component.
The {\em length of a set $S^v$ of $n$-dimensional vectors\/} is the
maximum component in the vector sum of all the vectors in $S^v$.
More formally,
\begin{eqnarray*}
l(\overline{v}) = \max_{1\leq k\leq n} \{ v[k] \} &\ts\ , \ts\ &
l(S^v)=\max_{1\leq k\leq n}\{\sum_{\overline{v}\in S^v}v[k]\}.
\end{eqnarray*}

The {\em performance ratio\/} of a scheduling algorithm is defined as the
ratio of the response time of the schedule it generates over that of
the optimal schedule.
All the scheduling problems addressed in this paper are 
non-trivial generalizations
of traditional multiprocessor scheduling~\cite{garey-johnson:79}
and, thus, they are clearly \np-hard.
Given the intractability of the problems, we develop polynomial-time
heuristics that are {\em provably near-optimal\/}, i.e., with a 
constant bound on the performance ratio.

Since the parallelization of rooted operators is pre-determined, our algorithms
are only concerned with the scheduling of floating operators.
Furthermore, they assume that
the degree of partitioned parallelism for all floating operators
has already been determined based on system-wide granularity 
constraints (Section~\ref{sec.quant}); for example, as  shown in 
Proposition~\ref{prop.degree}.

\subsection{Modeling Parallel Execution and Resource Sharing}
\label{sec.modelpar}
We present a set of extensions to the (one-dimensional) cost model of a traditional
DBMS based on the multi-dimensional resource usage formulation described in 
Section~\ref{sec.usage}.
Our extensions account for all forms of parallelism and quantify the effects of  
sharing \tsr\ and \ssr\ resources on the response time of a parallel execution.

\subsubsection{Partitioned and Independent Parallelism}
\label{sec.partindep}
For partitioned parallelism, the work and demand vectors of an operator 
$\mop_i = (\overline{W}_{\mop_{i}}, \overline{V}_{\mop_{i}})$
are partitioned among a set of $N_i$ independent operator clones, where
each clone executes on a single site and works on a portion of the 
operator's data.
%
%
Given such a partitioning 
$\{ (\overline{W}_{\mop_{i}}^{1},\overline{V}_{\mop_{i}}^{1}),(\overline{W}_{\mop_{i}}^{2},\overline{V}_{\mop_{i}}^{2}),\ldots,
(\overline{W}_{\mop_{i}}^{N_{i}},\overline{V}_{\mop_{i}}^{N_{i}}) \}$,
where $\sum_{k=1}^{N_{i}}\overline{W}_{\mop_{i}}^{k} = \overline{W}_{\mop_{i}}$
and $\sum_{k=1}^{N_{i}}\overline{V}_{\mop_{i}}^{k} = \overline{V}_{\mop_{i}}$,
a lower bound on the parallel execution time for $\mop_{i}$ is the maximum of the
sequential execution times of its $N_{i}$ clones; that is, the parallel execution 
time for $\mop_{i}$ is always greater than or equal to
\[
T^{max}(\mop_{i}, N_{i}) \ts=\ts
\max_{1\leq k\leq N_{i}}\{\ms T^{seq}(\overline{W}_{\mop_{i}}^{k}) \ms\}.
\]

By our definitions of the \tsr\ and \ssr\ resource classes, it is obvious that a collection
of operator clones\linebreak
$\{ (\overline{W}_{1},\overline{V}_{1}),(\overline{W}_{2},\overline{V}_{2}),\ldots,
(\overline{W}_m,\overline{V}_m) \}$
can be executed concurrently at some system site {\em only if\/}
$l(\sum_1^m\overline{V}_{i})\leq 1$; that is, the total \ssr\ requirements 
of the clones do not exceed the capacity of the site.
We call such clone collections {\em compatible\/}.
The following definition formalizes the notion of a schedule for independent operators
in the presence of \ssr\ resource constraints.
\begin{defn}
\rm
Given a collection of $M$ independent operators $\{ \mop_{i}, i=1, \ldots, M\}$
and their respective degrees of partitioned parallelism 
$\{N_{i}, i = 1, \ldots, M\}$, a {\em schedule\/}
is a partitioning of the $\sum_{i=1}^{M}N_{i}$ operator clones into a collection of compatible
subsets $S_1,\ldots, S_n$ followed by a mapping of these subsets to the set of available 
sites. \csbbox
\label{def.schedule}
\end{defn}

The effects of time-slicing the \tsr\  resources of a site among the clones in a 
compatible subset $S_i$ can be quantified as follows.
Let $S_i^W$ denote the set of work vectors for all clones in $S_i$.
Since all clones are running concurrently, the execution time for the clones in 
$S_i$ is determined by the ability to overlap the processing of \tsr\ resource 
requests by different clones.
Specifically,  under our model of preemptable resources described in Section~\ref{sec.usage},
the execution time for all the operator clones in $S_i$ is defined as
\[
T(S_i) \ts\ =\ts \max\{\ms \max_{\overline{W}\in S_i^W} \{ T^{seq}(\overline{W}) \}
             \ts\ ,\ts  l(S_i^W)  \ms\}.
\]
Intuitively, the above formula states that the execution time is determined by either the
slowest executing clone in $S_i$ ($\max_{\overline{W}\in S_i^W} \{ T^{seq}(\overline{W})$),
or the load imposed at the most heavily congested \tsr\  resource at the site
($l(S_i^W)$), whichever is greater.

\csxam
Consider two pairs of a \opselect\ and a \opbuild\  operator clones, whose work vectors
and stand-alone execution times are depicted in Figure~\ref{fig.shar}(a,b).
Assume that both clone pairs are compatible (i.e., their total memory requirement 
does not exceed the number of memory pages available at each site) and consider their
concurrent execution at a system site.
For the first \opselect/\opbuild\  pair, the total work  requirements can be ``packed'' 
within the response time of the \opselect\ clone; that is, because of the limited
overlap between CPU and disk processing during the \opselect, the CPU resource has 
enough ``slack''  to handle the concurrent \opbuild\ (Figure~\ref{fig.shar}(a)).
Thus, the  execution time of the first pair is simply the stand-alone response time
of the slower \opselect\ clone, i.e.,
$T(\{ \opselect, \opbuild \})  = T^{seq}(\overline{W}_{\tt select})$.
Note that, the rightmost part of Figure~\ref{fig.shar}(a) depicts the utilization
of the site's CPU and disk bandwidth over the period $T(\{ \opselect, \opbuild \})$,
under our ``uniform stretching'' assumption for \tsr\ resources (Assumption (A.2)).

On the other hand, when the second \opselect/\opbuild\  pair is executed concurrently, 
the CPU slack is not sufficient to handle the \opbuild\  within the response time
of the \opselect\ (Figure~\ref{fig.shar}(b)).
Thus, the CPU resource essentially becomes the bottleneck and the execution time
is determined by the total CPU work, i.e.,
$T(\{ \opselect, \opbuild \})  = l(\{\overline{W}_{\tt select} , \overline{W}_{\tt build} \}) 
 = W_{\tt select}[CPU] + W_{\tt build}[CPU]$. 
Again, the rightmost part of Figure~\ref{fig.shar}(b) depicts the ``stretching''
of the two operators' \tsr\ resource requirements over their  execution period.
Clearly, the bottleneck resource (i.e., CPU) is 100\% utilized.
\csexam

\begin{SingleSpace}

\epsfig{1.0}{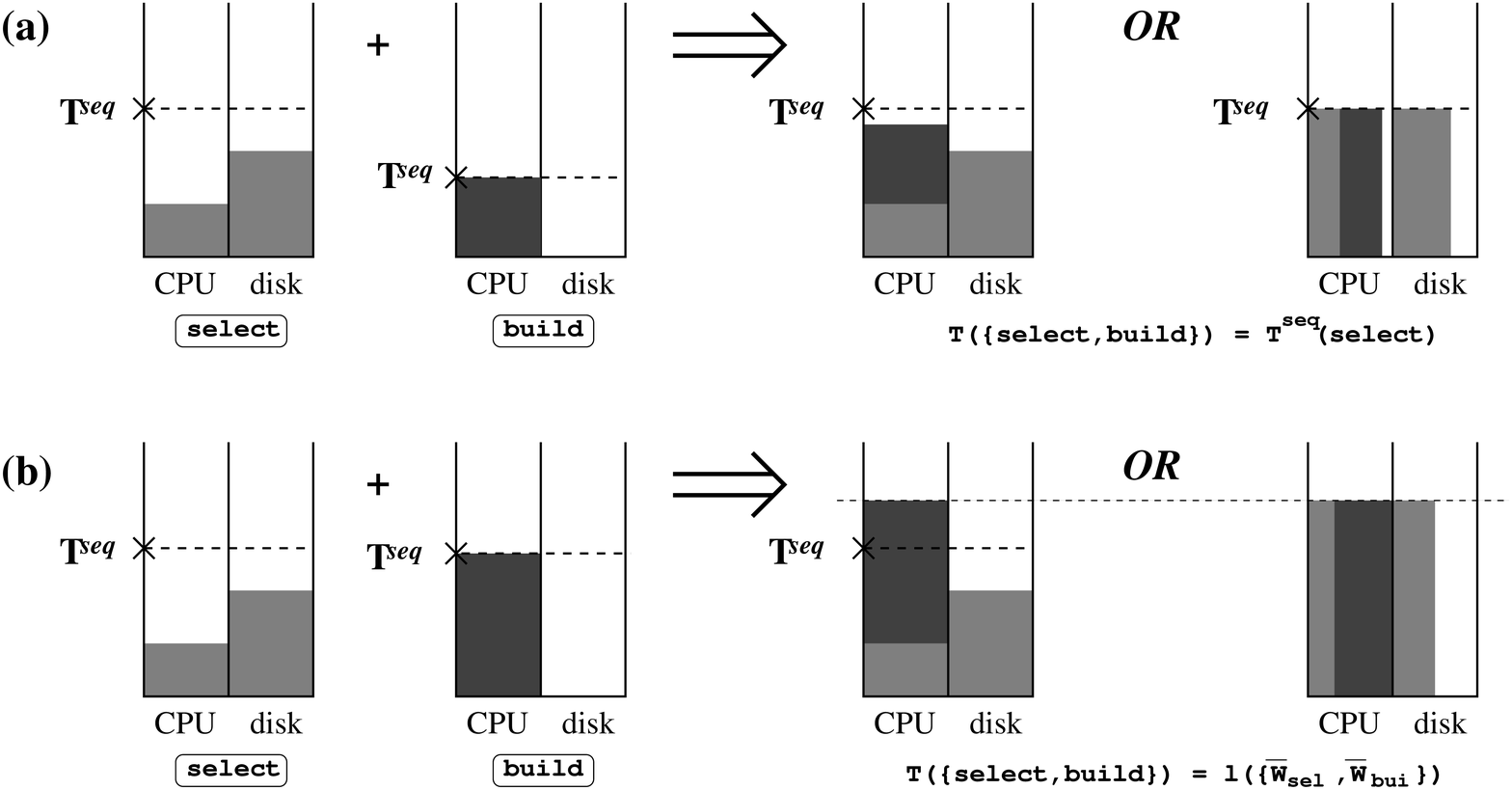}
{Packing the work vectors of compatible clones at a system site when
(a) the slowest execution time can accommodate the total \tsr\ requirements; and,
(b) a \tsr\ resource (CPU) becomes the bottleneck for the concurrent execution.}
{fig.shar}

\end{SingleSpace}

Let SCHED denote any given schedule for independent operators (as described 
in Definition~\ref{def.schedule}) and let $S(B_j)$ denote the collection of
compatible clone subsets mapped to system site $B_j, 1\leq j\leq P$ under SCHED.
Then, the total execution time for site $B_j$ is simply the sum of the execution
times for all compatible subsets in $S(B_j)$; that is,
%
%
\begin{equation}
T^{site}(B_j) \ts\ =\ts \sum_{S_i\in S(B_j)} T(S_i) \ts\ =\ts \sum_{S_i\in S(B_j)}
					       \max\{\ms \max_{\overline{W}\in S_i^W} \{ T^{seq}(\overline{W}) \} \ms\ ,\ms  l(S_i^W)  \ms\}.
\label{eqn.binheight}
\end{equation}
Finally, the response time of the schedule SCHED is obviously determined by the 
longest running system site; that is, 
$T^{par}(\msched, P) \ms\ =\ms\  \max_{1\leq j\leq P} \{\ms\  T^{site}(B_{j}) \ms\}$.

\subsubsection{Pipelined Parallelism}
\label{sec.pipelined}
Pipelined parallelism  introduces a {\em co-scheduling requirement\/} for 
query operators, requiring a collection of clones to execute in producer-consumer
pairs using fine-grain/lock-step synchronization.
The problems with load-balancing a pipelined execution have been
identified in previous work~\cite{graefe:surveys93}.
Compared to our model of a schedule for partitioned and independent parallelism
(Definition~\ref{def.schedule}), pipelined
execution constrains the placement and execution of compatible clone 
subsets to ensure that all the clones in a pipe run
concurrently --  they all start  and terminate at the same time~\cite{hm:vldb94}.
As we suggested in the previous section, this means that it is no longer
possible to schedule resources at one site independent of the others.
Compatible subsets containing clones from the same (physical) operator 
pipeline {\em must\/} run concurrently.
Furthermore, given that the scheduler does not modify the
query plan, scheduling a pipeline is an ``all-or-nothing'' affair:
either all clones will execute in parallel or none will.
The implications of pipelined parallelism for our scheduling problem
will be studied further in Section~\ref{sec.pipesched} where a 
near-optimal solution will be developed.

\subsection{Scheduling Independent Operators}
\label{sec.opsched}
%
In this section, we derive a lower bound on the optimal parallel execution time
of independent operators (i.e., operators not in any pipeline) that accounts 
for both \tsr\  resource requirements and \ssr\  resource constraints.
We then demonstrate that a heuristic based on Graham's LPT (Largest Processing
Time) {\em list scheduling\/} method~\cite{graham:siamjc69} 
can guarantee near-optimal schedules for such operators.

\begin{lem}
\rm
Let $\{\mop_i, i=1, \ldots, M \}$ be a collection of independent 
operators with respective degrees of partitioned parallelism 
$\{ N_i, i=1, \ldots, M \}$. 
Let $S$ be the corresponding collection of clones and let $T^{max}(S)$ be the
maximum stand-alone execution time among all clones in $S$; that is,
$T^{max}(S) = \max_{i=1,\ldots,M} \{ T^{max}(\mop_{i}, N_{i}) \}$.
If $T^{par}(\mopt, P)$ is the response time of the optimal execution on $P$
sites then $T^{par}(\mopt, P) \geq LB(S, P)$, where
\[
LB(S, P) \ts\ =\ts\
		    \max \{\ts
                         T^{max}(S)        \ts,\ts
                        \frac{l(S^{W})}{P} \ts,\ts
                        \frac{l(S^{TV})}{P}
                       \ts\} .  \ts\ts\csbbox
\]
\label{lem.newlb}
\end{lem}
As with all other theoretical results in this paper, 
the proof of Lemma \ref{lem.newlb} is given in Appendix~\ref{sec.proofs}.

The basic idea of our heuristic scheduling algorithm, termed \OpSched, is to construct
the partition of clones into compatible subsets (or, ``shelves'') incrementally, using a 
Next-Fit rule~\cite{cgj:survey84,cgjt:siamjc80}.
More formally, let $n_i$ denote the number of compatible clone subsets that have
already been mapped by \OpSched\  to site  $B_i$ and let  $S_{i,n_i}$ denote the
``topmost'' compatible subset in $B_i$, i.e., the shelf containing the most
recently scheduled clone at $B_i$.
(Initially, $n_i= 0$ and $S_{i,0} = \phi$ for all $i$.)
%
%
\OpSched\ scans the list of independent clones to be scheduled in 
non-increasing order of execution time.
At each step, the clone selected is placed in the site $B_i$ with the minimum
total execution time (i.e., bin ``height'') $T^{site}(B_i)$ (see 
Equation~(\ref{eqn.binheight})).
This placement is done as follows. 
If the clone can fit in the topmost compatible subset $S_{i,n_i}$ without 
violating \ssr\ capacity constraints, then add the clone to $S_{i,n_i}$ and
update $T^{site}(B_i)$ accordingly.
Otherwise, set $n_i = n_i +1$, place the clone by itself in a new topmost 
subset $S_{i,n_i}$, and set $T^{site}(B_i)$ accordingly. 
The following theorem establishes an absolute performance bound
of $d+2s+2$ for our  heuristic.

\begin{thm}
\rm
Given a set of {\em independent\/} operator clones $S$, \OpSched\ runs in 
time $O(|S|\log|S|)$ and produces a schedule $\msched$ with response time
\[
T^{par}(\msched, P) \ts < \ts d\cdot\frac{l(S^W)}{P} \ms+\ms 2s\cdot\frac{l(S^{TV})}{P}
             \ms+\ms 2\cdot T^{max}(S) \ts\leq\ts (d + 2s + 2)\cdot LB(S,P). \ts\ts\csbbox
\]
\label{thm.basels}
\end{thm}

Theorem~\ref{thm.basels} guarantees an 
{\em asymptotic performance bound\/}\footnote{An asymptotic performance bound
characterizes the behavior of an algorithm as the ratio of the optimal schedule
response time to the longest job processing time goes to infinity, i.e.,
when $\frac{T^{par}(\mopt,P)}{T^{max}(S)}\rightarrow\infty$.}
of $d+2s$ for \OpSched.
This bound gives us a feeling for the performance of the algorithm when
the optimal response time is much larger that the longest execution time of
all clones and is a better measure of performance when $|S|$ is 
large~\cite{cgjt:siamjc80,bs:siamjc83}.

Note that our scheduling algorithm combines the list scheduling method of 
Graham~\cite{graham:siamjc69} with the Next-Fit Decreasing Height (NFDH) shelf-based 
algorithm of Coffman et al.~\cite{cgjt:siamjc80}.
Essentially, the Next-Fit rule means that, at each step, only the topmost shelf
is examined for placing a new  clone.
Using the more exhaustive First-Fit Decreasing Height (FFDH) shelving 
strategy in place of NFDH, we can combine our proof methodology with that
of Coffman et al. to demonstrate a $d+1.7s$ asymptotic performance bound 
for independent operator scheduling.

\subsection{Scheduling with Pipelining Constraints}
\label{sec.pipesched}
The co-scheduling requirement of pipelined operator execution
introduces an extra level of complexity that \OpSched\ cannot
address, namely the problem of deciding whether a physical
operator pipeline is {\em schedulable\/} on a given number of sites.
Given a collection of clones in a pipeline, the schedulability  
question poses an \np-hard decision problem that essentially
corresponds to the decision problem of $s$-dimensional vector
packing~\cite{cgj:survey84}.
Thus, it is highly unlikely that efficient (i.e., polynomial time)
necessary and sufficient conditions for pipeline schedulability 
exist.
Note that no such problems were raised in the previous section, 
since the clones were assumed to be feasible (i.e., 1-granular)
and executing independently of each other.

In this section, we demonstrate  that $\lambda$-granularity 
(Definition~\ref{defn.lgranul}) with $\lambda < 1$ for all 
operator parallelizations can provide an easily
checkable sufficient condition for pipeline schedulability.
(We use the term {\em $\lambda$-granular pipeline\/} to describe an operator
pipeline in which all operator parallelizations are $\lambda$-granular.)
Once schedulability is ensured, balancing the work and memory requirements
of the pipeline across sites to minimize its response time still poses
an  \np-hard  optimization problem.
We  present a polynomial-time scheduling algorithm that 
is within a constant multiplicative factor of the response 
time lower bound for schedulable $\lambda$-granular pipelines.
Further, we demonstrate that, using a level-based approach, our 
methodology can be extended to provide a provably near-optimal 
solution for multiple, independent physical operator pipelines. 
%

\subsubsection{Scheduling a Single $\lambda$-granular Pipeline}
\label{sec.onepipe}
We present a near-optimal algorithm for scheduling a physical operator
pipeline $C$ consisting of $\lambda$-granular parallel operators,
where $\lambda < 1$.
Let $S_C$ denote the  collection of clones in $C$ and, as always,
let $S^W_C$ and $S^V_C$ be the corresponding sets of work and 
demand vectors, respectively.
Also, let $T^{max}(S_C)$ denote the maximum stand-alone execution time
among all clones in $S_C$.
Note that, by our definitions, the pipeline $C$ will require {\em at least\/}
$l(S^V_C)$ sites for its execution (remember that each site has a capacity
of $1$ along all \ssr\ dimensions).
The following lemma provides a sufficient condition for the schedulability
of  a $\lambda$-granular pipeline.
\begin{lem}
\rm
The number of sites required to schedule a $\lambda$-granular pipeline $C$
($\lambda < 1$) is always less than or
equal to $\frac{l(S^V_C)\cdot s}{1-\lambda}$.
Furthermore, this bound is tight.  \csbbox
\label{lem.ubnosites}
\end{lem}


Our heuristic, \PipeSched,  belongs to the family of list scheduling 
algorithms~\cite{graham:siamjc69}.
\PipeSched\ assumes that it is given a number of sites $P_C$ that is 
sufficient for the scheduling of $C$, according to the condition of
Lemma~\ref{lem.ubnosites}.
The algorithm considers the clones in $S_C$ in non-increasing order of their
{\em work density ratio\/} $\frac{l(\Wvec_i)}{l(\Vvec_i)}$.
At each step, the clone under consideration is placed in the site with the
least amount of work
that has sufficient \ssr\ resources to accommodate it;
that is, clone $(\overline{W}_i, \overline{V}_i)$ is mapped
to site $B$ such that $l(B^W)$ is minimal among all sites $B_j$ such
that $l(B^V_j\cup\{\Vvec_i\}) \leq 1$.
The full \PipeSched\ algorithm is depicted in Figure~\ref{fig.pipesched}.

\begin{SingleSpace}

\alg{
\vspace{2ex}
{\bf\underline{Algorithm \PipeSched($C$, $P_C$)}}

\small
{\bf Input:} A set of $\lambda$-granular pipelined operator 
	     clones $S_C$  and a set of
             $P_C$ sites, where $P_C\geq \frac{l(S^V_C)\cdot s}{1-\lambda}$
             (see Lemma~\ref{lem.ubnosites}).

{\bf Output:} A mapping of the clones to sites that does not violate \ssr\ resource
              constraints. (Goal: Minimize response time.)

\begin{enumerate}
\item
let $L = < (\Wvec_1, \Vvec_1),\ldots, (\Wvec_N, \Vvec_N) >$ be the list of 
all clones in {\em non-increasing\/} order of $\frac{l(\Wvec_{i})}{l(\Vvec_{i})}$.
\item
for $k = 1$ to $N$ do
\begin{enumerate}
\item[2.1.]
let $SB_k = \{ B_j: l(B^V_j\cup\{\Vvec_k\}) \leq 1 \}$, i.e, the set of
system sites with sufficient free \ssr\ resources to accommodate  the $k^{th}$ clone.
\item[2.2.]
let $B\in SB_k$ be a site such that
$l(B^W) = \min_{B_j\in SB_k}\{ l(B^W_j) \}$.
\item[2.3.]
place clone $(\Wvec_k, \Vvec_k)$ at site $B$ and 
set $B^W = B\cup \{ \Wvec_k \}$, $B^V = B\cup \{ \Vvec_k \}$.
\end{enumerate}
\end{enumerate}
\vspace{2ex}
}
{1.0}{Algorithm \PipeSched}{fig.pipesched}

\end{SingleSpace}

\csxam
Consider a simple physical operator pipeline $C$ comprising $N=4$ clones with 
work (CPU+disk)  and memory demands as follows:
$(\Wvec_1, \Vvec_1)= ([10,5], [0.2])$, $(\Wvec_2, \Vvec_2)= ([15,0], [0.3])$,
$(\Wvec_3, \Vvec_3)= ([7,9], [0.3])$, and $(\Wvec_4, \Vvec_4)= ([2,10], [0.35])$.
These clones are obviously $\lambda$-granular with $\lambda = 0.35$.
We now demonstrate how our \PipeSched\ algorithm would schedule this pipeline
on a set of $P_C = 2$ sites.
First, observe that $C$ is clearly schedulable on $2$ sites, since
$\frac{l(S^V_C)}{1-\lambda} = \frac{0.2+0.3+0.3+0.35}{1-0.35} = 1.77 \leq 2$.
Also, note that the clone subscripts denote the ordering of the clones in
non-increasing work density; that is,
$\frac{l(\Wvec_1)}{l(\Vvec_1)} = \frac{10}{0.2}=50 \geq
\frac{l(\Wvec_2)}{l(\Vvec_2)} = 50 \geq
\frac{l(\Wvec_3)}{l(\Vvec_3)} = 30 \geq
\frac{l(\Wvec_4)}{l(\Vvec_4)} = 28.58$.
(The tie between the first two clones is broken arbitrarily.)

Figure~\ref{fig.psched} depicts the series of steps taken by \PipeSched\ and
the final execution schedule produced.
Note that, even for this very simple example, \PipeSched\ is able to effectively
balance the work and memory requirements of the clones across the two system sites.
\csexam

\begin{SingleSpace}

\epsfig{1.0}{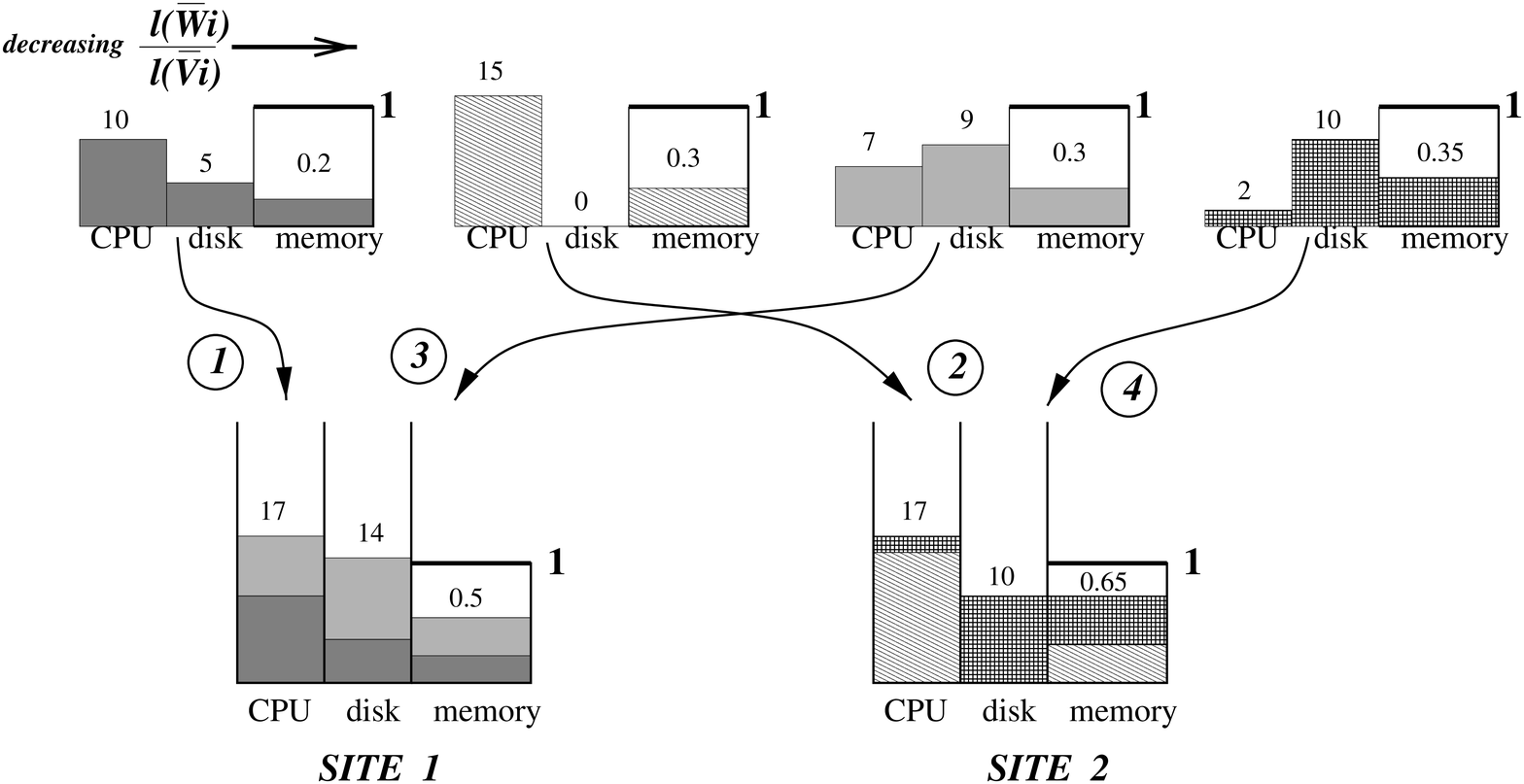}
{An example schedule produced by algorithm \PipeSched.}
{fig.psched}

\end{SingleSpace}

The following theorem establishes an asymptotic upper bound of
$d (1 + \frac{s}{1 - \lambda})$ on the worst-case
performance ratio of our algorithm.

\begin{thm}
\rm
Given a $\lambda$-granular pipeline $C$ ($\lambda < 1$) \PipeSched\ runs
in time $O(|S_C|\log|S_C|)$ and produces a schedule $\msched$ with response
time
\[
T^{par}(\msched, P_C) \ts\leq\ts d(1 + \frac{s}{1 - \lambda}) \cdot \frac{l(S^W_C)}{P_C} 
		      + T^{max}(S_C) 
		\ts\leq\ts [d(1 + \frac{s}{1 - \lambda}) +1] \cdot LB(S_C,P_C). \ts\ts\csbbox
\]
\label{thm.pipesched}
\end{thm}

Note that the volume term of our lower bound (Lemma~\ref{lem.newlb})
does not come into the expression for the performance bound of
\PipeSched.
This is because, by definition, all the clones in $S_C$ must execute
in parallel and thus $l(S^V_C) \leq P_C$. 
Thus, for a single pipeline, we always have 
$\frac{l(S^{TV}_C)}{P_C}\leq T^{max}(S_C)\cdot\frac{l(S^{V}_C)}{P_C}\leq T^{max}(S_C)$; that is, the maximum clone execution time always dominates the average
volume for a single physical operator pipeline.

The bound established in Theorem~\ref{thm.pipesched} clearly captures 
the granularity tradeoffs identified in Section~\ref{sec.partitioned}.
Increasing the degree of partitioned parallelism decreases both $T^{max}(S_C)$ and
$\lambda$, thus  allowing for a better asymptotic bound on the ratio and
a smaller additive constant.
On the other hand, it also increases
the total amount of work $l(S^W_C)$ because of the overhead of parallelism.
The importance of such work-space tradeoffs for parallel query processing
and optimization has been stressed in earlier work as well~\cite{hfv:sigmodrec96}.

\subsubsection{Scheduling Multiple Independent $\lambda$-granular Pipelines}
\label{sec.multipipe}
The basic observation here is that our \PipeSched\ algorithm presented
in the previous section can be used to schedule any collection
of independent pipelines as long as schedulability is guaranteed
by Lemma~\ref{lem.ubnosites}.

Our algorithm for scheduling multiple independent pipelines uses
a Next-Fit Decreasing Height (NFDH) policy~\cite{cgjt:siamjc80}
in conjunction with Lemma~\ref{lem.ubnosites} to identify pipelines
that can be scheduled to execute concurrently on $P$ sites 
(i.e., in one layer/``shelf''  of execution).
\PipeSched\ is then used for determining the execution schedule 
within each layer.
The overall algorithm, termed \LevelSched,  is formally outlined 
in Figure~\ref{fig.levelsched}.

\begin{SingleSpace}

\alg{
\vspace{2ex}
{\bf\underline{Algorithm \LevelSched($\{ C_1,\ldots, C_N\}$, $P$)}}

\small
{\bf Input:} A set of $\lambda$-granular operator pipelines 
             $\{ C_1,\ldots, C_N\}$ and a set of $P$ sites.

{\bf Output:} A mapping of clones to sites that does not violate \ssr\ 
             resource constraints or pipelining dependencies. 
	     (Goal: Minimize response time.)

\begin{enumerate}
\item
Sort the pipelines in non-increasing order of $T^{max}$, i.e.,
let $L = < C_1,\ldots, C_N >$, where
$T^{max}(S_{C_1})\geq\ldots\geq T^{max}(S_{C_N})$.
\item
Partition the list $L$ in $k$ {\em maximal schedulable sublists:}
$L_1 = <C_1,\ldots, C_{i_1}>$, $L_2 = <C_{i_1+1},\ldots, C_{i_2}>$,
$\ldots$, $L_k = <C_{i_{k-1}+1},\ldots, C_{N}>$ 
based on Lemma~\ref{lem.ubnosites}. 
That is,
\[
l( \cup_{C\in L_j} S^V_C ) \ms\leq\ms \frac{P (1-\lambda)}{s}
\mbox{\hspace{0.2cm} and \hspace{0.2cm}}
l( (\cup_{C\in L_j} S^V_C) \cup S^V_{C_{i_j+1}} )
                     \ms>\ms \frac{P (1-\lambda)}{s},
\mbox{\hspace{0.2cm} for all $j=1,\ldots,k-1$.}
\]
\item
for $j=1,\ldots, k$ do
\begin{enumerate}
\item[3.1.]
call \PipeSched( ($\cup_{C\in L_j}C$), $P$ )
\end{enumerate}
\end{enumerate}
\vspace{2ex}
}
{1.0}{Algorithm \LevelSched}{fig.levelsched}

\end{SingleSpace}

The following theorem gives an upper bound on the worst-case performance
ratio of \LevelSched.
Note that the co-scheduling requirement for the clones in a pipe implies
that the total volume for all the clones in $\{ C_1,\ldots, C_N\}$ is
$l(S^{TV}) = l( \sum_{1}^{N} T^{max}(S_{C_i})\cdot
                             \sum_{\vvec\in S^V_{C_i}}\vvec )$,
since any clone in $C_i$ will require its share of ss resources for
at least $T^{max}(S_{C_i})$ time.
The lower bound in Lemma~\ref{lem.newlb} holds using the above definition
of volume.

\begin{thm}
\rm
Given a collection of $N$ independent $\lambda$-granular pipelines 
($\lambda < 1$) comprising a set of clones $S$,
\LevelSched\ runs in time $O(N|S|\log P|S|)$ and produces a schedule
$\msched$ with response time
\[
T^{par}(\msched, P) \ts<\ts d^2(1 + \frac{s}{1 - \lambda}) \cdot \frac{l(S^W)}{P}
              \ms+\ms \frac{2s^2}{1-\lambda}\cdot \frac{l(S^{TV})}{P}
              \ms+\ms T^{max}(S)
	      \ts\leq\ts [d^2(1 + \frac{s}{1 - \lambda}) + \frac{2s^2}{1-\lambda} +1]
		   \cdot LB(S, P). \ts\ts\csbbox
\]
\label{thm.levelsched}
\end{thm}

It is important to note that the lower bound estimated in Lemma~\ref{lem.newlb}
will, in most cases, significantly underestimate the optimal response time since
it assumes that 100\% utilization of system resources is always possible 
{\em independent of the given task list\/}.
Thus, the quadratic multiplicative constants in Theorem~\ref{thm.levelsched}
reflect only a worst case that is rather far from the average.
This fact has been verified through extensive experimentation using
cost model computations in our earlier work~\cite{gi:vldb97}.

\subsection{Handling Blocking Constraints and On-Line Task Arrivals}
\label{sec.treesched}
Our discussion so far has focused on the case of scheduling a collection of 
one or more {\em independent\/} physical operator pipelines.
In this section, we propose heuristic extensions that allow our scheduling
algorithms to handle {\em blocking constraints\/} (and the corresponding 
execution dependencies (Section~\ref{subsec.definitions})) that arise in 
{\em logical\/} operator (e.g., hash-join) pipelines and bushy query 
execution plans.
Our proposed extensions deal with the two key aspects of query operator
scheduling, namely (1) determining the degree of partitioned parallelism 
(i.e., number of clones) for interdependent operators, and  
(2) mapping operator clones to sites so that execution dependencies are 
respected.

\vspace{.4em}\noindent 
$\bullet$ {\bf Determining the Degree of Partitioned Parallelism.}
Blocking parent-child operator pairs with a disk- or memory-materialization 
dependence are always constrained to run on the exact same set of system 
sites.
This is the case, for example, with a \opstore\ operator at the end of a query 
task  and the \opscan\ (or \opselect) operator in its parent query task 
(disk-materialization dependence) or  a \opbuild\  operator and its parent 
\opprobe\ (memory-materialization dependence).
In such scenarios, determining the degree of parallelism for the child operator
independently of its  parent may result in poor parallelization decisions.
As an example, the amount of work for a \opbuild\  depends on the inner/build
relation size whereas the amount of work for a \opprobe\  depends primarily 
on the size of  the outer/probe relation.
Fixing the degree of parallelism for a hash-join based on just one of these two 
operations is probably a bad idea, since the inner and outer relation sizes can
be vastly different.

Given such a pair of execution-dependent operators $\mop_1$ (parent) and $\mop_2$ (child), 
we propose setting their (common) degree of partitioned parallelism based on an extension
of the definition for $\lambda$-granular \cgf\ executions (Definition~\ref{def.lcgf}).
The basic idea is to use the total processing area $\procarea(\mop_1) + \procarea(\mop_2)$ 
and total communication area $\commarea(\mop_1, N) + \commarea(\mop_2, N)$  of the operator
pair in defining the communication granularity of the execution,  while ensuring that the
memory granularity constraint $\lambda$ is enforced for the (floating) child 
operator $\mop_2$.
More formally, given the system-wide granularity parameters $f$ and $\lambda$, 
we set the degree of parallelism of $\mop_1$ and $\mop_2$ to the maximal $N$
such that 
$(\commarea(\mop_1, N) + \commarea(\mop_2, N)) \leq f' \cdot 
(\procarea(\mop_1) + \procarea(\mop_2))$, where $f'$ is the minimum value larger than or
equal to $f$ such that $\maxmem(\mop_2,N)\leq\lambda$.
Combining the work demands allows more effective parallelization decisions to be made
for execution-dependent operator pairs.

\vspace{.4em}\noindent 
$\bullet$ {\bf Mapping Clones to Sites.}
Scheduling operator clones from arbitrary bushy query execution plans must ensure
that all execution dependencies specified by the blocking edges in the corresponding 
query task tree (Figure~\ref{fig.opertree}(c)) are satisfied.
Our \LevelSched\ algorithm can be readily extended to handle 
{\em disk-materialization dependencies\/} by simply ensuring that the (sorted) ready
list of tasks $L$ always contains the collection of query tasks that are ready for 
execution, i.e., they are not blocked waiting for the materialization of some intermediate
result on disk from some other
(descendant) task in the task tree.
For {\em memory-materialization dependencies}, care must be taken to ensure that 
when memory-dependent collections of query tasks are present in adjacent levels of
the task tree (e.g., a pipeline of \opprobe s  and all its child \opbuild\ pipelines)
then (1) the clones of all child tasks  are executed in parallel, and (2) the clones
of all parents are executed {\em immediately afterwards\/}  using the exact same 
sites and memory resources.
Given this very tight coupling of memory-dependent physical operators, we propose 
treating such parent-child pairs as a single, ``combined'' operator with a work 
vector that captures the total \tsr\ requirements for both phases.
(Of course, the \ssr\ demand vector remains the same throughout the execution of
the combined operator.)
This heuristic method tries to account for the tight execution dependencies between 
memory-dependent operator pairs and achieve better overall load-balancing by
essentially ``merging'' the two corresponding execution shelves into one.
As an example, in the query task of Figure~\ref{fig.opertree}(c), our proposed
method merges the \opbuild\ pipelines T3 and T4 into their parent \opprobe s
in T5.
%
%
All this is achieved  by modifying \LevelSched\ to produce a new
algorithm \TreeSched\ as follows 
(see Figure~\ref{fig.levelsched}):
\begin{enumerate}
\item
Any sibling physical operator pipelines with a {\em memory-materialization dependence\/} 
to their parent in the task tree are merged into their respective parent operators in 
the parent task (as described above).
This essentially gives rise to a ``coarser'' parent pipeline that is treated as 
a unit, i.e., the way individual pipelines are treated in \LevelSched.
For the purposes of this algorithm, assume that the term `pipeline' is interpreted
as such a unit.
%

\item
Initially, the input set of pipelines $\{C_1, C_2, \ldots \}$
contains exactly the tasks at the leaf nodes of the query task tree.

\item
After Step 3.1, determine the set of tasks {\bf C}  that have been enabled 
(i.e., are no longer blocked) because of the last invocation of
\PipeSched.
If {\bf C}$\neq \emptyset$, then merge the tasks in {\bf C} into the
ready list $L$ and go to Step 2.
Otherwise, continue with the next invocation of \PipeSched.
\end{enumerate}

\vspace{.4em}
The exact same idea of dynamically updating and partitioning the ready list
$L$ can be used to handle {\em on-line\/} task arrivals in a dynamic or 
multi-query environment.
Basically, newly arriving query tasks are immediately merged  into $L$ to 
participate in the partitioning of $L$ into schedulable sublists right
after the completion of the current execution layer.
Thus, our layer-based approach provides a uniform scheduling framework 
for handling intra-query as well as inter-query parallelism.

Deriving theoretical performance bounds for our scheduling algorithms 
in the presence of blocking constraints and execution dependencies is 
a very difficult problem that continues to  elude our efforts.
The difficulty stems from the interdependencies between different execution
layers: 
scheduling decisions made at earlier layers can impose data placement and
execution constraints on the layers that follow.
We leave this problem open for future research.

%% file: zsim.tex
\section{Performance Evaluation Using Detailed Simulation}
\label{sec.zsim}
In our earlier work~\cite{gi:vldb97}, we have presented the results of several
experiments conducted using  cost model computations for various
query operators in order to examine the {\em average-case performance\/}
of our algorithms compared to the optimal solution.
These results have demonstrated that the average-case performance
of our multi-resource scheduling algorithms is  much closer to optimal
than what would be predicted by the pessimistic worst-case bound 
of Theorem~\ref{thm.levelsched}.
Thus, on the average, our algorithms produce very effective, near-optimal
packings of memory and work requirements over the system sites.

In this paper, we present a performance study of  our parallel query 
scheduling  algorithms conducted using a detailed simulation model for 
a shared-nothing parallel database system~\cite{brown:prpl94}.
The simulator is written in the CSIM/C++ process-oriented simulation 
language~\cite{schwetman:mcc90} and is based on the architecture of the
Gamma parallel database machine~\cite{dgsb:tkde90}.

Figure~\ref{fig.expts.zsim}(a)  gives a high-level overview of 
our experimentation procedure.
Briefly, the scheduling algorithms investigated in our simulation study
receive as input
a set of query execution plans along with  schema and system configuration
information (e.g., declustering of base relations, sizes of intermediate 
results, disk and CPU characteristics) and produce as output an execution 
schedule (i.e., a mapping of query plan operators onto system sites).
The scheduling algorithm makes its mapping decisions off-line,
as would be the case if it were part of a complete query optimizer,
based on its input and (possibly) on cost model computations.
For example, our multi-resource scheduler  makes its scheduling choices
using a cost model to determine the work and demand vectors for operator 
clones and taking the componentwise sum of vectors mapped to a specific
site to estimate the expected \tsr\ and \ssr\ resource loads at that
site.
The effectiveness of the execution schedule produced is determined by 
``feeding''  the schedule  to our simulator which {\em actually executes\/}
it on top of our detailed model of a shared-nothing  parallel database 
system and outputs the response time of the execution.


\begin{SingleSpace}

\epsfig{1}{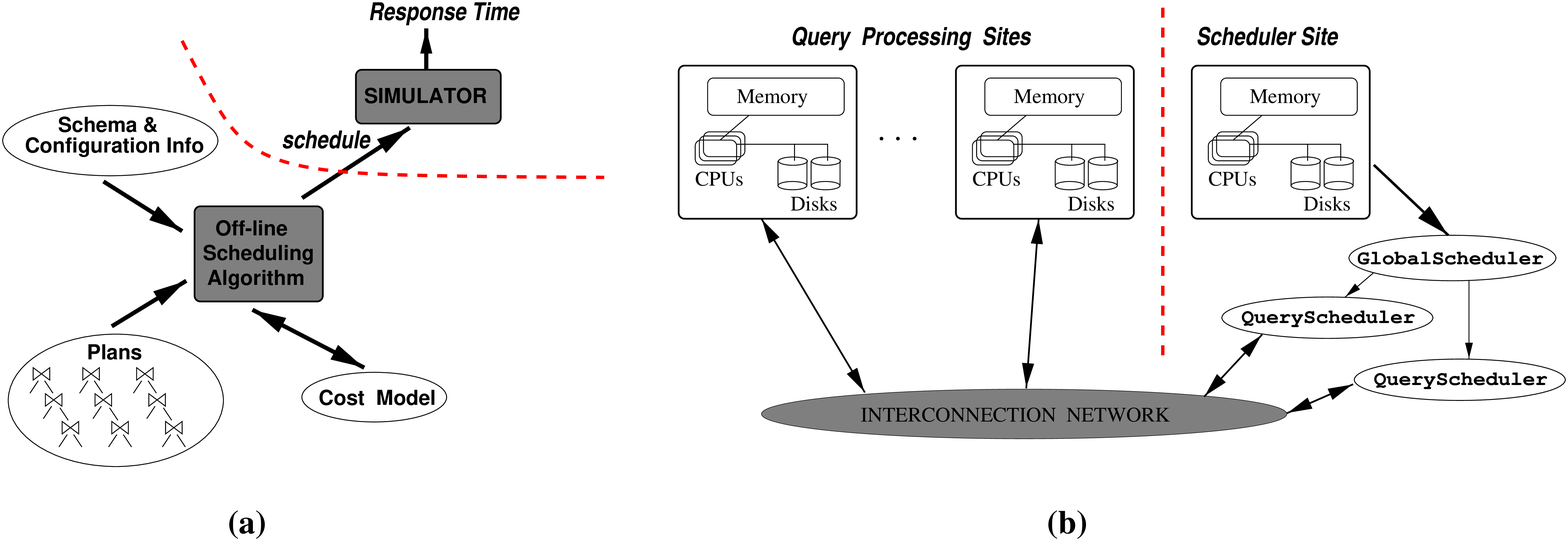}{(a) Experimentation procedure. 
(b) Simulator query processing architecture.}
{fig.expts.zsim}

\end{SingleSpace}

The following section describes the various components of our simulation 
model in detail.
We then go on to describe our experimental testbed and discuss the
simulation results.

\subsection{Execution Environment}
\label{sec.env.zsim}

\subsubsection{Simulator Query Processing Architecture}
\label{sec.qproc.zsim}
Our simulator models a typical shared-nothing system that consists of a 
collection of {\em query processing sites}, each comprising one or more 
CPUs, memory, and one or more disk drives.
The sites use an interconnection network for all communication.
Query plans are submitted for execution from external terminals to a  \gsched\
process running on a dedicated {\em scheduler site\/} that lies outside the set of 
query processing sites in the system.
The \gsched\  determines a mapping of plan operators to sites, allocates 
the appropriate resources for the plan, and finally initiates a \qsched\  process
on the scheduler site that is responsible for executing the plan to completion.
For each operator to be executed, the \qsched\  creates an execution thread 
on every site executing a clone of the operator. 
For each such thread,  appropriate communication channels are set up for the exchange 
of control messages between the \qsched\  and the thread (e.g., transition from
the \opbuild-phase to the \opprobe-phase of a join pipeline, split table broadcasting, and
``clone done'' messages)  and data messages between threads of neighboring operators in
the plan.
When a plan is run to completion the corresponding \qsched\  dies and a reply 
message is sent to the submitting terminal.
The \gsched\  may then choose to activate another query plan that had been 
waiting for some system resources to be released.
Figure~\ref{fig.expts.zsim}(b)  provides a high-level view of the query processing 
architecture in our simulation environment.
It is the same one assumed for our analytical results (Figure 
\ref{fig.sysmodel}) with respect to where plans are to run (to the left
of the vertical dashed line), complemented with one more site where the
scheduler runs to produce the appropriate schedule (to the right of the
vertical dashed line).

\subsubsection{Hardware and Operating System Characteristics}
\label{sec.hw.zsim}
Query processing sites comprise one or more CPUs, a buffer pool of 8 KByte pages,
and one or more disk drives.
CPU cycles are allotted  using a round-robin policy with a 5 msec timeslice.
The buffer pool models  a set of main memory page frames whose replacement is controlled 
by a 3-level LRU policy extended with {\em ``love/hate''\/} hints~\cite{hclm:tkde90}.
In addition, a memory reservation mechanism under the control of the \gsched\ process
allows memory to be  reserved for a particular query operator.
This mechanism is employed to ensure that hash table page frames allocated to join operators 
will not be stolen  by other operators.

The simulated disks model a Seagate Cheetah 36  Model ST136403FC 
(36.4 GByte, 3.5'')~\cite{cheetah:spec00}, a ``state-of-the-art'' SCSI disk drive.
This disk drive provides a 1024~Kbyte cache that is divided into sixteen 64~KByte cache
contexts to effectively support up to sixteen concurrent sequential prefetches.
In the disk model, which slightly simplifies the actual operation of the disk,
the cache is managed as follows. 
Each I/O request, along with the required page number, specifies whether or not 
prefetching is desired.
If prefetching is requested, eight additional pages  are read from the disk into a 
cache context as part of transferring the originally requested page
from disk to memory.
Subsequent requests for one of the prefetched pages can then be satisfied without
incurring an I/O operation.
A simple round-robin policy is used  to allocate cache contexts if the 
number of concurrent prefetch requests exceeds the number of available cache
contexts.
The disk queue is managed using an elevator algorithm~\cite{silberschatz-galvin:98}.

Our simulator models the interconnection network as a set of point-to-point connections
between every pair of system sites.
The speed of each point-to-point link is set to 200~Mbits/sec, which ensures that 
network speed is not the bottleneck in our experiments.
%
%
(This is in agreement with most real  parallel systems (e.g., iPSC/2, Hypercube, Paragon),
where the interconnect rarely is the bottleneck resource during parallel processing.)
However, the details of the particular interconnect technology are not modeled.
The cost of message-passing is captured as follows: our simulator charges for the
time-on-the-wire (depending on network latency, message size, and network bandwidth) 
as well as CPU instructions for networking protocol operations, which consist of
(1) a fixed cost per message, and (2) a per-byte cost based on the size of the message.
The CPU costs for messages are paid both at the sender and at the receiver.
Table~\ref{tab.params.zsim} summarizes the execution environment
parameters employed in our simulations.
The CPU instruction counts for various database operations are based on 
earlier simulation studies~\cite{mehta:phdthesis94,pi:vldb96}.

\begin{SingleSpace}

\begin{table}[ht]
\vskip 5pt
\hbox{
{\small
\begin{minipage}{0.5\hsize}
\sloppy
\hfil
\begin{center}
\begin{tabular}{||l|r||} \hline\hline
{\bf Configuration/Site Parameters}  	&      {\bf Value} 	\\ \hline\hline
Number of Sites                	&  16 -- 96  	\\ \hline
CPUs per Site                  	&  1 $\times$ 100 MIPS  \\ \hline
Disks per Site			&  2 $\times$ Cheetah-36 \\ \hline
Memory per Site                	&  64, 96, 128 MB  \\ \hline
Page Size		      	&  8 KB		\\ \hline
Network Bandwidth               &  200 Mbits/sec\\ \hline
Minimum Wire Delay		&  0.1 msec 	\\ \hline
Disk Seek Factor		&  0.1414	\\ \hline
Disk Rotation Time		&  5.98 msec	\\ \hline
Disk Settle Time		&  0.5 msec	\\ \hline
Disk Transfer Rate		&  29 MB/sec	\\ \hline
Disk Cache Context Size		&  8 pages	\\ \hline
Disk Cache Size 		&  16 contexts  \\ \hline
Disk Cylinder Size		&  560 pages	\\ \hline
Buffer Manager			&  3-level LRU	\\ \hline\hline
\end{tabular}
\end{center}
\hfil
\end{minipage}

\begin{minipage}{0.5\hsize}
\sloppy
\hfil
\begin{tabular}{||l|r||} \hline\hline
{\bf CPU Cost Parameters}             &    {\bf No. Instr.} \\ \hline\hline
Initiate Select			& 20000 \\ \hline
Initiate Join			& 40000 \\ \hline
Initiate Store			& 10000 \\ \hline
Terminate Select		& 5000 	\\ \hline
Terminate Join			& 10000 \\ \hline
Terminate Store			& 5000 	\\ \hline
Read Tuple			& 300 	\\ \hline
Write Tuple into Output Buffer	& 100	\\ \hline
Probe Hash Table		& 200	\\ \hline
Insert Tuple in Hash Table	& 100	\\ \hline
Hash Tuple Using Split Table	& 500	\\ \hline
Apply a Predicate		& 100	\\ \hline
Copy 8K Message to Memory	& 10000 \\ \hline
Message Protocol Costs		& 1000  \\ \hline\hline
\end{tabular}
\hfil
\end{minipage}
}
}
\caption{Simulation Parameter Settings}
\label{tab.params.zsim}
\vskip 5pt
\end{table}

\end{SingleSpace}

\subsection{Experimental Testbed and Methodology}
\label{sec.testbed.zsim}
\noindent
{\bf Scheduling Algorithms.}
Our experiments have focused on three different scheduling algorithms:
\begin{itemize}
\item
\TreeSched: Our level-based, multi-resource  scheduling algorithm for query 
task trees, observing blocking constraints and execution dependencies across 
execution layers (Section~\ref{sec.treesched}).

\item
\hier: A one-dimensional, processor allocation algorithm based on the 
hierarchical method developed by Wolf et al.~\cite{wtcy:sigmetrics94,wtcy:tpds95}
for scheduling multiple parallel queries on a set of processors.
(Since the original method did not account for pipelining dependencies or
memory constraints, we had to adapt it appropriately.)
Briefly, \hier\  ``collapses'' each right-deep join pipeline into a single 
parallel job and incrementally explores different possible allotments of 
processors to individual joins starting from a ``minimum work'' allotment.
At each step, a list schedule based on the current processor allotments is
constructed and is used to identify the pipeline with the maximum 
``processor waste''.
Then, the degree of parallelism for the maximum-waste job is increased by one
by giving one more processor to its most  time-consuming join.
This process is repeated until a list schedule with zero waste is produced,
and the best (i.e., shortest makespan) schedule found so far is chosen.

\item
\zsched: The ``standard'', simple query scheduling algorithm that has been
built-in the \gsched\  process of our simulator.
\zsched\ simply runs each join operator at the set of sites where its
left input (i.e., \opbuild-relation) is declustered.
\end{itemize}

Note that the search process for the \hier\ algorithm is always guaranteed to
terminate, since it will eventually reach an allotment where every pipeline is 
allotted all sites in the system, in which case the ``processor waste'' is 
zero for all jobs.
This, however, also means that, in the worst case, the \hier\  scheduler has
to build $O(N\cdot P)$ list schedules, where $N$ is the number of join
pipelines and $P$ is the number of sites.
Thus, the running time complexity of \hier\ can be significantly higher than 
that of our \TreeSched\ algorithm.

\vspace*{.5em}\noindent
{\bf Data Placement Strategies.} 
For all three scheduling algorithms, we experimented with various
possible data placement  strategies for the base relations of our
workload queries:

\begin{itemize}
\item
{\bf Declust}: Every base relation is horizontally partitioned
across all sites in the system.
\item
{\bf Declust-1/4}: Every base relation is horizontally partitioned
across the same subset of the system sites, comprising 1/4th of the 
entire system.
\item
{\bf NoDeclust}: A base relation resides on a {\em single\/} system
site, chosen randomly for every relation from the set of all sites.
\item
{\bf NoDeclust-1/4}: A base relation resides on a {\em single\/} system
site, chosen randomly for every relation  from the same subset of the
system sites, comprising 1/4th of the entire system.
\item
{\bf Random}: Both the degree of declustering and the actual sites 
storing a relation are selected randomly from the underlying set of
sites.
To avoid partitioning small relations over too many sites (which  is
typically avoided in real systems), we placed an upper bound on the degree
of declustering of a relation that is proportional to its size.
(The upper bound for the largest relation in the system is the total
number of sites.)
The actual degree of declustering was then randomly chosen between 1 
and that upper bound.
\item
{\bf QueryBasedDeclust}: The degree of declustering for a base relation is 
equal to the minimum number of sites required to hold the entire \opbuild\
hash table in main memory.
Furthermore, the actual set of sites given to a \opbuild\ relation is carefully
selected so that {\em it does not overlap\/} with any of the homes of the other
\opbuild\ relations in the same query.
\end{itemize}
Obviously, {\bf Declust} and {\bf NoDeclust} represent the two extreme choices
in the space of data declustering strategies.
The {\bf 1/4}-versions of these strategies were chosen to represent situations in 
which input queries will experience hotspots due to data placement choices.
Although the choice of the fraction of the system used for mapping data 
(1/4) was arbitrary, we believe that our experimental observations will remain 
valid for other system fractions.
The {\bf Random} placement strategy tries to relieve query hotspots by mapping 
relation fragments to randomly selected sites.
Such hotspots may still occur, however, based on the actual relations accessed 
by a query.
For example, it may very well happen that the homes of the four  \opbuild\
relations in a 4-join right-deep query  overlap at one or more system sites.
These sites will then be the bottlenecks for the \opbuild\ phase of the pipe.
It is exactly such bottlenecks that our final and more sophisticated 
data placement strategy, {\bf QueryBasedDeclust}, tries to avoid by explicitly
examining the set of queries to be executed.
Of course, determining the exact query workload may not always be possible (e.g.,
in an ad-hoc querying environment).

\vspace*{.5em}\noindent
{\bf Query Workloads.}
We executed all three scheduling algorithms over the same set of input 
right-deep hash-join query plans (submitted from the external terminals
at time 0), for each of the aforementioned  base relation placement
strategies.
Although our analytical results are completely general, we focused our
simulations on hash-joins only, since it is well-known that they offer
several advantages over other join algorithms in a parallel execution
environment, particularly because of the opportunities they present for
pipelining~\cite{clyy:vldb92,dgsb:tkde90,hcy:sigmod94,lcry:sigmod93}.
For each workload, we generated right-deep query plans of 
sizes 40, 60, and 80 over randomly selected base relations.
Given the total number of joins in the workload, different right-deep plan 
size combinations were tested (e.g., our 40-join workloads included 
one workload  with five 4-join pipelines and ten 2-join pipelines, one workload
with ten 4-join plans, etc.).
For each workload combination, we executed five randomly generated workloads (by
randomly selecting the participating base relations) and recorded the observed 
simulation times for each of the five runs.
The results presented in the next section are indicative of the results obtained 
for all randomly generated workloads and different combinations of pipeline sizes.
For the experiments presented in this section, there was 
{\em no base relation shared across query plans}, i.e., each 
relation was part of at most one join in a workload.
The purpose of this was to mitigate the effects of the buffer management
algorithm on our scheduling results.
%
%
We plan to explore these effects in future work.

Our simulator implements a hybrid hash-join processing algorithm for right-deep
trees, so it is possible
to schedule each join in a pipe to be executed with as little as the square root
of its maximum memory demand~\cite{schneider:phdthesis90,shapiro:tods86}.
Of course, using less memory implies that the join pipeline is executed in
multiple passes.
This is never the case for \LevelSched, since demand vectors are always
determined using the maximum (i.e., one-pass) memory demand for a join.
Since our initial results showed that multi-pass join pipelining can really
hurt the overall response time performance for a workload, we also tried to
avoid multi-pass pipeline executions for \hier\ by starting with a ``minimum work''
allotment that gives each join  a sufficient number of processors for keeping
its entire \opbuild\ hash table in memory.
Note, however, that this may not be possible under certain (``small'') configurations,
since \hier\  does {\em not\/}  allow joins in a pipeline to share processors.
In such scenarios, we allocated sites to joins in the pipeline in proportion
to their total memory requirement.
For \zsched, we implemented a scheduling switch that forces the \gsched\ process
not to start the execution of a query until its maximum memory requirement can
be satisfied.
Once again, however, this may be impossible given the degree of declustering
of the \opbuild\ relations.

All our right-deep plans included a final step of storing the result relation across
all sites using a single disk at each site.
We decided to use this default strategy for \opstore's rather than treating
them as floating operators since for many real-world queries the results
do not actually need to be stored on disk,  so we wanted to minimize the impact
of \opstore's on the relative quality of the execution schedules.
We also employed the simulator's built-in assumption of uniform join attribute
distributions and decided to leave the study the effectiveness of the
schedules generated by the various algorithms in the presence of data
skew to future work.
However, we firmly believe that skew can only increase the gains
of our scheduling algorithm over both \hier\ and \zsched.
The reason is that, unlike \TreeSched,
(a) \hier's list scheduling scheme is based on the assumption of uniform
partitioning of an operator's work among its clones~\cite{wtcy:sigmetrics94,wtcy:tpds95},
and (b) by its definition, \zsched is not allowed to re-balance the 
tuples of a \opbuild\ relation across sites.
%

Our basic comparison metric was the observed {\em response time\/}
for a  given query workload. 
The values of our database schema and workload parameters are summarized
in Table~\ref{tab.wloadparams.zsim}.
Both the  coarse granularity parameter $f$ and the memory granularity parameter
$\lambda$ were varied over a wide range of possible values in an effort to
determine their impact on the performance of \LevelSched.
(The range of values for the $f$ parameter was decided based on 
the experimental observation of some actual values for operator processing and 
parallelization costs during the operation of our simulator.)
Although ``fine tuning'' $f$ and $\lambda$ is a non-trivial problem, we will
discuss some of the insights obtained from our experimental findings in 
the next section.
The {\em join probability\/} is  the factor multipled by the cardinality 
of the right input to determine the result cardinality of a join,
and the {\em projection factor\/} indicates the percentage of the sum 
of the left and right input tuple widths that should be retained in the 
result.
We kept these join parameters fixed at 1.0 and 0.5, respectively during all 
our experiments.

\begin{SingleSpace}

\begin{table}[ht]
\vskip 10pt
{\small
\begin{center}
\begin{tabular}{||l|r||}\hline\hline
{\bf Schema/Workload Parameters}   	&	{\bf Value}	\\ \hline \hline
Tuple Size			& 	200 bytes	\\ \hline
Base Relation Size		&	$10^4$ -- $10^6$ tuples \\ \hline
Join Attribute Distribution	&	Uniform		\\ \hline\hline
Workload Size (Total No. of Joins) &	40, 60, 80 joins\\ \hline
Right-Deep Query Size		& 	2, 4, 8 joins	\\ \hline
Coarse Granularity Parameter ($f$) &	0.001 -- 0.1	\\ \hline
Memory Granularity Parameter($\lambda$)	&  0.1 -- 0.75	\\ \hline
Join Probability		&	1.0 		\\ \hline
Projection Factor		&	0.5		\\ \hline\hline
\end{tabular}
\end{center}
}
\caption{Database Schema and Workload Parameters}
\label{tab.wloadparams.zsim}
\vskip 5pt
\end{table}

\end{SingleSpace}

\subsection{Experimental Results}
\label{sec.results.zsim}

The objective of our empirical study was twofold: 
(1) to examine the impact of the clone granularity parameter values ($f$ and $\lambda$)
on the performance of our \TreeSched\  algorithm, and 
(2) to demonstrate the benefits of our multi-resource scheduling strategies compared
to earlier one-dimensional or ``naive'' approaches.
The major findings of our study can be summarized as follows.

\begin{itemize}
\item
{\bf Insensitivity of \TreeSched\ to Clone Granularity Parameters.}
The problem of ``optimally'' tuning the clone granularity parameters 
($f$ and $\lambda$) is obviously very complex and depends on a multitude
of factors.
However, our results demonstrate that the performance of the schedules
produced by our \TreeSched\  algorithm is relatively insensitive to the exact
choice of values for the granularity parameters, as long as certain extreme
or pathological scenarios for operator parallelization are avoided.

\item
{\bf Improved Schedule Quality Compared to \hier\ and \zsched.}
The quality of the schedules produced by our \TreeSched\ algorithm is,
in general, superior to that obtained by either the ``naive'' \zsched\
scheduler or the one-dimensional \hier\ algorithm.
This demonstrates  \TreeSched's ability to effectively balance the query workload
across the system sites and verifies the benefits of \TreeSched's multi-dimensional
approach for balancing the utilization of different resources within the same
site.
As our results indicate,  the need for effective multi-resource scheduling
is  more evident in smaller, resource-limited system configurations. 

\end{itemize}

We believe that our simulation results establish the practical importance 
of multi-resource scheduling for effective compile-time parallelization of
query execution plans.
Further, the fact that the quality of the schedules produced by our algorithm
is relatively insensitive to the exact values of its input granularity parameters
implies that our techniques should be easy to adapt and tune for different 
real-life parallel query optimizers.

\subsubsection{Effect of Clone Granularity Parameters on the Performance of \TreeSched}
\label{sec.tuning.zsim}
Selecting ``optimal''  values for the coarse granularity parameter $f$ and the 
memory granularity parameter $\lambda$ of the \TreeSched\  algorithm is a obviously
a very difficult problem.
This is because, the optimal choice depends  on a number of parameters and 
characteristics of the underlying hardware and query processing architecture,
such as the CPU overhead of messages and clone synchronization, clone startup costs, 
and the speed of the interconnection network.
Further, even for a fixed set of architectural parameters the impact of the 
granularity parameters on the observed workload response time can depend on
characteristics of the workload itself, such as the size of the workload or
the initial placement of base relation data.
As an example, higher parallelization overheads (e.g., larger values for $f$) are 
obviously less desirable as the level of multiprogramming (i.e., number of distinct
queries) in the system increases.
Finally, as explained in Section~\ref{sec.quant}, it is important to strike a 
balance between the overhead of parallelism and the ability to effectively 
``pack'' memory demands at  system sites.
The implication is that it is probably a very hard problem to discover settings 
for $f$ and $\lambda$  that are ``universally optimal''.
As our experimental results demonstrate, however, for our simulator configuration,
the performance of our \TreeSched\  algorithm is stable and relatively 
insensitive to the exact choice of values for the granularity parameters.
Based on our experimentation with other system configurations, we believe that
this is a more general phenomenon. 
That is, we expect the performance of our multi-resource scheduling algorithm 
to be reasonably stable over a range of choices for the $f$ and $\lambda$
parameters, as long as certain pathological scenarios for operator parallelization
are avoided.
We try to provide some general guidelines for avoiding such scenarios at the end
of this section.
%

Figure~\ref{fig.effect-f.zsim} depicts the response times of the 
execution schedules obtained by \TreeSched\  under four of the six data 
placement policies considered in our simulations.
(The results for {\bf Declust-1/4} and {\bf NoDeclust-1/4} are very similar
to those for {\bf Declust} and {\bf NoDeclust}, respectively.)
Results are shown for a 60-join workload consisting of fifteen 4-join right-deep
queries, for various values of the coarse granularity parameter $f$, 
with 64 MBytes of main memory per site, and a memory granularity parameter 
$\lambda = 0.2$.
Nearly identical results were also obtained with different values of $\lambda$ and 
different memory sizes.
Our results clearly show that
the performance of the schedules produced by our \TreeSched\ 
algorithm remains essentially unaffected by the specific choice of
the coarse granularity parameter $f$.
(The differences observed are typically in the range of 1-3\%.)
There are two main reasons for this.
First, in our implementation of \TreeSched, we use the earlier results of
Wilschut et al.~\cite{wfa:vldb92} to estimate and enforce an upper bound 
on the degree of effective partitioning for each floating operator.
Thus, even for very large values of $f$,  our \TreeSched\  implementation
makes sure that  the assigned degree of operator  parallelism  never 
exceeds the ``optimal number of clones'' for the operator,
as estimated by the formula of Wilschut et al.~\cite{wfa:vldb92}.
%
%
Second, and perhaps most importantly, the startup and synchronization
overhead of parallelism turns out to be relatively small,
and thus higher numbers of join clones do not imply
performance penalties for the schedule.
This is mainly due to the high speed (100~MIPS) of site CPUs in our simulated
system, which basically implies that the CPU overhead of setting up and synchronizing
a few tens of clones is insignificant compared to the amount of query processing
work.
(This effect is also evident from the good performance of \zsched\  schedules
under the fully-declustered  {\bf Declust} data placement policy even
for ``large'' simulator   configurations (Section~\ref{sec.dplcmt.zsim}).)
Note that, in experiments for configurations with lower CPU speeds
(10~MIPS) we found that such  overheads can become 
significant when the number of sites exceeds 32.
For such scenarios, more careful tuning of the coarse granularity 
parameter $f$ may be required.

\begin{SingleSpace} 
 
\quadepsfig{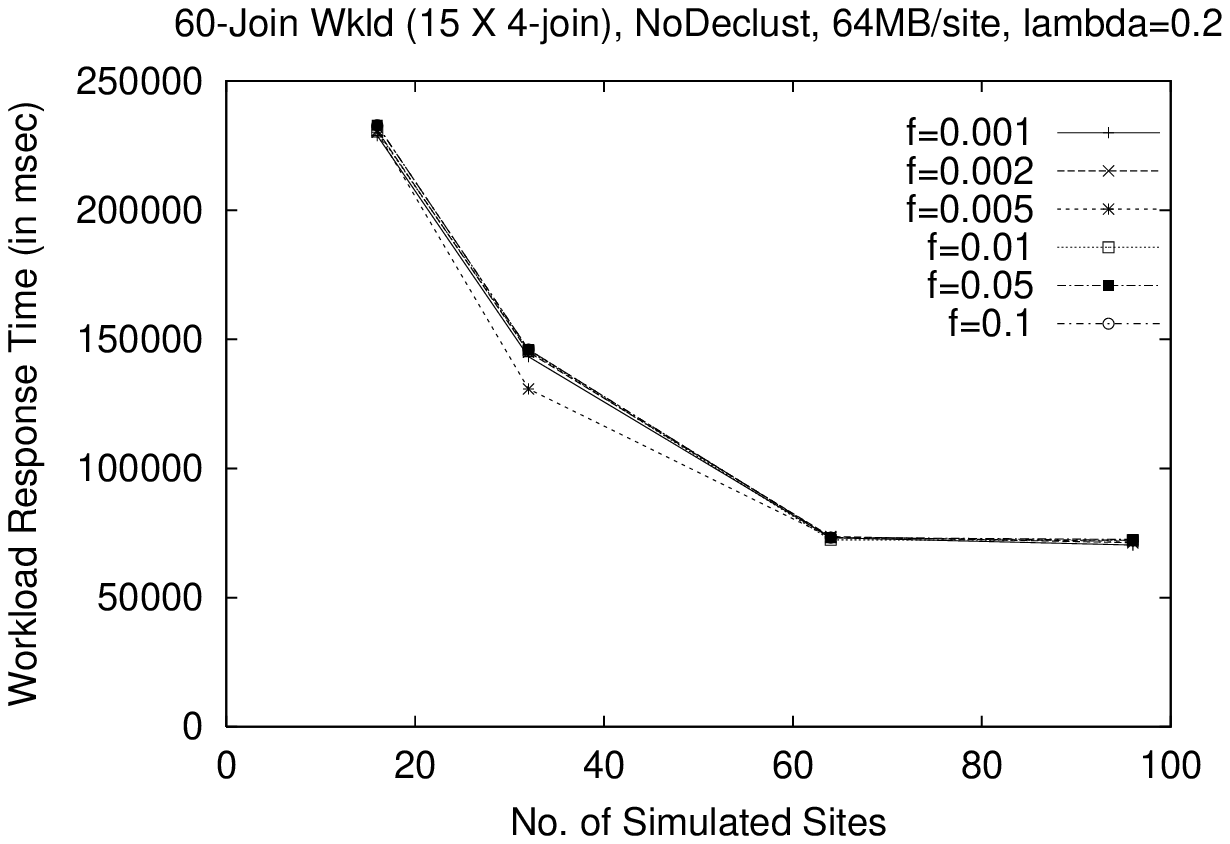}
           {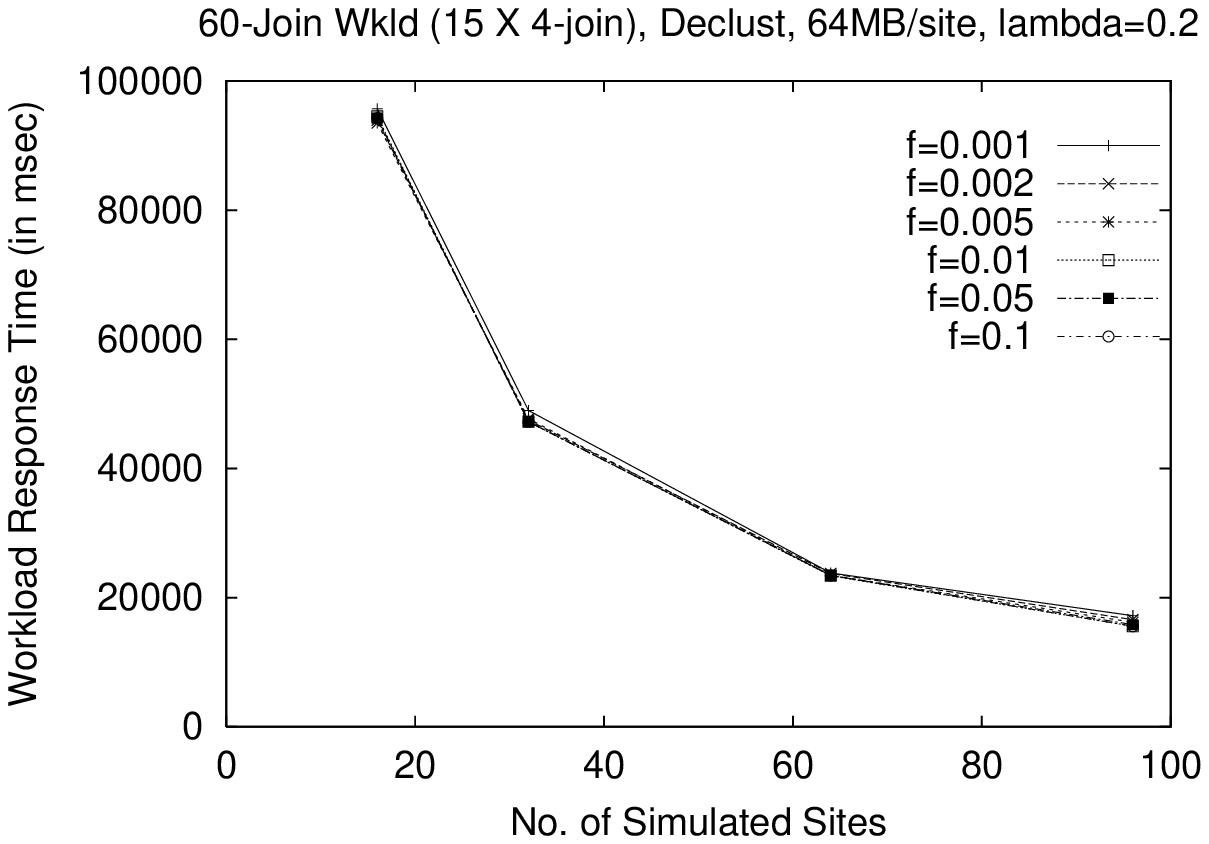}
           {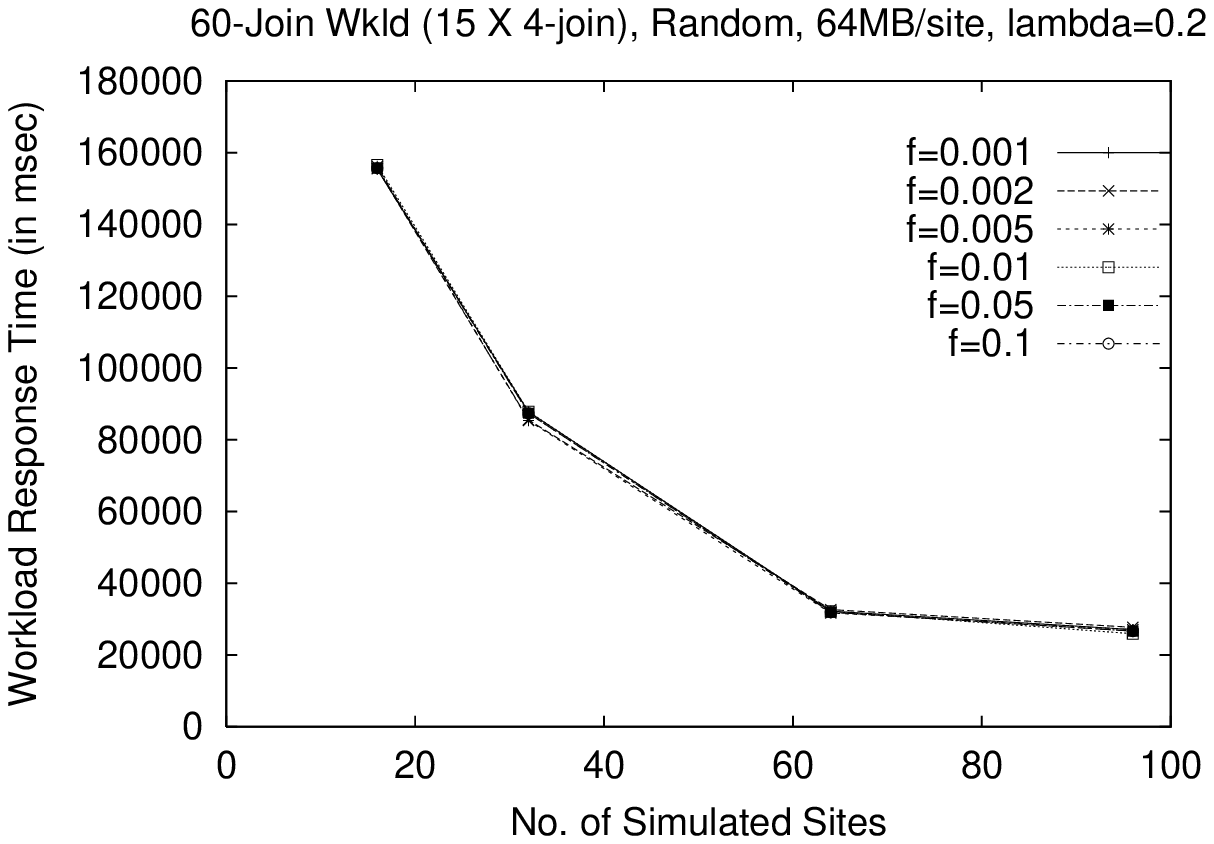}
           {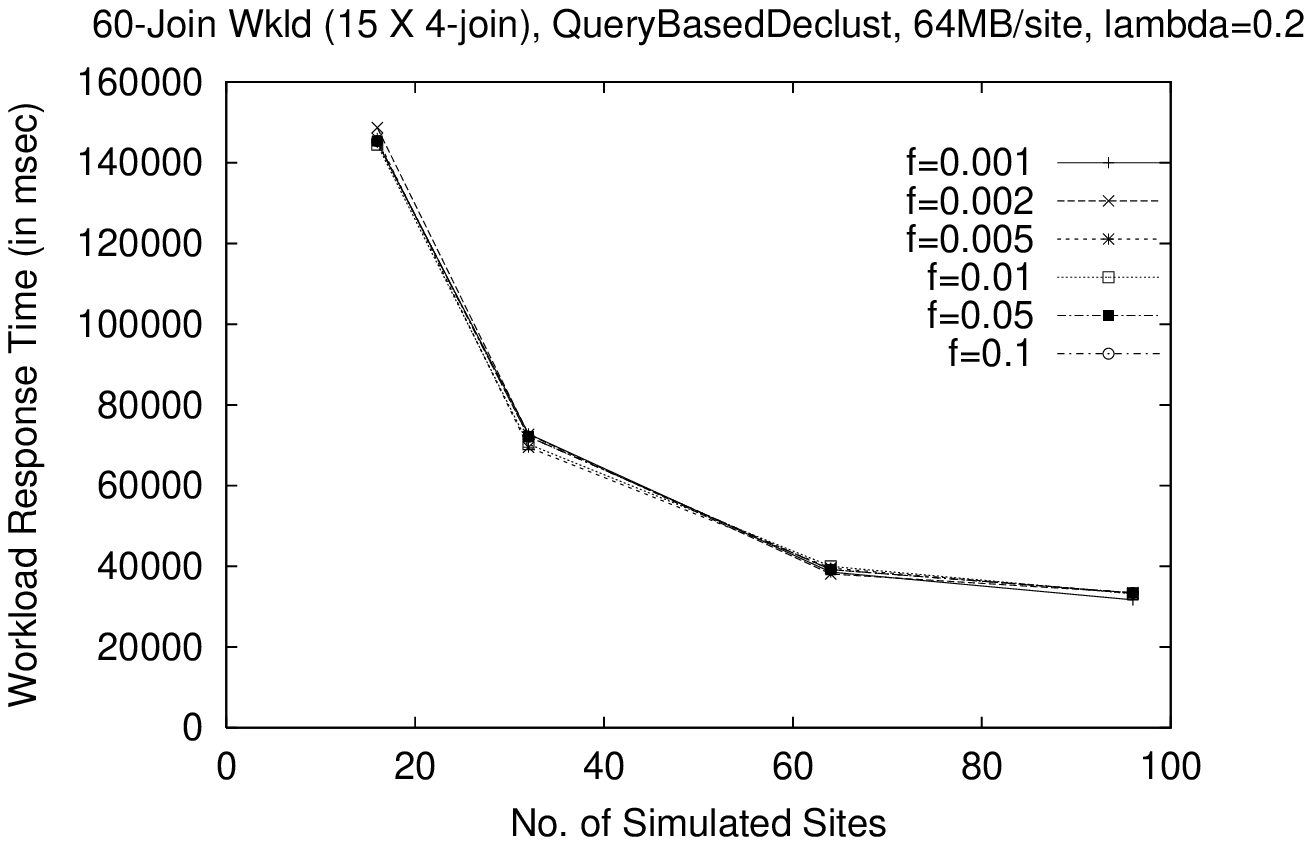}
{1.0}
{Effect of the coarse granularity parameter $f$ on the performance of \TreeSched\ for a
(15 X 4-join) workload under {\bf NoDeclust}, {\bf Declust}, {\bf Random}, 
and {\bf QueryBasedDeclust} data placement with 64 MBytes per site. ($\lambda = 0.2$)}
{fig.effect-f.zsim}

\end{SingleSpace}

The effect of different values for the memory granularity parameter $\lambda$
on the performance of \LevelSched\ for the same 60-join workload is depicted
in Figure~\ref{fig.effect-l.zsim} for {\bf NoDeclust}, {\bf Declust}, 
{\bf Random}, and {\bf QueryBasedDeclust} data placement,  a coarse
granularity parameter of $f=0.005$, and 64 MBytes of main memory per site.
The results for {\bf Declust-1/4} and {\bf NoDeclust-1/4} are almost identical
to those for {\bf Declust} and {\bf NoDeclust}, respectively, while very 
similar results were also obtained for different values of $f$ and different
memory sizes.
The main observation again is that the performance of our \TreeSched\ algorithm
basically remains stable under different settings of the memory granularity 
parameter $\lambda$.
(The observed differences are, once again, typically in the 1-3\% range.)


\begin{SingleSpace}

\quadepsfig{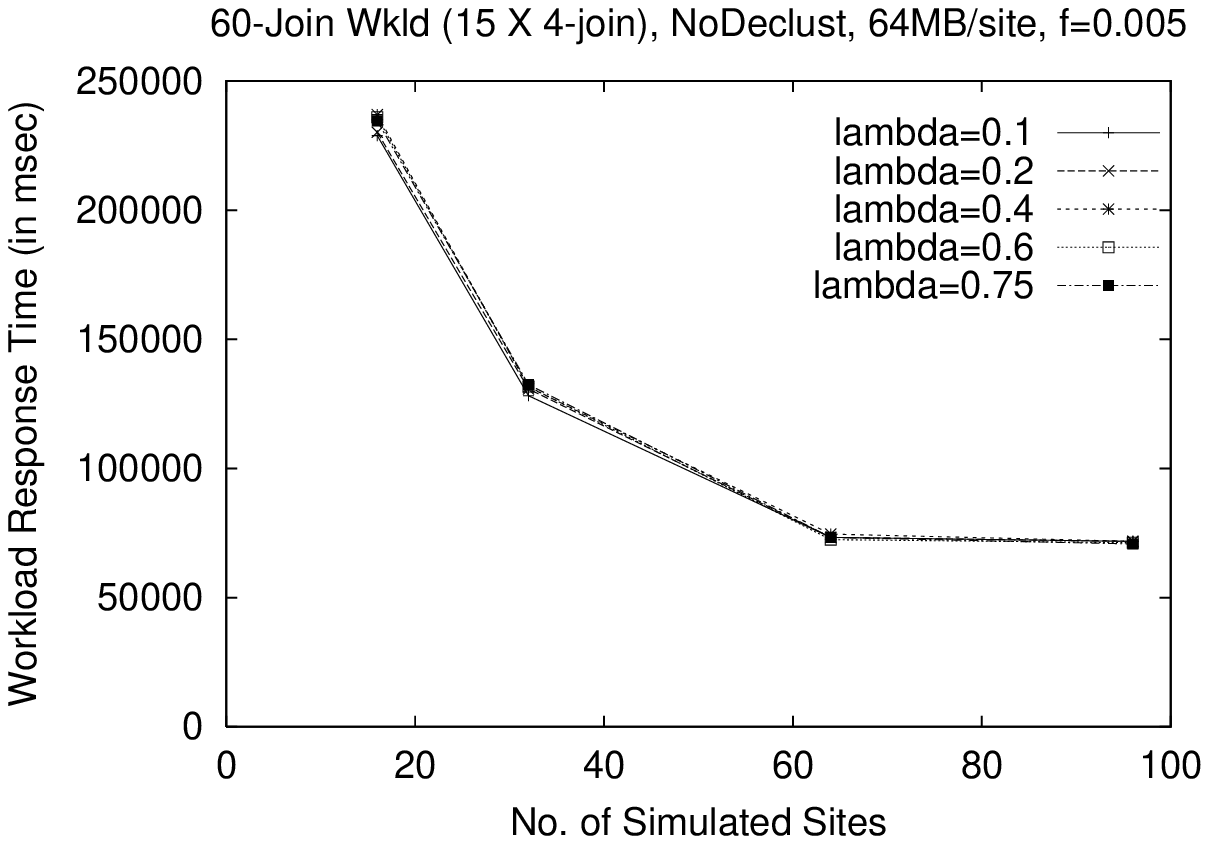}
           {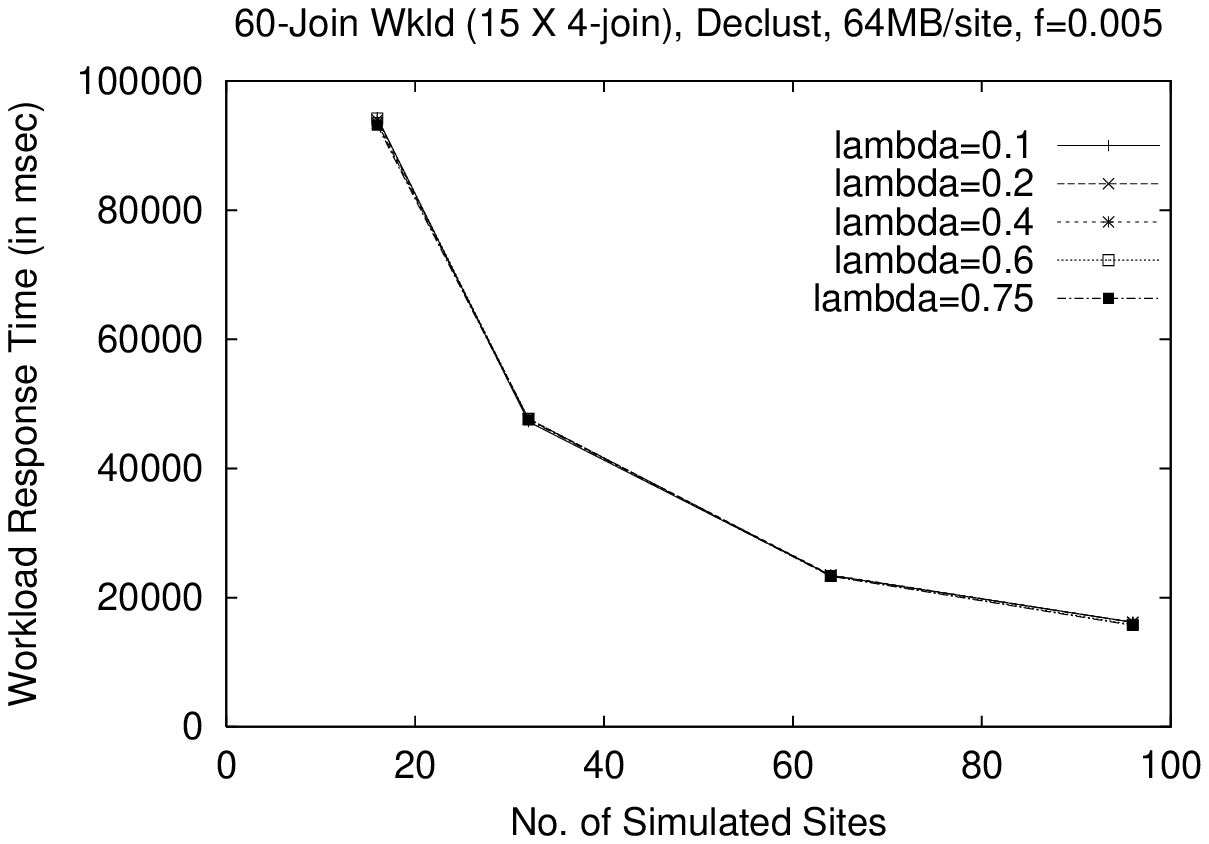}
           {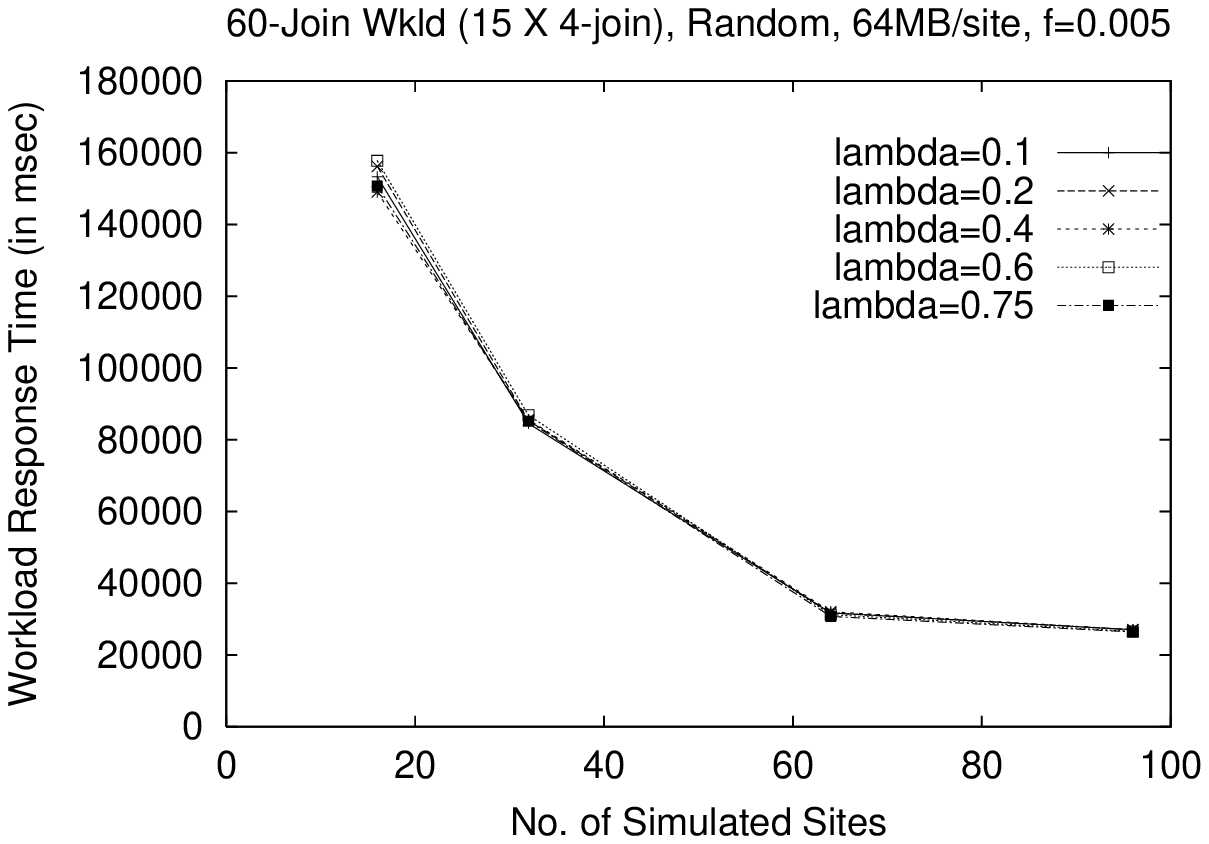}
           {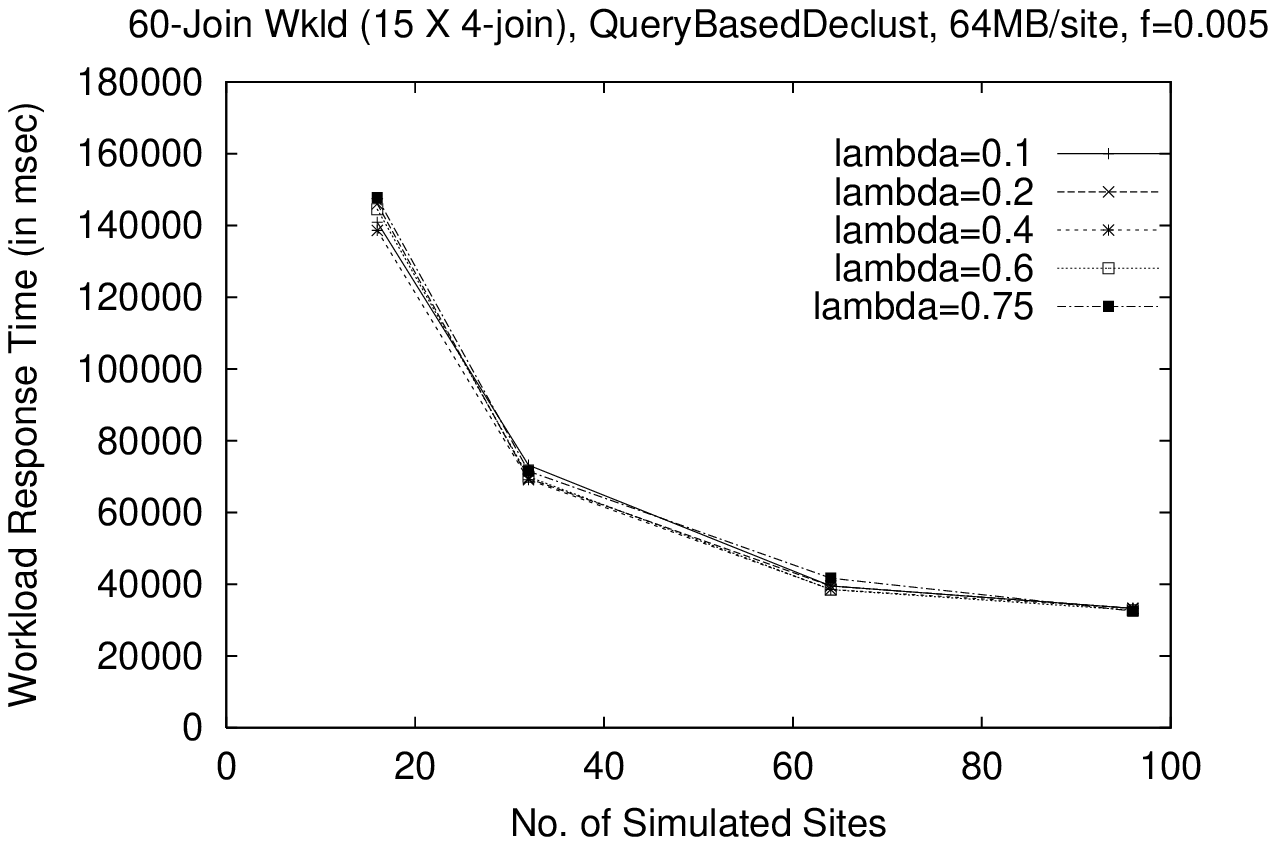}
{1.0}
{Effect of the memory granularity parameter $\lambda$ on the performance of \TreeSched\ for a
(15 X 4-join) workload under {\bf NoDeclust}, {\bf Declust}, {\bf Random},
and {\bf QueryBasedDeclust} data placement with 64 MBytes per site. ($f = 0.005$)}
{fig.effect-l.zsim}

\end{SingleSpace}

\vspace*{.5em}\noindent
{\bf General Guidelines for Clone Granularity.}
Our simulation results show that the performance of our \TreeSched\ algorithm 
is  stable and relatively insensitive to the exact choice of granularity 
parameters, for a reasonably wide range of possible values.
%
%
There are, however,  some general guidelines that should be adhered to 
when selecting $f$ and $\lambda$ values in order to avoid certain pathological 
scenarios with respect to operator parallelization.
Abstractly, $f$ tries to put an upper bound on the relative overhead of 
parallelization (compared to an operator's work), whereas 
$\lambda$ determines how ``thin'' the memory demands of clones are going
to be.
Based on our definition of $\lambda$-granular \cgf\ executions, we identify
two ``dangerous'' scenarios  that should probably be avoided when trying to
choose reasonable (ranges of) values for $f$ and $\lambda$.
(Of course, such ``reasonable'' choices depend on system parameters
and can be determined with some preliminary experimentation.)
\begin{itemize}
\item
{\bf Very large $f$ or very small $\lambda$} typically imply that operators
are ``overpartitioned'' to very thin clones, resulting in high parallelization
overheads and possible slowdowns in execution time.
This effect was not observed for our default simulator configuration setting
(due to the reasons outlined above), but it did show up for certain configurations
with lower CPU speeds.

\item
{\bf Very large $\lambda$ and very small $f$} typically imply that operators
are assigned very low degrees of parallelism which may be too conservative to 
exploit the available system resources.
An example of such an extreme case is depicted in Figure~\ref{fig.extr.zsim}
for two distinct data placement strategies ({\bf Declust} and {\bf Declust-1/4})
and our default simulator settings with $f=0.00001$ and $\lambda = 0.75$.
As expected, under this fairly restrictive combination of values for
$f$ and $\lambda$, \TreeSched\ is unable to appropriately leverage the 
additional resources as the system size grows.
\end{itemize}

\begin{SingleSpace}

\depsfig{1.0}{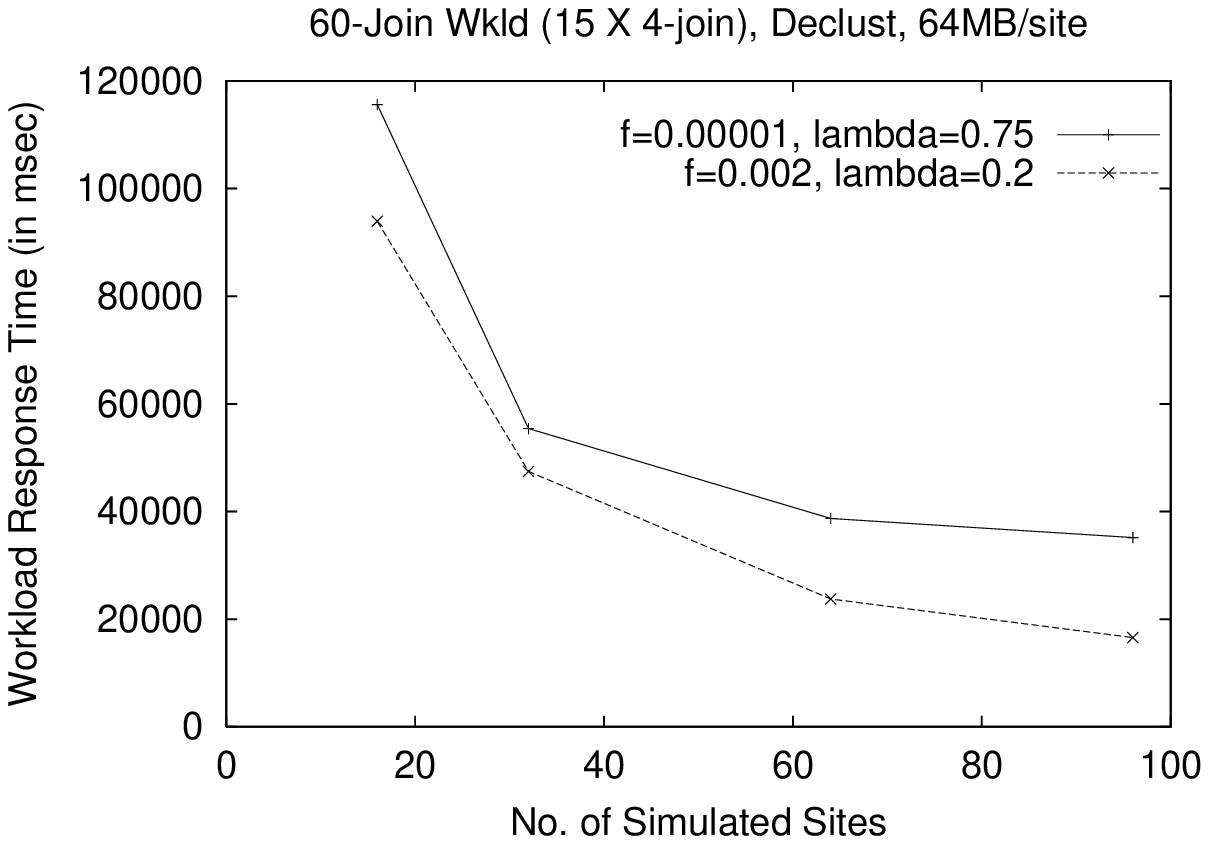}
        {1.0}{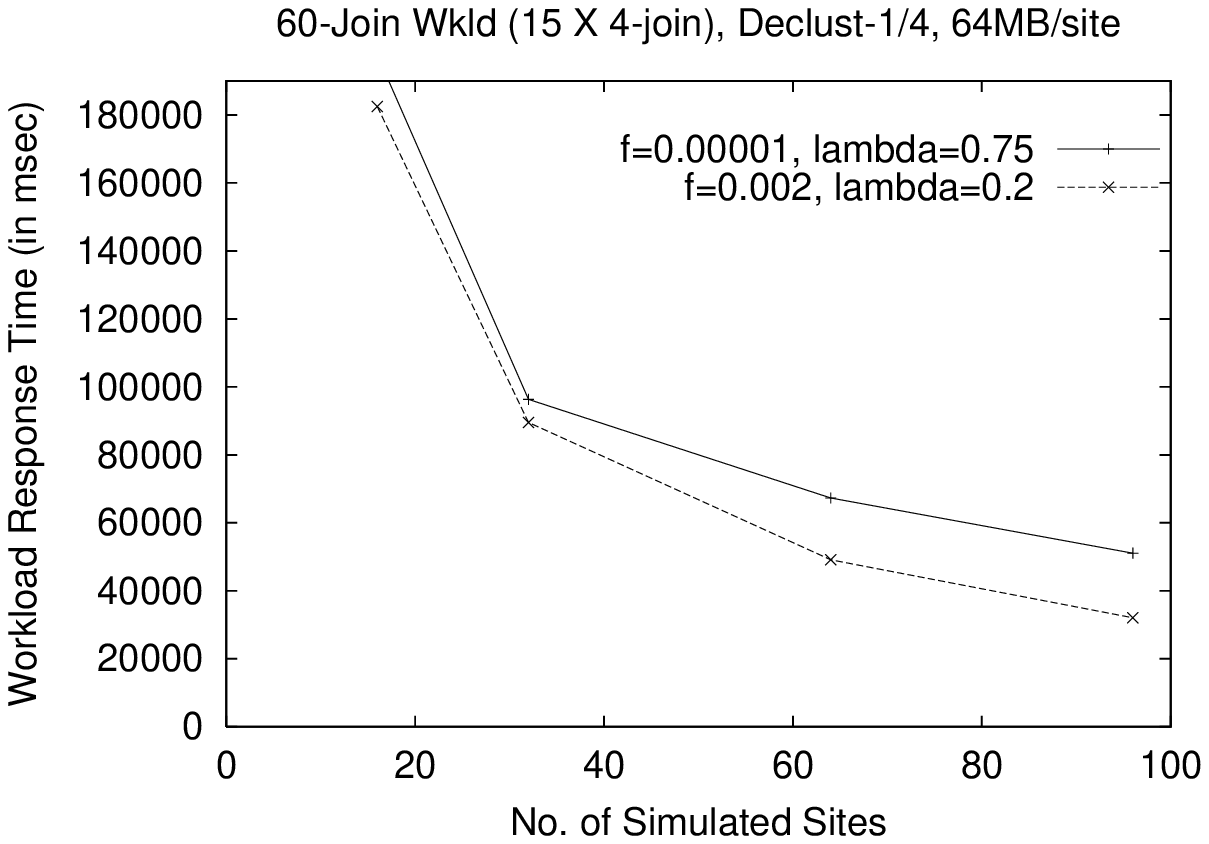}
{Effect of overly conservative combination of granularity parameters
($f=0.00001$ and $\lambda = 0.75$) on the performance of \TreeSched\ 
for (a) {\bf Declust} and (b) {\bf Declust-1/4}.}
{fig.extr.zsim}

\end{SingleSpace}



\subsubsection{Comparison of Scheduler Performance}
\label{sec.dplcmt.zsim}
Figures~\ref{fig.nodecl.zsim}--\ref{fig.ssched.zsim} depict the performance
of the three scheduling algorithms (\TreeSched, \hier,  and \zsched)  for the
various data placement strategies considered in our simulation study and site 
memory sizes of 64 and 96 MBytes.
These results were obtained for a 60-join workload consisting of five
8-join queries and ten 2-join queries, assuming the default granularity
parameter values of $f = \fdef$ and $\lambda = \ldef$.

Figure~\ref{fig.nodecl.zsim} gives the response times of the 
schedules produced by the \TreeSched\ and \hier\ algorithms for the
{\bf NoDeclust} and {\bf NoDeclust-1/4}  data placement policies.
As expected, both algorithms were clear winners over the naive
\zsched\ strategy for our two ``No-Declustering''  policies ---
the response times for the \zsched\  schedules ranged from $4$
to $16$ times the corresponding response times for our \TreeSched\
scheduler, and are therefore omitted from the
plots in Figure~\ref{fig.nodecl.zsim}.
The main reason for this large gap in performance is that, unlike
\zsched, both \TreeSched\ and \hier\ manage to balance the memory demand
for \opbuild\ relations across the system without being restricted by 
the specific sites at which base relations reside on disk.
Between the two ``smarter'' schedulers, \TreeSched\ exploits its multi-dimensional
model of operator costs to better balance the utilization of CPU and disk 
resources across system sites leading to improvements of up to 35-40\% for
certain configurations over the one-dimensional \hier\ schedules.
Note that the performance differences between the two schedulers are
typically diminished as the size of the system is increased. 
Obviously, effective resource scheduling  becomes less critical
once there is an abundance of resources in the system.

\begin{SingleSpace}

\depsfig{1.0}{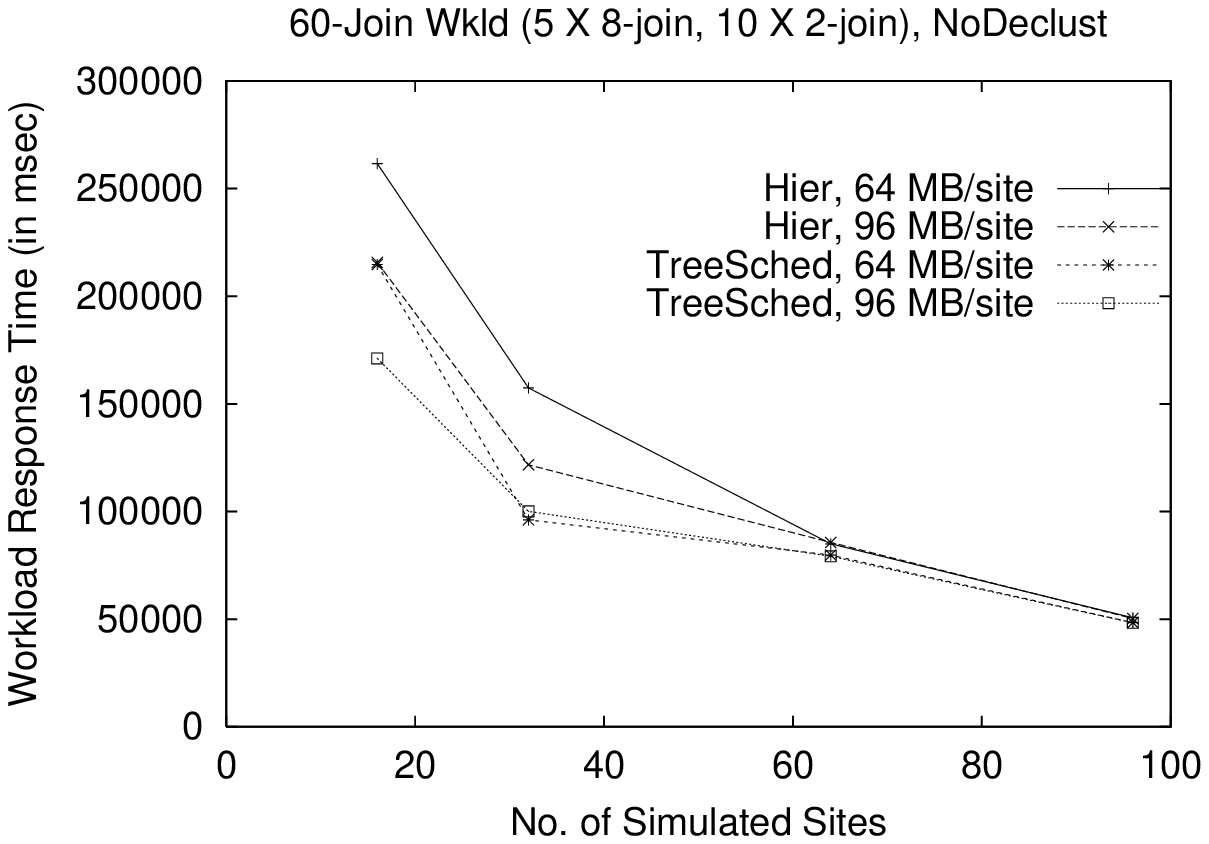}
	{1.0}{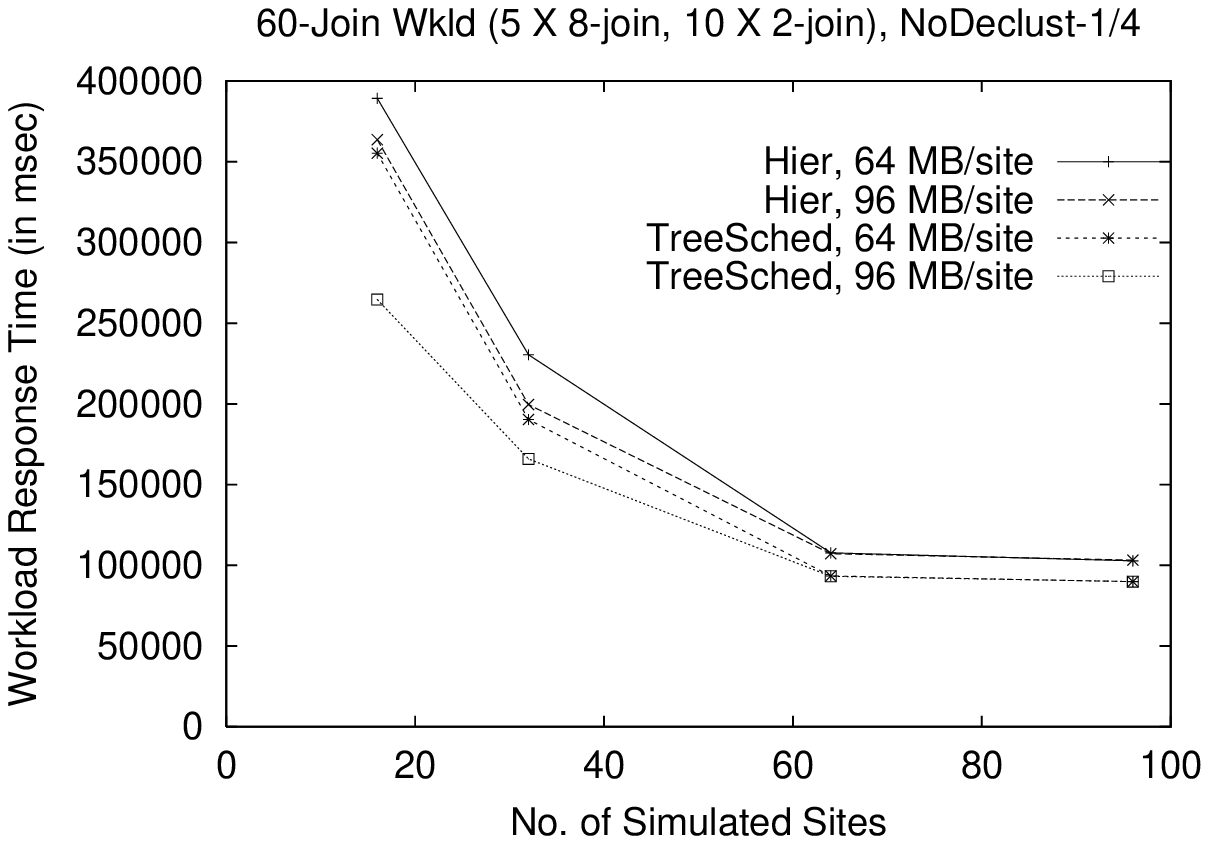}
	{Performance of \TreeSched\ and \hier\ for (a) {\bf NoDeclust} and (b) {\bf NoDeclust-1/4}.
	($f = \fdef$, $\lambda = \ldef$)}
	{fig.nodecl.zsim}

\end{SingleSpace}

The performance of the three scheduling algorithms under our 
fully-declustered data placement policy  {\bf Declust} is depicted 
in Figure~\ref{fig.fdecl.zsim}.
As our results demonstrate, \TreeSched, \hier, and \zsched\  perform
almost identically under {\bf Declust}.
%
%
This was something we expected for the following reasons.
First, given that all base relation scans for a pipeline occur concurrently 
at all sites and all result stores are again defaulted to execute on all
sites, \TreeSched\ really has no opportunities to exploit its multi-dimensional
cost model in order to balance CPU and I/O processing across system sites.
Second, since the startup and synchronization overhead of parallelism is 
rather small in our simulator configuration, cloning every join across all sites does
not penalize performance, at least for the range of system sizes considered
in our experiments.
Thus, both the one-dimensional \hier\ and the naive \zsched\  perform 
well, managing to consistently stay very close to \TreeSched.
(The slight degradation in the performance of \zsched\ for small 
configurations is due to the fact that some pipelines are executed
in multiple passes, incurring intermediate I/O.)
Note that, earlier work (e.g., \cite{mehta:phdthesis94,md:vldb95,md:vldbj97})
has shown that the naive strategy of cloning every query operator 
across all sites can result in very poor execution schedules for system 
configurations with either higher parallelization overheads or larger 
numbers of sites than our assumed simulator configuration.
In fact, we have also experimented with simulator configurations with
lower CPU speeds and  we have found that, in many situations, 
the response times for \zsched\  actually {\em increase\/} as the
number of sites exceeds $32$.

%

\begin{SingleSpace}

\depsfig{1.0}{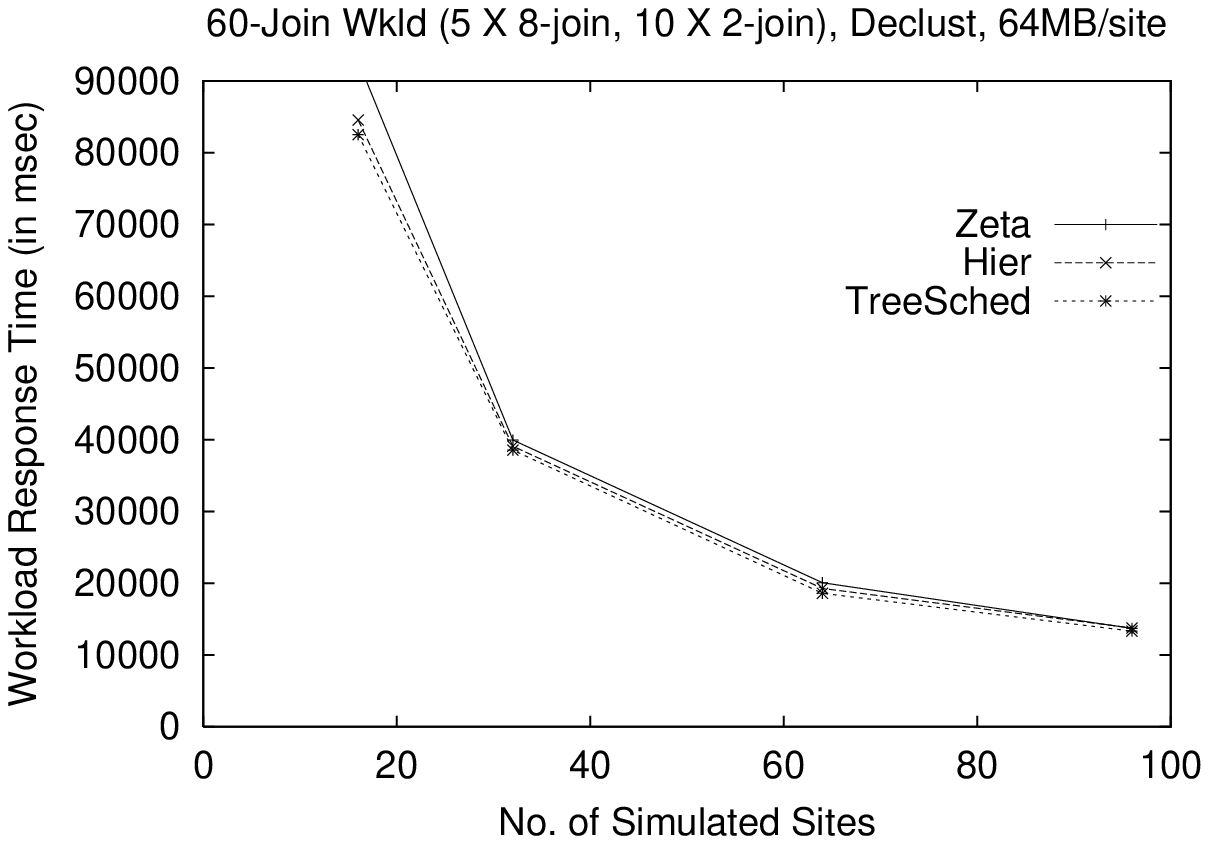}
        {1.0}{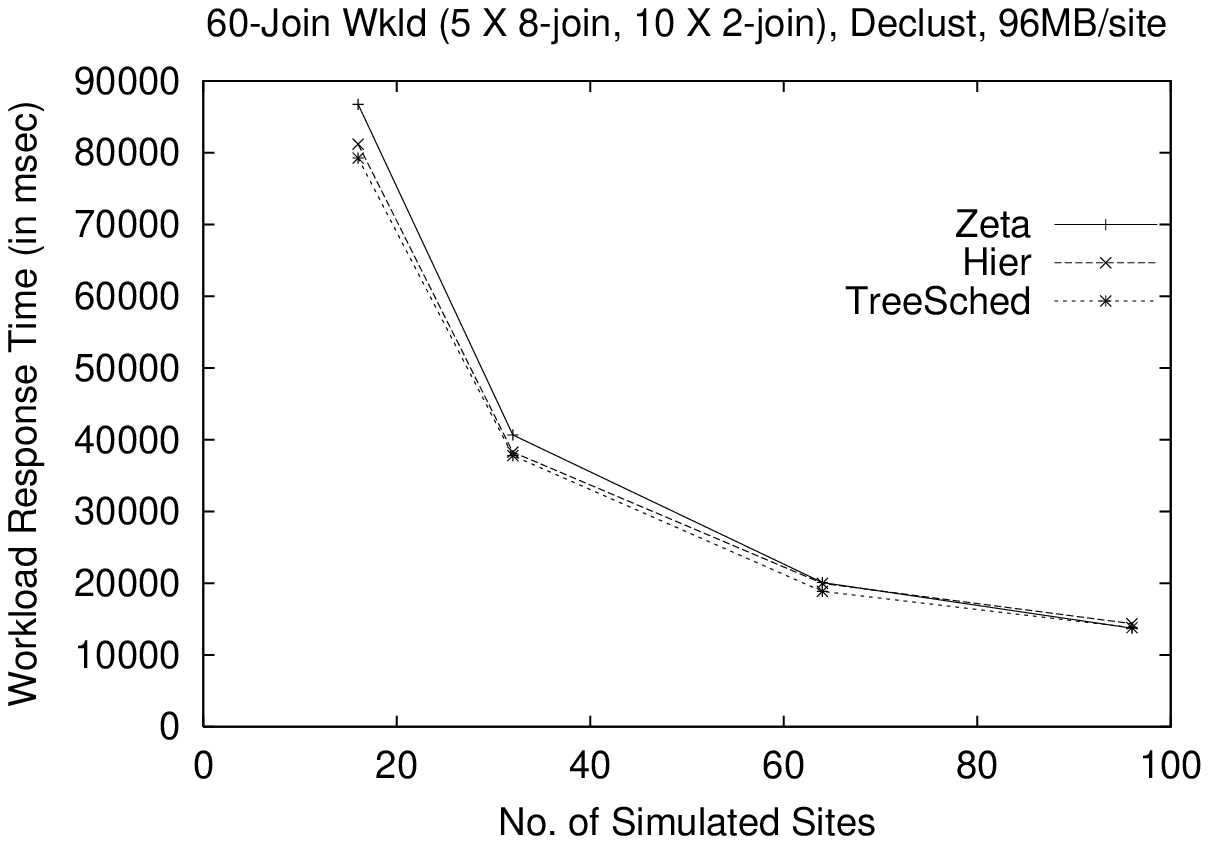}
{Performance of \TreeSched, \hier,  and \zsched\ for {\bf Declust}.
($f = \fdef$, $\lambda = \ldef$)}
{fig.fdecl.zsim}

\end{SingleSpace}

Figure~\ref{fig.fdecl-random.zsim} depicts the response times of the
schedules produced by  \TreeSched, \hier, and \zsched\ under our 
{\bf Declust-1/4} and {\bf Random} data placement policies.
Once again, both \TreeSched\ and \hier\  beat \zsched\  by a wide margin,
since they are able to balance the load across the system without being
restricted by data placement decisions.
Furthermore, \TreeSched\ is also able to exploit its multi-dimensional
scheduling model to better balance CPU and disk requirements at sites 
and provide significant response time gains of up to 30\% over the  
one-dimensional \hier\ scheduler.
Note that, as with the ``no-declustering'' results, our multi-dimensional
scheduler gets its most convincing victories over \hier\   with
resource-limited configurations (e.g., $16$ sites with $64$ MBytes/site).
It is for such resource-limited scenarios that effective resource 
scheduling  becomes a major concern for query processing and
optimization.

\begin{SingleSpace}
 
\quadepsfig{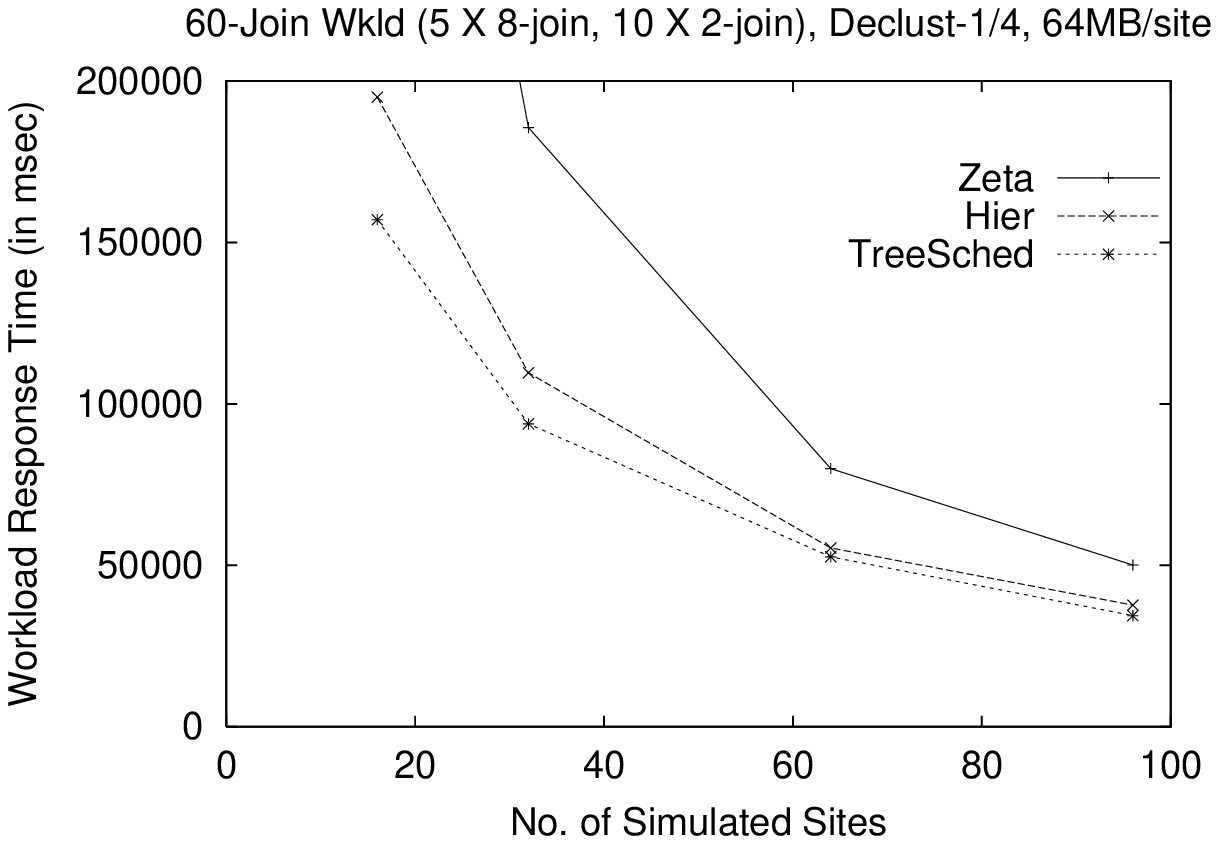}
           {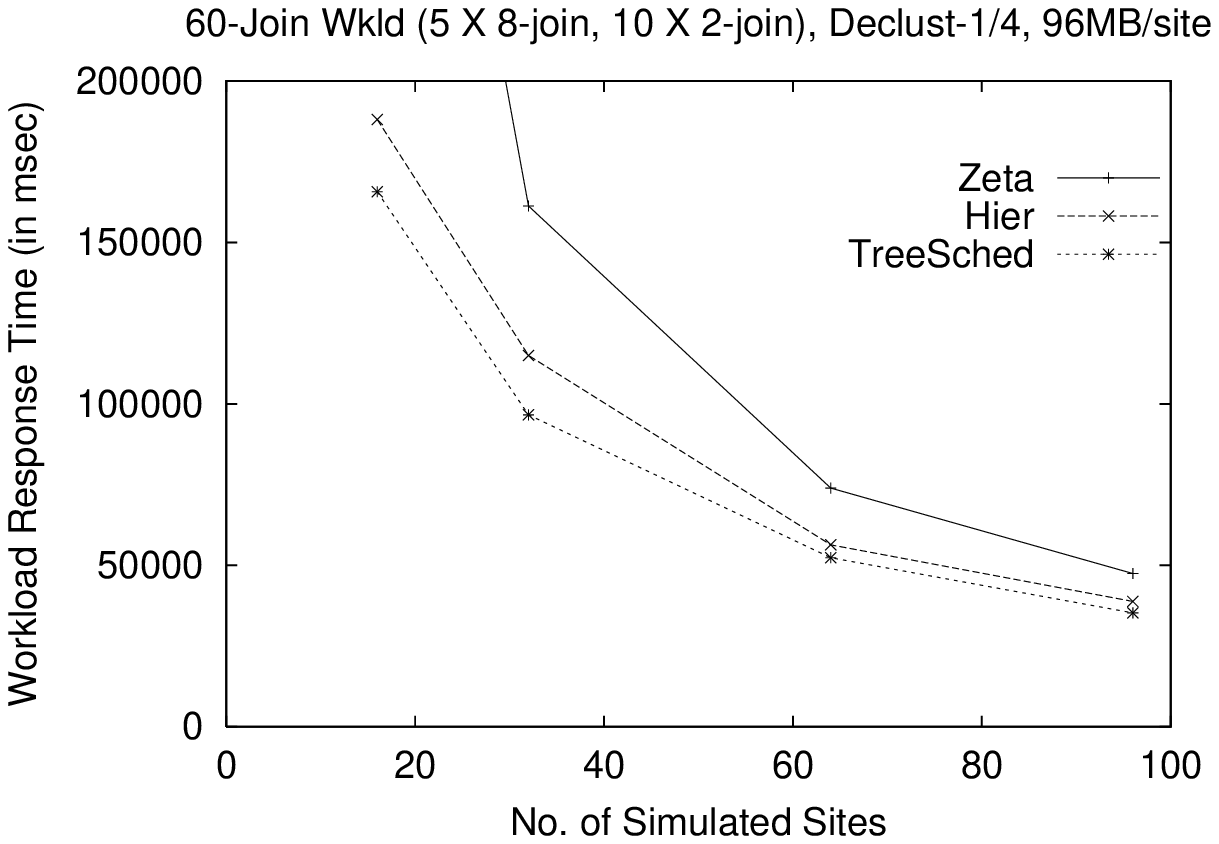}
	   {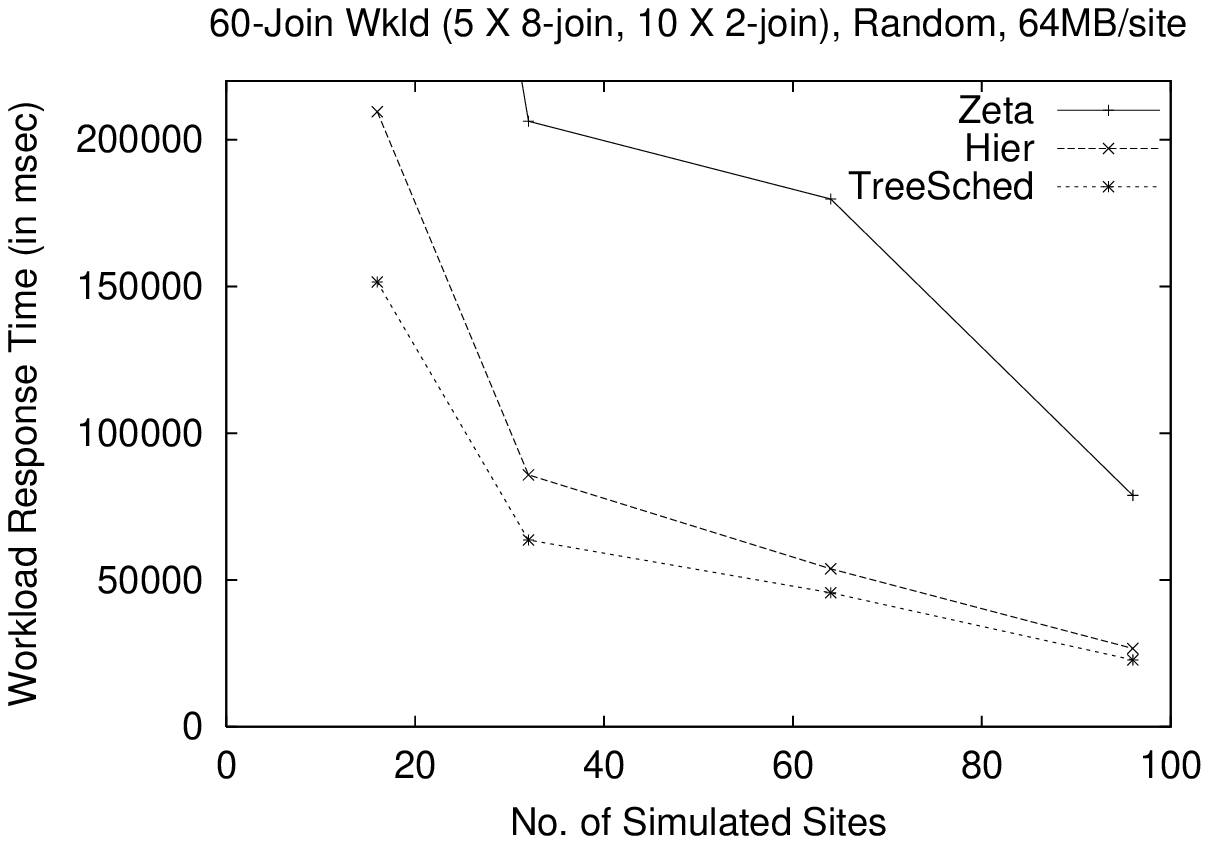}
	   {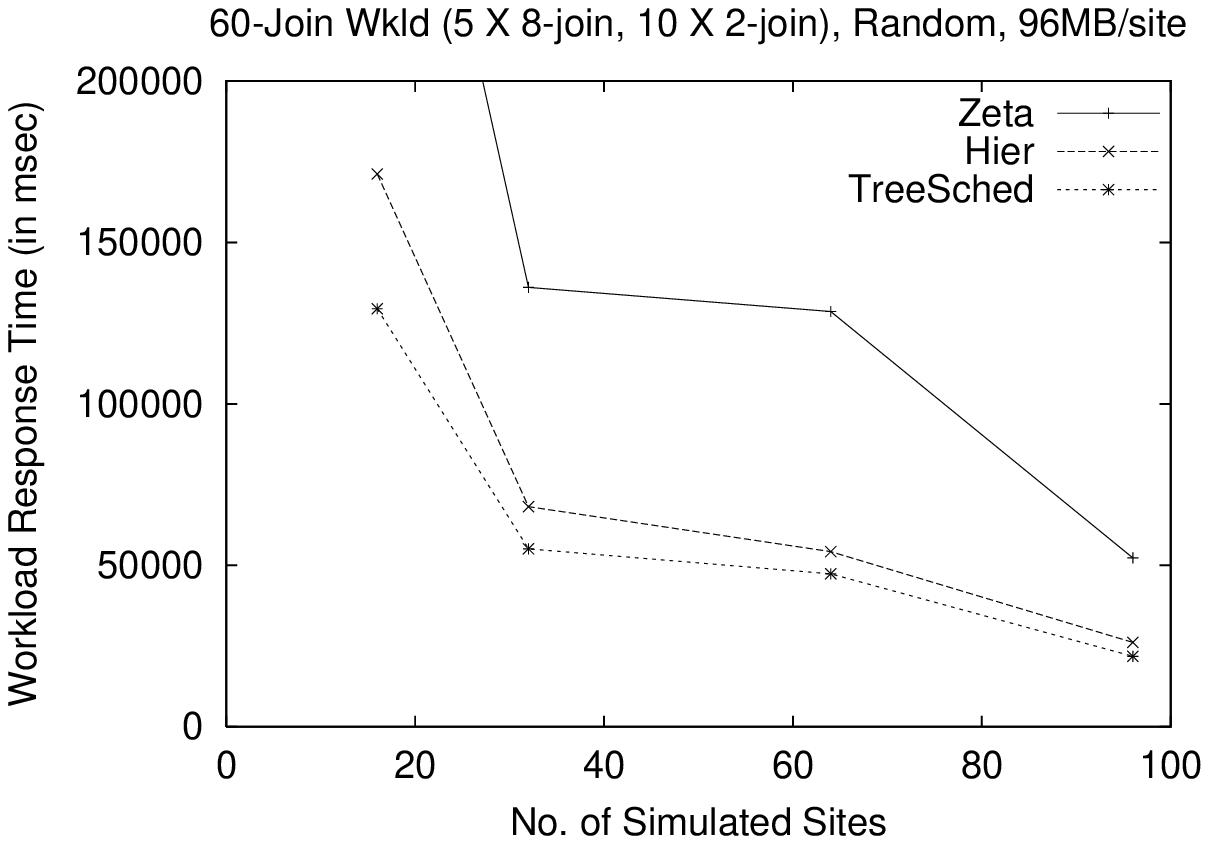}
{1.0}
{Performance of \TreeSched, \hier,  and \zsched\ for (a) {\bf Declust-1/4} (top half),
 and (b) {\bf Random} (bottom half).  ($f = \fdef$, $\lambda = \ldef$)}
{fig.fdecl-random.zsim}
 
\end{SingleSpace}


Finally, the performance of the three scheduling algorithms under our
{\bf QueryBasedDeclust} data placement policy is shown in
Figure~\ref{fig.ssched.zsim}.
In this particular experiment, we activated the  ``maximum memory'' flag for the
\zsched\ algorithm which ensures that a join will only be executed with its maximum
memory allocation.
We found that this really helped the response time performance of
\zsched\ under {\bf QueryBasedDeclust}.
Still, both \TreeSched\  and \hier\   manage to outperform
\zsched\ by a significant margin across our entire system size 
range.
A main reason for this phenomenon is that, by the {\bf QueryBasedDeclust}
data placement rule, \zsched\  essentially decides the degree of parallelism
for a join by looking only at the \opbuild\ relation, which is not always
a good predictor for the amount of work that needs to be done.
For example, consider a very small \opbuild\ relation that is joined with
a very large \opprobe\ input.
\zsched\  will probably assign only one site to that relation,
which means that it will very likely be a bottleneck during the \opprobe\
phase of the pipeline.
On the other hand, both \TreeSched\ and \hier\  consider both the memory 
{\em and} work requirements of a join when making their parallelization 
decisions.
\TreeSched\ is once again able to exploit its multi-dimensional model to
better balance CPU and I/O processing and get some performance gains over
the one-dimensional \hier\ schedules.
Nevertheless, the benefits in this case are somewhat limited (up to 15\%) 
since, by virtue of the {\bf QueryBasedDeclust} data placement policy,
the I/O workload has already been somewhat ``uniformly'' distributed across
the system sites.

\begin{SingleSpace}

\depsfig{1.0}{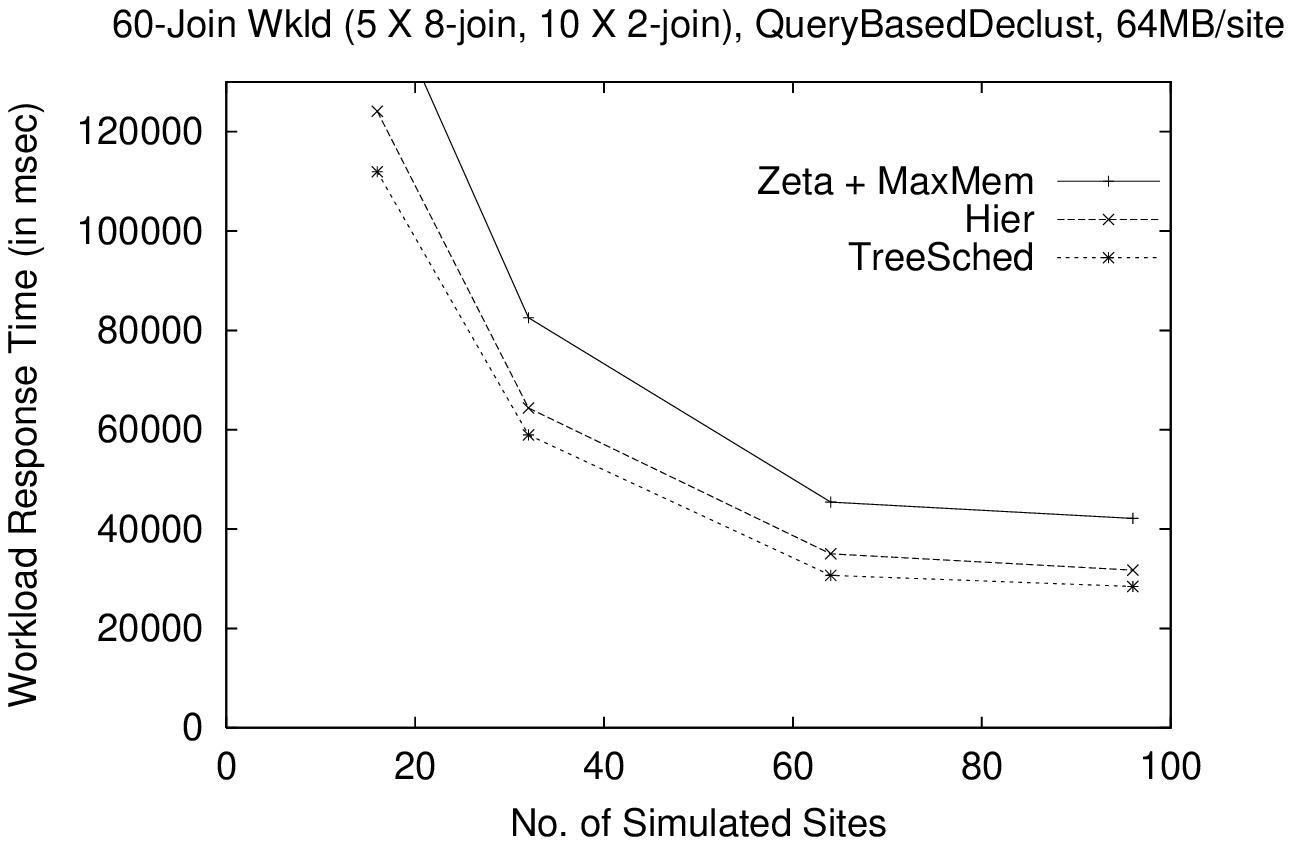}
        {1.0}{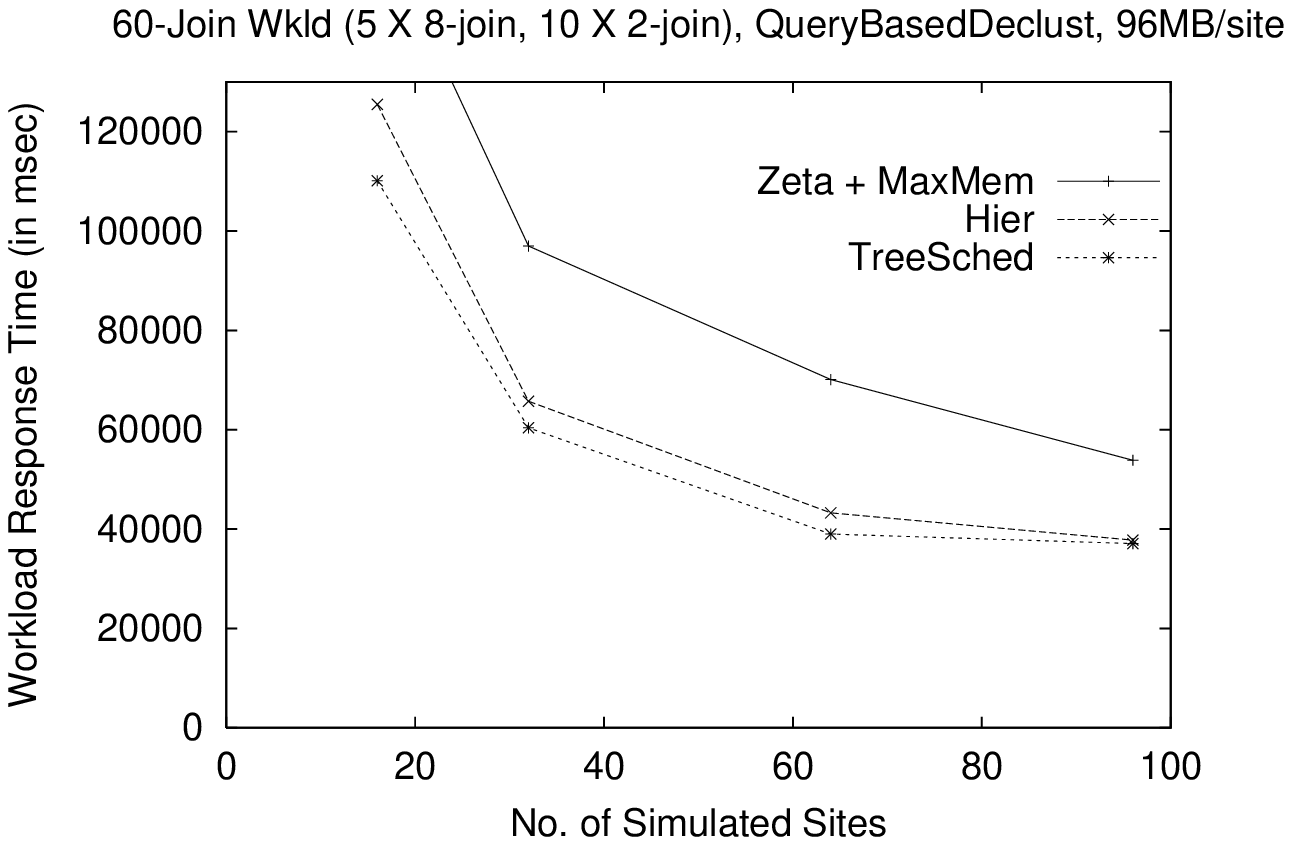}
{Performance of \TreeSched, \hier,  and \zsched\ for {\bf QueryBasedDeclust}.
($f = \fdef$, $\lambda = \ldef$)}
{fig.ssched.zsim}

\end{SingleSpace}

%% file: pqopt.tex
\section{Parallel Query Optimization}
\label{sec.pqopt}
In this section, we study the implications of the analytical and experimental
results presented in this paper for the open problem of designing 
efficient cost models for parallel query optimization~\cite{dg:cacm92}.

As noted in Section \ref{sec.intro},  the use of {\em response time\/} as
optimization metric
implies that a parallel query optimizer cannot afford to ignore resource
scheduling during the optimization process.
Prior work has demonstrated that a {\em two-phase\/} 
approach~\cite{hs:pdis91,hong:sigmod92}
using the traditional work metric during the plan generation phase 
often results in plans that are inherently sequential
and often unable to exploit the available 
parallelism~\cite{jps:apads93,bfgh:ibmsj95}.
On the other hand, using a detailed resource scheduling model
during plan generation (as advocated by the {\em one-phase\/} 
approach~\cite{se:pdis93,jps:apads93,lvz:vldb93}) can have a tremendous 
impact on optimizer complexity and optimization cost.
For example, a Dynamic Programming (DP) algorithm must use  much
stricter pruning criteria that account for the use of system
resources~\cite{ghk:sigmod92,lvz:vldb93}.
This leads to a combinatorial explosion in the state that must be
maintained while building the DP tree, rendering the algorithm
impractical even for small query sizes.

As in centralized query optimization, the role of the optimizer cost
model is to provide an abstraction of the underlying execution system.
In this respect, the one- and two-phase approaches lie at the two
different ends of a spectrum, incorporating either detailed
knowledge (one-phase) or no knowledge (two-phase) of the parallel execution
environment in the optimizer cost metric.
The goal is to devise cost metrics that are more realistic than total
resource consumption, in the sense that they are cognizant of the
available parallelism, and at the same time are sufficiently
efficient to keep the optimization process tractable.

In earlier work, Ganguly et al.~\cite{ggs:pods96} suggested the use 
of a novel scalar cost metric for parallel query optimization.
Their metric was defined as the maximum of two ``bulk parameters'' of a 
parallel query plan, namely the critical path length of the plan tree and
the average work per site.
Although the model used in the work of Ganguly et al. was one-dimensional,
it is clear that the ``critical path length'' corresponds to the maximum, over
all root-to-leaf paths, sum of $T^{max}$'s in  the task tree, whereas the
``average work'' corresponds to $\frac{l(S^W)}{P}$ with $S$ being all 
operator clones in the plan.

Based on our analytical and experimental
results, there clearly exists 
a third parameter, namely the {\em average volume per site\/} 
$\frac{l(S^{TV})}{P}$  that is  an essential component of  query plan quality.
The importance of this third parameter stems from  the fact that it 
is the only one capturing the constraints on parallel execution that
derive from \ssr\ (i.e., memory) resources.

We believe that the triple {\em (critical path, average work, average volume)\/}
captures all the crucial aspects characterizing the expected response time 
of a parallel query execution plan. 
Consequently, we feel that these three components can provide the basis 
for an efficient and accurate cost model for parallel query optimizers.
They can be used exactly as Ganguly et al.~\cite{ggs:pods96} suggested, i.e.,
combined through a $\max\{\}$ function to produce a scalar metric.
Alternatively, they can be used as a 3-dimensional vector together with a
3-dimensional ``less than'' to compare plans (e.g., to prune the 
search space in a DP-based parallel optimizer~\cite{ghk:sigmod92}).
Clearly, using only three dimensions turns the Partial Order DP (PODP) approach
of Ganguly et al.~\cite{ghk:sigmod92} into a feasible and efficient paradigm 
for DP-based parallel query optimization.

%% file: related.tex
\section{Related Work}
\label{sec.related}
The problem of scheduling complex query plans on parallel machines has
attracted a lot of attention from the database research
community.
Hasan and Motwani~\cite{hm:vldb94} study the tradeoff between pipelined
parallelism and its communication overhead and develop near-optimal
heuristics for scheduling a star or a chain of pipelined relational operators
on a multiprocessor architecture.
Chekuri et al.~\cite{chm:pods95} extend these results to arbitrary
pipelined operator trees.
The heuristics proposed in these papers ignore both independent and
partitioned parallelism.
Ganguly and Wang~\cite{gw:rutgers93} describe the design of
a parallelizing scheduler for a tree of coarse grain operators.
Based on a {\em one-dimensional\/} model of  query operator costs (that 
combines the operators'  CPU and disk work in a scalar 
``execution time'' metric),
the authors show their scheduler to be near-optimal for a limited
space of query plans (i.e., left-deep join trees with a single
materialization point in any right subtree).
Ganguly et al.~\cite{ggw:cismod95} obtain similar results for the
problem of partitioning independent pipelines without the coarse
granularity restriction.
The benefits of resource sharing and the multi-dimensionality
of query operators are not addressed in these papers.
Furthermore, no experimental results are reported.
Lo et al.~\cite{lcry:sigmod93} develop optimal schemes for assigning
processors to the stages of a pipeline of hash-joins in a shared-disk
environment.
Their schemes are based on a {\em two-phase minimax\/}
formulation of the
problem that ignores communication costs and prevents
processor sharing among stages.
Moreover, no methods are proposed for handling multiple join
pipelines (i.e., independent parallelism).

With the exception of the papers mentioned above, most
efforts are experimental in nature and offer no theoretical
justification for the algorithms that they propose.
In addition, many proposals have simplified the scheduling issues
by ignoring independent (bushy tree) parallelism; these include
the right-deep trees of Schneider~\cite{schneider:phdthesis90}
and the segmented right-deep trees of
Chen et al.~\cite{clyy:vldb92}.
Nevertheless, the advantages offered by such parallelism, especially
for large queries, have been demonstrated in prior
research~\cite{cyw:icde92}.

Tan and Lu~\cite{tl:ipl93} and Niccum et al.~\cite{nshl:dimacs95}
consider the general problem of scheduling
bushy join plans on parallel machines exploiting all forms of
intra-query parallelism and suggest heuristic methods of splitting the
bushy plan into non-overlapping {\em shelves\/} of concurrent joins.
For the same problem, Hsiao et al.~\cite{hcy:sigmod94} propose a
processor allocation scheme based on the concept of {\em synchronous
execution time\/}: the  set of processors allotted to a parent join
pipeline are recursively partitioned among its subtrees in such a
way that those subtrees can be completed at approximately the same time.
For deep execution plans, there exists a point beyond which further
partitioning  is detrimental or even impossible, and serialization
must be employed for better performance.
Wolf et al.~\cite{wtcy:sigmetrics94,wtcy:tpds95} present a one-dimensional
hierarchical algorithm for scheduling multiple parallel queries on a set of
processors.
Their main idea is to collapse each query plan to a single ``large'' 
parallel job and then apply the known results for independent jobs.
This has the serious drawback that some obvious, critical co-scheduling
may be lost.
For example, although it may be highly desirable to combine CPU-bound
and IO-bound tasks from different plans~\cite{hong:sigmod92}, this 
may not be possible after the collapse.
Wilschut et al.~\cite{wfa:sigmod95} present a comparative performance
evaluation of various multi-join execution strategies on the PRISMA/DB
parallel main-memory database system.
Mehta and DeWitt~\cite{md:vldb95} and Rahm and Marek~\cite{rm:vldb95}
present experimental evaluations of various heuristic strategies for 
determining the degree of intra-operation parallelism and assigning
processors in shared-nothing DBMSs.
Both of these papers avoid dealing with complex query scheduling issues
by assuming workloads consisting of simple binary joins and/or
OLTP transactions.
Bouganim et al.~\cite{bfv:vldb96} propose methods for optimizing
load-balancing on each site of a hierarchical architecture at run-time
so that inter-site data transfers are minimized.
In their model, the optimizer still has to determine the assignment of
operators to sites and the run-time environment has to make up 
for optimizer inaccuracies.
The issue of how the high-level mapping should be done at the
optimizer is not addressed.

A common characteristic of all approaches described above
is that they consider a one-dimensional model of resource
allocation based on a scalar cost metric (e.g., ``work''),
which ignores any possibilities for effective resource sharing among
concurrent operations.
Perhaps the only exception is Hong's method  for exploiting independent
parallelism in the XPRS shared-memory database system~\cite{hong:sigmod92}.
His approach is based on dynamically balancing resource use between
one I/O-bound and one CPU-bound operator pipeline to ensure that the system 
always executes at its IO-CPU balance point.
However, the substantial cost of communication renders such a scheduling 
method impractical for shared-nothing architectures.
In more recent work, Nippl and Mitschang~\cite{nm:vldb98} discuss the 
design of TOPAZ, a rule-driven, multi-phase parallelizer with a cost model
that takes into account CPU, disk, and communication costs, as well as
memory usage.
However, the target architecture of the TOPAZ parallelizer is based on 
the shared-disk  paradigm and the proposed strategies are primarily concerned 
with determining the degree of parallelism for operators and not with
mapping operator clones to distributed resource sites.

Moving away from the database field, there is a significant body of
work on parallel task scheduling in the field of deterministic
scheduling theory.
Since the problem is \np-hard in the strong sense~\cite{dl:siamdm89},
 research efforts have concentrated on providing fast heuristics with 
provable worst case bounds on the suboptimality of the solution.
However, scheduling query plans on shared-nothing parallel architectures
requires a significantly richer model of parallelization than what
is assumed in the classical \cite{graham:belltj66,gg:siamjc75,gllk:annalsDM79}
or more 
recent \cite{bb:icpp90,bb:ipps91,km:toc92,twy:spaa92,wc:siamjc92,st:ipl99,cm:spaa96}
efforts in that field.
To the best of our knowledge, there have been no theoretical results in the literature
on parallel task scheduling that consider multiple \tsr\ system resources
and explore sharing of such resources among concurrent tasks, or study the
implications of pipelined parallelism and data communication
costs.
Perhaps most importantly, even the very recent results of Chakrabarti and
Muthukrishnan~\cite{cm:spaa96} and Shachnai and Turek~\cite{st:ipl99}
on multi-resource scheduling are based on the assumption that all resources are
{\em globally accessible\/} to all tasks.
In contrast, our target architectures are characterized by a 
physical distribution of resource units and an
{\em affinity\/} of system resources to sites:
an operation scheduled at a particular site can only make use
of the resources locally available to that operation.
To the best of our knowledge, there are no previous theoretical
results on multi-resource scheduling in this context.

Finally, it is worth noting that the two resource classes (i.e., \tsr\
and \ssr) considered in this paper
have been identified in prior work, e.g., the ``stretchable'' and 
``non-stretchable'' resources of Pirahesh et al.~\cite{pmcls:dbpds90} and 
Ganguly et al.~\cite{ghk:sigmod92}.
The idea of lower bounding the makespan of a schedule based on a resource-time
product (like our ``volume'' bound) dates back to the seventies and the
pioneering work of Garey and Graham on multi-resource 
scheduling~\cite{gg:siamjc75}.
Yu and Cornell~\cite{yc:vldbj93} have  used the memory-time product 
of a join operator to define its ``return on consumption'' which, as 
they demonstrate, can be used to effectively allocate memory among 
multiple competing single-join queries.
Other work has also suggested heuristic strategies for allocating
memory statically or dynamically to complex query 
plans~\cite{bkv:cikm98,nd:cikm98}.
We should stress, however, that the general problem of scheduling operator
graphs on a parallel system with both \tsr\ and \ssr\ resources has not been
addressed in prior work on databases or deterministic scheduling theory.

%% file: concl.tex
\section{Conclusions}
\label{sec.concl}

\noindent
The problem of scheduling complex queries in shared-nothing parallel database
systems of multiple time-shared and space-shared resources has been open for
a long time both within the database field and the deterministic scheduling
theory field.
Despite the importance of such architectures in practice, the difficulties of
the problem have led researchers in making various assumptions and
simplifications that are not realistic.
In this paper, we have provided what we believe is the first comprehensive
formal approach to the problem.
We have established a model of resource usage that allows the scheduler to
explore the possibilities for concurrent operations sharing both \tsr\ and
\ssr\ resources and quantify the effects of this sharing on the parallel
execution time.
The inclusion of both types of resources has given rise to interesting
tradeoffs with respect to the degree of partitioned parallelism, which are
nicely exposed within our analytical models and results, and for which we
have provided some effective resolutions.
We have provided efficient, near-optimal heuristic algorithms for
query scheduling in such parallel environments, paying special attention
to various constraints that arise from the existence of \ssr\ resources,
including the co-scheduling requirements of pipelined operator execution, which
has been the most challenging to resolve.
Our set of results apply to all types of query plans and even sets of plans
that are either provided all at the beginning or arrive dynamically for
scheduling.
As a side-effect of our effort, we have identified the importance of a parameter
that captures one aspect of parallel query execution cost, which should play
an important role in obtaining realistic cost models for parallel query
optimization.
Finally, we have presented a set of experimental results from the
implementation of our scheduling algorithms on top of a detailed
simulation environment for shared-nothing database systems based on
the Gamma parallel database machine.
These results have verified the effectiveness of our scheduling
algorithms in a realistic system setting.

Our effort suggests several directions for future research.
First, it is a challenging problem to extend our multi-dimensional scheduling
algorithms and results to the more general {\em malleable\/} scheduling scenario,
in which  the scheduler is allowed to trade off some types of resources for 
others.
For example, the number of pages reserved for a hash-join operator can be
anywhere within a range of possible allocations with smaller allocations
typically implying more disk I/O.
Scheduling multiple such malleable operators is a very 
hard problem.
%
%
Second, we would like to investigate the effect of our results
on parallel query optimization and see if they lead to efficient and
accurate optimizers.
Finally, given the generality of our scheduling framework,
it would be interesting to investigate its applicability to 
other similar environments, such as multimedia storage servers.
These questions form the basis of our current and future research.

%% file: appendix.tex
\section{Proofs of Theoretical Results}
\label{sec.proofs}

\noindent
\underline{\bf Proof of Lemma~\ref{lem.newlb}:}\hspace{1em}
Given our earlier results for the case of only \tsr\ resources~\cite{gi:sigmod96},
we only need to establish the result concerning the new ``volume'' term in the
$\max$ function.
Let $T_{ij}$, $\overline{W}_{ij}$, and $\overline{V}_{ij}$ denote,
respectively, the stand-alone time, work vector, and demand vector
of  the $j$-th clone of $\mop_{i}$ ($j = 1,\ldots, N_{i}$).
Note that the $j$-th clone of $\mop_{i}$ will require resource
fractions $\overline{V}_{ij}$ for {\em at least\/} $T_{ij}$
time.
(It may be longer if the clone experiences contention on some
preemptable resource(s).)
Thus,  for {\em any\/} schedule SCHED, the total resource-time product
for the parallel execution must exceed these requirements for all
non-preemptable resources.
That is, the parallel execution time of SCHED on $P$ sites
should satisfy the following (componentwise) inequality:
\[
T^{par}(\msched, P)\ms P\ms \overline{1}  \ms\geq_{s}\ms
                       \sum_{i,j}T_{ij} \overline{V}_{ij},
\]
where $\overline{1}$ denotes an $s$-dimensional vector of 1's.
Taking this inequality for the maximum component of the
right hand side gives the desired result.
\hfill\csbbox\\
\vthm

\noindent
\underline{\bf Proof of Theorem~\ref{thm.basels}:}\hspace{1em}
Let $n_i$ denote the number of compatible clone subsets in site $B_i$ and
let $S_{ij}$ be the $j$-th such subset.
Define $S^V_{ij}$, $S^W_{ij}$, $S^{TV}_{ij}$, and $T^{max}(S_{ij})$ in
the usual manner (see Section~\ref{sec.notation}).
Consider any two consecutive compatible subsets $S_{ij}$ and
$S_{i,j+1}$ in site $B_i$.
Let $c_m = (T_m, \overline{W}_m, \overline{V}_m)$ be the first clone
placed in  $S_{i,j+1}$.
By the operation of the algorithm we know that $l(S^V_{ij}) + l(\overline{V}_m) >1$.
Further, by the order of clone placement we know that for all clones
$c_k$ placed in $S_{ij}$ we will have $T_k \geq T_m = T^{max}(S_{i,j+1})$.
So, considering the total volume packed on the two shelves we have:
\[
l(S^{TV}_{ij}) + l(S^{TV}_{i,j+1}) \geq T^{max}(S_{i,j+1})\cdot l(S^V_{ij}) +
                                 T^{max}(S_{i,j+1})\cdot l(\overline{V}_m) > 
							       T^{max}(S_{i,j+1}).
\]

Thus, taking the sum of $T^{max}$'s over all compatible subsets in  $B_i$ we have:
\[
\sum_{j=1}^{n_i} T^{max}(S_{ij}) < T^{max}(S_{i,1}) + \sum_{j=2}^{n_i}l(S^{TV}_{i,j-1}) +
                                       \sum_{j=2}^{n_i}l(S^{TV}_{ij})
                       \leq T^{max}(S) + 2\cdot \sum_{j=1}^{n_i}l(S^{TV}_{ij}).
\]

Consider the placement of the final clone
$c_N = (T_N, \overline{W}_N, \overline{V}_N)$.
Without loss of generality, assume that the clone is placed in $B_1$.
%
%
%
By the operation of our list scheduling heuristic we know that all the other
bins must be of height larger than or equal to the height of $B_1$.
Thus, using the notation of Section~\ref{sec.partindep}, we have
\[
P\cdot T^{site}(B_1) \leq \sum_{i,j}T(S_{ij})  = 
				 \sum_{i,j} \max\{ T^{max}(S_{ij}) , l(S^W_{ij}) \}
                                   \leq \sum_{i,j} T^{max}(S_{ij})  \ts+\ts
                                        \sum_{i,j} l(S^W_{ij}).
\]
And using our previous inequality, we have:
\begin{equation}
P\cdot T^{site}(B_1) <  P\cdot T^{max} + 2\cdot \sum_{i,j}S^{TV}_{ij} + 
			\sum_{i,j}l(S^W_{ij}).
\label{eqn.tsite}
\end{equation}

The following ancillary lemma establishes an fundamental property of the
``vector length'' ($l()$) function.

\begin{lem}
Let $S$ be a set of $d$-dimensional vectors, and
let $\pi=\{S_{1},\ldots,S_{k}\}$ be {\bf any}
partition of $S$.
Then,
\[
\frac{\sum_{i=1}^{k}l(S_{i})}{d} \ms\leq\ms l(S) \ms\leq\ms \sum_{i=1}^{k}l(S_{i})
\]
\label{lem.lproperty}
\end{lem}
\noindent
{\bf Proof:}\hspace{.5em}
The inequality $l(S) \leq \sum_{i=1}^{k}l(S_{i})$ is obvious
by the definition of $l()$.
For the other side of the inequality, let
$\overline{v}_{i} = \sum_{\overline{w}\in S_{i}}\overline{w}$
for each $i=1,\ldots,k$.
Also, define the function
$m(\overline{w}) = \min_{1\leq j\leq d}\{j: l(\overline{w}) = w[j]\}$, for
each $d$-dimensional vector $\overline{w}$.

\noindent
Rearranging the sum $\sum_{i=1}^{k}l(S_{i})$, we obtain
\begin{eqnarray*}
\sum_{i=1}^{k}l(S_{i})&=& \sum_{i=1}^{k}\max_{1\leq j\leq d}\{ v_{i}[j] \} \\
     &=&
\sum_{\overline{v}_{i}:m(\overline{v}_{i}) = 1}
    \max_{1\leq j\leq d}\{ v_{i}[j] \} + \ldots +
\sum_{\overline{v}_{i}:m(\overline{v}_{i}) = d}
    \max_{1\leq j\leq d}\{ v_{i}[j] \}
\end{eqnarray*}
Observe that, by the definition of $m()$, the $\sum$ and $\max$
functions can be legally interchanged in each of the above $d$ terms.
Therefore,
\begin{eqnarray*}
\sum_{i=1}^{k}l(S_{i})&=&
\max_{1\leq j\leq d}\{
\sum_{\overline{v}_{i}:m(\overline{v}_{i}) = 1}v_{i}[j]
                    \}
                    +\ldots+
\max_{1\leq j\leq d}\{
\sum_{\overline{v}_{i}:m(\overline{v}_{i}) = d}{v}_{i}[j]
                    \}\\
 &\leq&
 d \max_{1\leq j\leq d}\{ \sum_{i=1}^{k}v_{i}[j] \} \ms=\ms
 d \max_{1\leq j\leq d}\{ \sum_{i=1}^{N}w_{i}[j] \} \ms=\ms d l(S)
\end{eqnarray*}
Which, of course, implies $l(S)\geq \frac{\sum_{i=1}^{k}l(S_{i})}{d}$.
\hfill\csbbox\\

We can now combine Inequality~\ref{eqn.tsite} and  Lemma~\ref{lem.lproperty} 
to get:
\[
T^{site}(B_1) < T^{max}(S) + 2s\cdot\frac{l(S^{TV})}{P} + d\cdot\frac{l(S^W)}{P}.
\]
Finally, note that the response time of the schedule SCHED obtained by our
heuristic will certainly satisfy 
$T^{par}(\msched, P) \leq T^{site}(B_1) + T_{N} \leq T^{site}(B_1) + T^{max}(S)$.
Combining this with the above inequality for $T^{site}(B_1)$ gives the result.
\hfill\csbbox\\
\vthm

\noindent
\underline{\bf Proof of Lemma~\ref{lem.ubnosites}:}\hspace{1em}
Assume the claim is false.
This implies that there exists at least one \ssr\ vector $\vvec$ in $S^V$ that cannot
fit in any of the $\frac{l(S^V)\cdot s}{1-\lambda}$ sites used thus far.
Since $l(\vvec)\leq \lambda$, this means that the \ssr\ ``length'' $l(B^V_i)$ of
all these bins $B_i$ will be $l(B_i^V) > 1-\lambda$.
Summing over all bins this gives:
\[
\sum_{i}l(B^V_i) > (1-\lambda)\cdot \frac{l(S^V)\cdot s}{1-\lambda} =
                 l(S^V)\cdot s
\]
Since $\sum_{i}l(B^V_i)\leq s\cdot l(S^V-\{\vvec\})$ (by Lemma~\ref{lem.lproperty}), 
the above inequality gives:
$l(S^V-\{\vvec\}) > l(S^V)$, which is impossible.
This completes the proof of the stated upper bound.

This result basically states that, given an arbitrary collection $S^V$
of $\lambda$-granular $s$-dimensional vectors and an arbitrary rule for
partitioning these vectors into $s$-dimensional bins of unit capacity 
such that each bin is filled to a height  of at least $1-\lambda$,
then the number of bins produced cannot exceed 
$\frac{l(S^V)\cdot s}{1-\lambda}$.
To show that this upper bound  is tight, we demonstrate a 
{\em worst-case collection\/} of $s$-dimensional demand vectors 
$S^V$ and a {\em worst-case partitioning\/} of these vectors for 
which the stated  upper bound on the required number of sites is 
actually reached.
Consider a collection $S^V$ of $P\cdot(k-1)$  $s$-dimensional vectors,
where $P$ and $k$ are positive integers.
Suppose that each vector has a non-zero value of $\frac{1}{k-\epsilon}$ 
($\epsilon>0$ a small positive number) in only one of its components
and a value of zero in all other components.
(Note that, all these vectors are $\lambda$-granular with 
$\lambda = 1/(k-\epsilon)$.)
Furthermore, for every vector component $i$, there is an equal number
(i.e., $\frac{P}{s}\cdot (k-1)$) of vectors with a non-zero value at
component $i$, for $i=1,\ldots,s$.
Without loss of generality, we assume that $P/s$ is an integer.
Now, consider the (worst-case) partitioning of these vectors that
places only vectors with the {\em same  non-zero dimension\/} in 
the same bin.
(Note that such an ``inefficient'' packing rule could be imposed by
other scheduling criteria, for example, balancing work requirements 
across bins.)
It is fairly easy to see that such a packing will require exactly $P$
bins, since each bin can accommodate at most $k-1$ vectors with the
same non-zero component of $1/(k-\epsilon)$.
We will now show that this essentially coincides with the upper bound
stated in this lemma. 
Our upper bound for this case is:
\[
\frac{l(S^V)\cdot s}{1-\lambda} = 
\frac{\frac{P}{s}\cdot (k-1)\cdot \frac{1}{k-\epsilon} \cdot s}
     {1 - \frac{1}{k-\epsilon}} =
P\cdot \frac{k-1}{k-1-\epsilon},
\]
which obviously converges to $P$ as $\epsilon\rightarrow 0$.
\hfill\csbbox\\
\vthm

\noindent
\underline{\bf Proof of Theorem~\ref{thm.pipesched}:}\hspace{1em}
\noindent
First, note that by Lemma~\ref{lem.ubnosites} \PipeSched\ will be
able to pack $C$ in $P_C$.
We now  prove the following proposition that holds for  any
packing of  $S^W$ into $P_C$ $d$-dimensional bins.

\begin{prop}
\rm
There exists an index $j\in \{1,\ldots, P_C\}$ such that
\[
\sum_{\wvec\in B_j^W} l(\wvec) \leq d\cdot\frac{l(S^W)}{P_C}
\]
\label{prop.exist}
\end{prop}
\noindent
{\bf Proof:}\hspace{.5em}
Assume that, to the contrary, we have:
\[
\sum_{\wvec\in B_j^W} l(\wvec) >  d\cdot\frac{l(S^W)}{P_C}
\mbox{, for all $j = 1,\ldots,P_C$.}
\]
Summing over all $j$, this gives
\[
\sum_{\wvec\in S^W} l(\wvec)  >  d\ms l(S^W)
\]
which contradicts Lemma~\ref{lem.lproperty}.
\hfill\csbbox\\

Consider the work vector  packing produced by \PipeSched, where,
without loss of generality, we assume that the sites have been
renumbered in non-increasing order of total work; that is,
$l(B_{1}^W) \geq l(B_{2}^W) \geq\ldots\geq l(B_{P_C}^W)$
(Figure~(\ref{fig.packing})).

\begin{SingleSpace}

\epsfig{0.8}{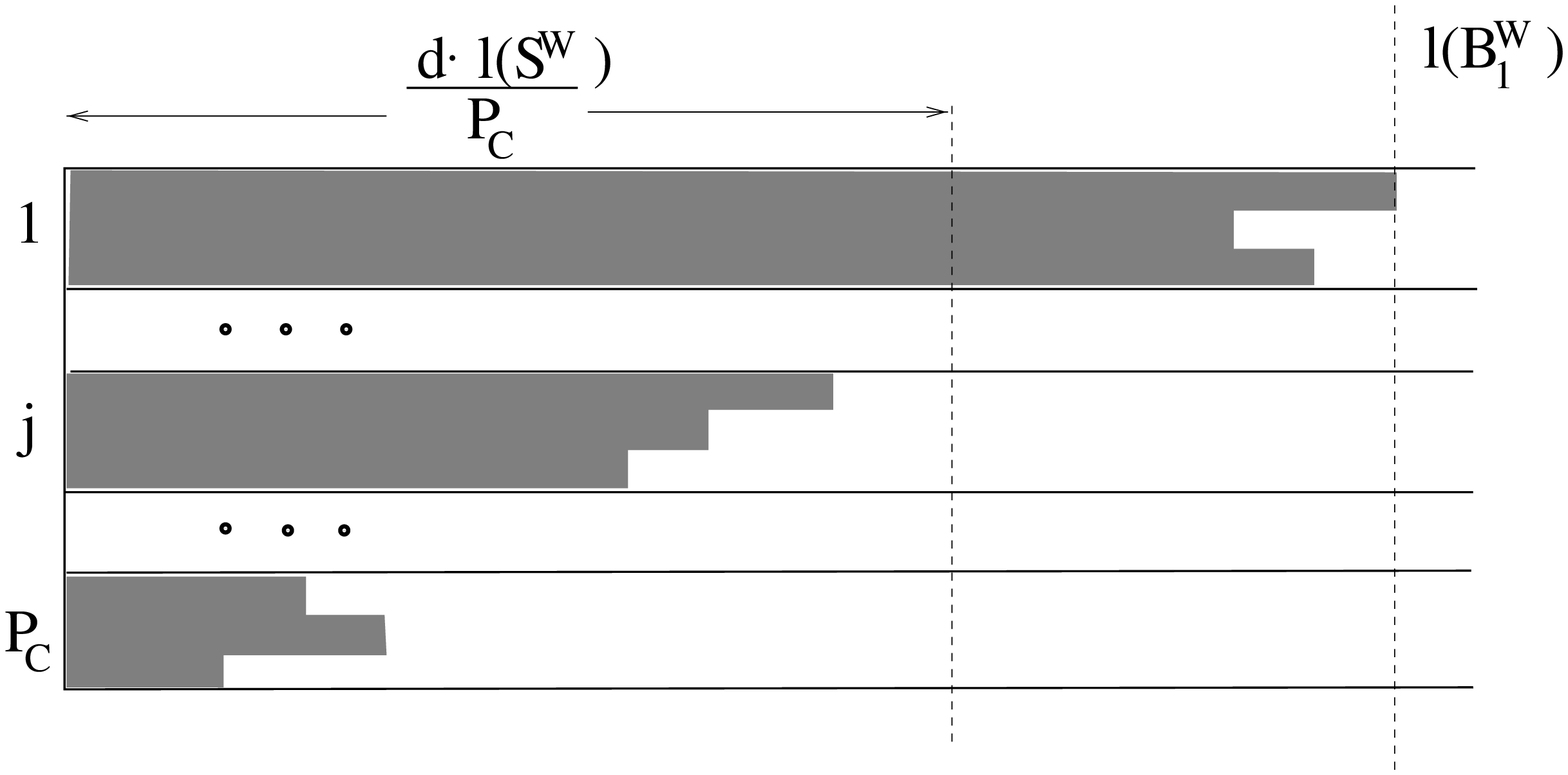}{A packing of $S$'s  work vectors  in $P$ sites.}
{fig.packing}

\end{SingleSpace}

If $l(B_{1}^W) \leq d\cdot\frac{l(S^W)}{P_C}$, the theorem obviously
holds.
Assume $l(B_{1}^W) > d\cdot\frac{l(S^W)}{P_C}$
Let $\overline{w}_{i_{0}}$ be the first work vector to ``push'' the
total work of site $B_1$ over $d\cdot\frac{l(S^W)}{P_C}$.
Also let $W_e = < \overline{w}_{i_{0}},\ldots, \overline{w}_{i_{M}} >$,
$m\geq 0$, be the time-ordered list of vectors packed at site
 $B_1$ after that moment (including $\overline{w}_{i_{0}}$).

By Proposition~\ref{prop.exist}, we know that in the packing produced
by \PipeSched\ there exists a site $B_j$ such that
$\sum_{\wvec\in B_j^W} l(\wvec) \leq d\cdot\frac{l(S^W)}{P_C}$.
By the logic of \PipeSched\ we know that site $B_j$ was not allowed
to pack any of the vectors $\overline{w}_{i_{k}}$, $k=1,\ldots,M$
only because of ss resource constraints.
This implies that when the packing of $\wvec_{i_1}$ was taking place
we had:
\[
\sum_{\vvec\in B_j^V} l(\vvec) \geq l(B_j^V) > 1 - \lambda.
\]

\noindent
Also, by the order of packing we know that at the time of packing
$\wvec_{i_1}$ we had:
\[
\frac{l(\wvec_{i_1})}{l(\vvec_{i_1})} \leq
             \frac{\sum_{\wvec\in B_j^V} l(\wvec)}{\sum_{\vvec\in B_j^V} l(\vvec)},
\]
which by the above inequality and Proposition~\ref{prop.exist} gives:
\[
\frac{l(\wvec_{i_1})}{l(\vvec_{i_1})} \leq
                    \frac{d\cdot l(S^W)}{P_C\cdot(1 - \lambda)}.
\]

\noindent
Again, by the order of packing we know that
$\frac{l(\wvec_{i_1})}{l(\vvec_{i_1})} \geq \frac{l(\wvec_{i_2})}{l(\vvec_{i_2})}
\geq\ldots\geq\frac{l(\wvec_{i_M})}{l(\vvec_{i_M})}$,
which in turn implies that:
\[
\frac{l(\wvec_{i_1})}{l(\vvec_{i_1})} \geq
\frac{\sum_{k=1}^{M} l(\wvec_{i_k})}{\sum_{k=1}^{M} l(\vvec_{i_k})}.
\]

\noindent
Combining the last two inequalities and using the fact that
$\sum_{k=1}^{M} l(\vvec_{i_k}) \leq s\cdot l(\sum_{k=1}^{M}\vvec_{i_k}) \leq s$
(since all these ss vectors ``fit'' in one site), we have:
\[
\frac{\sum_{k=1}^{M} l(\wvec_{i_k})}{s} \leq
                        \frac{d\cdot l(S^W)}{P_C\cdot(1 - \lambda)},
\]
or,
\[
\sum_{k=1}^{M} l(\wvec_{i_k}) \leq
              \frac{d\cdot s}{1 - \lambda} \cdot
                   \frac{l(S^W)}{P_C}.
\]

\noindent
Thus, the response time of the schedule produced by  \PipeSched\ can be
bounded as follows (see Figure~\ref{fig.packing}):
\[
T_H \leq d\cdot\frac{l(S^W)}{P_C} + \sum_{k=1}^{M} l(\wvec_{i_k})
                                  + l(\wvec_{i_0})
     \leq d\cdot (1 + \frac{s}{1 - \lambda}) \cdot \frac{l(S^W)}{P_C} + T^{max}.
\]
This completes the proof.
\hfill\csbbox\\
\vthm

\noindent
\underline{\bf Proof of Theorem~\ref{thm.levelsched}:}\hspace{1em}
Let $S^W_j = \cup_{C\in L_j} S^W_C$ for all $j= 1,\ldots,k$,
with $S^V_j$, $S^{TV}_j$ defined similarly.
Also, define $T^{max}_j = \max_{C\in L_j} T^{max}_C$.
Finally, let $H_j$ denote the parallel execution time of the
$j^{th}$ layer (i.e., the clones in $L_j$) as determined by
\PipeSched.
From Theorem~\ref{thm.pipesched} we know that
$H_j \ms\leq\ms d(1 + \frac{s}{1 - \lambda}) \cdot
                   \frac{l(S^W_j)}{P} + T^{max}_j$,
for all $j= 1,\ldots,k$.
Thus, for the overall execution time we have:
\begin{equation}
T_H = \sum_{j=1}^{k} H_j \ts\leq\ts
      d(1 + \frac{s}{1 - \lambda}) \cdot \sum_{j=1}^{k}\frac{l(S^W_j)}{P}
      \ms+\ms \sum_{j=1}^{k} T^{max}_j.
\label{eqn.levelschedtime}
\end{equation}

By the ordering of the pipelines in $L$ we know that the total volume
packed in layer $j$ is
$l(S^{TV}_j) \geq T^{max}_{j+1} \cdot l(S^{V}_j)$ for all
$j=1,\ldots, k-1$.
Furthermore, by the condition used in the layering of the pipes
(Step 2 of \LevelSched), we have:
\[
l(S^{TV}_j) + l(S^{TV}_{j+1}) \ts\geq\ts
      T^{max}_{j+1}\cdot [l(S^{V}_j) + l(S^{V}_{C_{i_j +1}})]
      \ts>\ts T^{max}_{j+1}\cdot \frac{P (1-\lambda)}{s}
\]
for all $j=1,\ldots, k-1$.
Summing over all $j$ this gives
\[
2\cdot\sum_{1}^{k} l(S^{TV}_j) \ts>\ts
   \sum_{1}^{k-1} [ l(S^{TV}_j) + l(S^{TV}_{j+1}) ]
   \ts>\ts \frac{P (1-\lambda)}{s} \sum_{2}^{k}T^{max}_{j}.
\]

\noindent
Combining this with Inequality~\ref{eqn.levelschedtime} we
get:
\[
T_H \ts<\ts d(1 + \frac{s}{1 - \lambda})
                  \cdot \sum_{j=1}^{k}\frac{l(S^W_j)}{P}
            \ms+\ms \frac{2s}{1-\lambda}\cdot
                         \sum_{j=1}^{k} \frac{l(S^{TV}_j)}{P}
            \ms+\ms T^{max}_1.
\]
Note that by the ordering of $L$, $T^{max}_1 = T^{max}$ (the overall
maximum).
Using the fundamental propery of $l()$ for the two summations in the
above inequality we have:
\[
T_H \ts<\ts d^2(1 + \frac{s}{1 - \lambda}) \cdot \frac{l(S^W)}{P}
            \ms+\ms \frac{2s^2}{1-\lambda}\cdot \frac{l(S^{TV})}{P}
            \ms+\ms T^{max}.
\]

\hfill\csbbox\\


%% file: paper.bbl
\begin{thebibliography}{GLLRK79}

\bibitem[BB90]{bb:icpp90}
Krishna~P. Belkhale and Prithviraj Banerjee.
\newblock {``Approximate Algorithms for the Partitionable Independent Task
  Scheduling Problem''}.
\newblock In {\em Proceedings of the 1990 International Conference on Parallel
  Processing}, pages I72--I75, August 1990.

\bibitem[BB91]{bb:ipps91}
Krishna~P. Belkhale and Prithviraj Banerjee.
\newblock {``A Scheduling Algorithm for Parallelizable Dependent Tasks''}.
\newblock In {\em Proceedings of the Fifth International Parallel Processing
  Symposium}, pages 500--506, 1991.

\bibitem[BFG{\etalchar{+}}95]{bfgh:ibmsj95}
C.K. Baru, G.~Fecteau, A.~Goyal, H.~Hsiao, A.~Jhingran, S.~Padmanabhan, G.P.
  Copeland, and W.G. Wilson.
\newblock {``DB2 Parallel Edition''}.
\newblock {\em {IBM} Systems Journal}, 34(2):292--322, 1995.

\bibitem[BFMV00]{bfmv:icde00}
Luc Bouganim, Fran\c{c}oise Fabret, C.~Mohan, and Patrick Valduriez.
\newblock {``Dynamic Query Scheduling in Data Integration Systems''}.
\newblock In {\em Proceedings of the Sixteenth International Conference on Data
  Engineering}, pages 425--434, San Diego, California, March 2000.

\bibitem[BFV96]{bfv:vldb96}
Luc Bouganim, Daniela Florescu, and Patrick Valduriez.
\newblock {``Dynamic Load Balancing in Hierarchical Parallel Database
  Systems''}.
\newblock In {\em Proceedings of the 22nd International Conference on Very
  Large Data Bases}, pages 436--447, Mumbai(Bombay), India, September 1996.

\bibitem[BKV98]{bkv:cikm98}
Luc Bouganim, Olga Kapitskaia, and Patrick Valduriez.
\newblock {``Memory-Adaptive Scheduling for Large Query Execution''}.
\newblock In {\em Proceedings of the Seventh International Conference on
  Information and Knowledge Management}, pages 105--115, Bethesda, Maryland,
  November 1998.

\bibitem[Bro94]{brown:prpl94}
Kurt Brown.
\newblock {``PRPL: A Database Workload Specification Language (Version 1.4)''}.
\newblock Unpublished Manuscript, March 1994.

\bibitem[BS83]{bs:siamjc83}
Brenda~S. Baker and Jerald~S. Schwarz.
\newblock {``Shelf Algorithms for Two-Dimensional Packing Problems''}.
\newblock {\em {SIAM} Journal on Computing}, 12(3):508--525, August 1983.

\bibitem[CGJ84]{cgj:survey84}
E.G. Coffman, Jr., M.R. Garey, and D.S. Johnson.
\newblock {``Approximation Algorithms for Bin-Packing -- An Updated Survey''}.
\newblock In {\em {``Algorithm Design for Computing System Design''}}, pages
  49--106. Springer-Verlag, New York, 1984.

\bibitem[CGJT80]{cgjt:siamjc80}
E.G. Coffman, Jr., M.R. Garey, D.S. Johnson, and R.E. Tarjan.
\newblock {``Performance Bounds for Level-Oriented Two-Dimensional Packing
  Algorithms''}.
\newblock {\em {SIAM} Journal on Computing}, 9(4):808--826, November 1980.

\bibitem[CHM95]{chm:pods95}
Chandra Chekuri, Waqar Hasan, and Rajeev Motwani.
\newblock {``Scheduling Problems in Parallel Query Optimization''}.
\newblock In {\em Proceedings of the Fourteenth {ACM SIGACT-SIGMOD-SIGART}
  Symposium on Principles of Database Systems}, pages 255--265, San Jose,
  California, May 1995.

\bibitem[CLYY92]{clyy:vldb92}
Ming-Syan Chen, Ming-Ling Lo, Philip~S. Yu, and Honesty~C. Young.
\newblock {``Using Segmented Right-Deep Trees for the Execution of Pipelined
  Hash Joins''}.
\newblock In {\em Proceedings of the Eighteenth International Conference on
  Very Large Data Bases}, pages 15--26, Vancouver, Canada, August 1992.

\bibitem[CM96]{cm:spaa96}
Soumen Chakrabarti and S.~Muthukrishnan.
\newblock {``Resource scheduling for parallel database and scientific
  applications''}.
\newblock In {\em Proceedings of the Eighth Annual {ACM} Symposium on Parallel
  Algorithms and Architectures}, pages 329--335, Padua, Italy, June 1996.

\bibitem[CYW92]{cyw:icde92}
Ming-Syan Chen, Philip~S. Yu, and Kun-Lung Wu.
\newblock {``Scheduling and Processor Allocation for Parallel Execution of
  Multi-Join Queries''}.
\newblock In {\em Proceedings of the Eighth International Conference on Data
  Engineering}, pages 58--67, Phoenix, Arizona, February 1992.

\bibitem[DG92]{dg:cacm92}
David~J. DeWitt and Jim Gray.
\newblock {``Parallel Database Systems: The Future of High Performance Database
  Database Systems''}.
\newblock {\em Communications of the {ACM}}, 35(6):85--98, June 1992.

\bibitem[DGS{\etalchar{+}}90]{dgsb:tkde90}
David~J. DeWitt, Shahram Ghandeharizadeh, Donovan~A. Schneider, Allan Bricker,
  Hui-I Hsiao, and Rick Rasmussen.
\newblock {``The Gamma Database Machine Project''}.
\newblock {\em {IEEE} Transactions on Knowledge and Data Engineering},
  2(1):44--62, March 1990.

\bibitem[DL89]{dl:siamdm89}
Jianzhong Du and Joseph Y-T. Leung.
\newblock {``Complexity of Scheduling Parallel Task Systems''}.
\newblock {\em {SIAM} Journal on Discrete Mathematics}, 2(4):473--487, November
  1989.

\bibitem[EGH95]{egh:sigmodrec95}
Susanne Englert, Ray Glasstone, and Waqar Hasan.
\newblock {``Parallelism and its Price: A Case Study of NonStop SQL/MP''}.
\newblock {\em {ACM SIGMOD} Record}, 24(4):61--71, December 1995.

\bibitem[GG75]{gg:siamjc75}
M.R. Garey and R.L. Graham.
\newblock {``Bounds for Multiprocessor Scheduling with Resource Constraints''}.
\newblock {\em {SIAM} Journal on Computing}, 4(2):187--200, June 1975.

\bibitem[GGS96]{ggs:pods96}
Sumit Ganguly, Akshay Goel, and Avi Silberschatz.
\newblock {``Efficient and Accurate Cost Models for Parallel Query
  Optimization''}.
\newblock In {\em Proceedings of the Fifteenth {ACM SIGACT-SIGMOD-SIGART}
  Symposium on Principles of Database Systems}, Montreal, Quebec, June 1996.

\bibitem[GGW95]{ggw:cismod95}
Sumit Ganguly, Apostolos Gerasoulis, and Weining Wang.
\newblock {``Partitioning Pipelines with Communication Costs''}.
\newblock In {\em Proceedings of the 6th International Conference on
  Information Systems and Data Management {(CISMOD'95)}}, pages 302--320,
  Bombay, India, November 1995.

\bibitem[GHK92]{ghk:sigmod92}
Sumit Ganguly, Waqar Hasan, and Ravi Krishnamurthy.
\newblock {``Query Optimization for Parallel Execution''}.
\newblock In {\em Proceedings of the 1992 {ACM SIGMOD} International Conference
  on Management of Data}, pages 9--18, San Diego, California, June 1992.

\bibitem[GI96]{gi:sigmod96}
Minos~N. Garofalakis and Yannis~E. Ioannidis.
\newblock {``Multi-dimensional Resource Scheduling for Parallel Queries''}.
\newblock In {\em Proceedings of the 1996 {ACM SIGMOD} International Conference
  on Management of Data}, pages 365--376, Montreal, Quebec, June 1996.

\bibitem[GI97]{gi:vldb97}
Minos~N. Garofalakis and Yannis~E. Ioannidis.
\newblock {``Parallel Query Scheduling and Optimization with Time- and
  Space-Shared Resources''}.
\newblock In {\em Proceedings of the 23rd International Conference on Very
  Large Data Bases}, pages 296--305, Athens, Greece, August 1997.

\bibitem[GI{\"O}98]{gio:vldb98}
Minos~N. Garofalakis, Yannis~E. Ioannidis, and Banu {\"O}zden.
\newblock {``Resource Scheduling for Composite Multimedia Objects''}.
\newblock In {\em Proceedings of the 24th International Conference on Very
  Large Data Bases}, New York, USA, August 1998.

\bibitem[GJ79]{garey-johnson:79}
M.R. Garey and D.S. Johnson.
\newblock {\em {``Computers and Intractability: A Guide to the Theory of
  NP-Completeness''}}.
\newblock W.H. Freeman, 1979.

\bibitem[GLLRK79]{gllk:annalsDM79}
R.L. Graham, E.L. Lawler, J.K. Lenstra, and A.H.G. Rinnooy~Kan.
\newblock {``Optimization and Approximation in Deterministic Sequencing and
  Scheduling: A Survey''}.
\newblock {\em Annals of Discrete Mathematics}, 5:287--326, 1979.

\bibitem[GMSY93]{gmsy:orsaj93}
Shahram Ghandeharizadeh, Robert~R. Meyer, Gary~L. Schultz, and Jonathan Yackel.
\newblock {``Optimal Balanced Assignments and a Parallel Database
  Application''}.
\newblock {\em {ORSA} Journal on Computing}, 5(2):151--167, Spring 1993.

\bibitem[Gra66]{graham:belltj66}
R.L. Graham.
\newblock {``Bounds for Certain Multiprocessing Anomalies''}.
\newblock {\em The Bell System Technical Journal}, 45:1563--1581, November
  1966.

\bibitem[Gra69]{graham:siamjc69}
R.L. Graham.
\newblock {``Bounds on Multiprocessing Timing Anomalies''}.
\newblock {\em {SIAM} Journal on Computing}, 17(2):416--429, March 1969.

\bibitem[Gra93]{graefe:surveys93}
Goetz Graefe.
\newblock {``Query Evaluation Techniques for Large Databases''}.
\newblock {\em ACM Computing Surveys}, 25(2):73--170, June 1993.

\bibitem[GW93]{gw:rutgers93}
Sumit Ganguly and Weining Wang.
\newblock {``Optimizing Queries for Coarse Grain Parallelism''}.
\newblock Technical Report LCSR-TR-218, Rutgers University, Dept. of Computer
  Sciences, October 1993.

\bibitem[Has95]{hasan:phdthesis95}
Waqar Hasan.
\newblock {\em {``Optimization of SQL Queries for Parallel Machines''}}.
\newblock PhD thesis, Stanford University, December 1995.

\bibitem[HCL{\etalchar{+}}90]{hclm:tkde90}
Laura~M. Haas, Walter Chang, Guy~M. Lohman, John McPherson, Paul~F. Wilms,
  George Lapis, Bruce~G. Lindsay, Hamid Pirahesh, Michael~J. Carey, and
  Eugene~J. Shekita.
\newblock {``Starburst Mid-Flight: As the Dust Clears''}.
\newblock {\em {IEEE} Transactions on Knowledge and Data Engineering},
  2(1):143--160, March 1990.

\bibitem[HCY94]{hcy:sigmod94}
Hui-I Hsiao, Ming-Syan Chen, and Philip~S. Yu.
\newblock {``On Parallel Execution of Multiple Pipelined Hash Joins''}.
\newblock In {\em Proceedings of the 1994 {ACM SIGMOD} International Conference
  on Management of Data}, pages 185--196, Minneapolis, Minnesota, May 1994.

\bibitem[HFV96]{hfv:sigmodrec96}
Waqar Hasan, Daniela Florescu, and Patrick Valduriez.
\newblock {``Open Issues in Parallel Query Optimization''}.
\newblock {\em {ACM SIGMOD} Record}, 25(3):28--33, September 1996.

\bibitem[HM94]{hm:vldb94}
Waqar Hasan and Rajeev Motwani.
\newblock {``Optimization Algorithms for Exploiting the
  Parallelism-Communication Tradeoff in Pipelined Parallelism''}.
\newblock In {\em Proceedings of the 20th International Conference on Very
  Large Data Bases}, pages 36--47, Santiago, Chile, August 1994.

\bibitem[Hon92]{hong:sigmod92}
Wei Hong.
\newblock {``Exploiting Inter-Operation Parallelism in XPRS''}.
\newblock In {\em Proceedings of the 1992 {ACM SIGMOD} International Conference
  on Management of Data}, pages 19--28, San Diego, California, June 1992.

\bibitem[HS91]{hs:pdis91}
Wei Hong and Michael Stonebraker.
\newblock {``Optimization of Parallel Query Execution Plans in XPRS''}.
\newblock In {\em Proceedings of the First International Conference on Parallel
  and Distributed Information Systems}, Miami Beach, Florida, December 1991.

\bibitem[HS93]{hs:dpdbs93}
Wei Hong and Michael Stonebraker.
\newblock {``Optimization of Parallel Query Execution Plans in XPRS''}.
\newblock {\em Distributed and Parallel Databases}, 1:9--32, 1993.

\bibitem[Ioa93]{ioan:vldb93}
Yannis~E. Ioannidis.
\newblock {``Universality of Serial Histograms''}.
\newblock In {\em Proceedings of the Nineteenth International Conference on
  Very Large Data Bases}, pages 256--267, Dublin, Ireland, August 1993.

\bibitem[JMP97]{jmp:debull97}
Anant Jhingran, Timothy Malkemus, and Sriram Padmanabhan.
\newblock {``Query Optimization in DB2 Parallel Edition''}.
\newblock {\em {IEEE} Data Engineering Bulletin}, 20(2):27--34, June 1997.
\newblock (Special Issue on Commercial Parallel Systems).

\bibitem[JPS93]{jps:apads93}
Anant Jhingran, Sriram Padmanabhan, and Ambuj Shatdal.
\newblock {``Join Query Optimization in Parallel Database Systems''}.
\newblock In {\em Proceedings of the {IEEE} Workshop on Advances in Parallel
  and Distributed Systems}, 1993.

\bibitem[KM92]{km:toc92}
Ramesh Krishnamurti and Eva Ma.
\newblock {``An Approximation Algorithm for Scheduling Tasks on Varying
  Partition Sizes in Partitionable Multiprocessor Systems''}.
\newblock {\em {IEEE} Transactions on Computers}, 41(12):1572--1579, December
  1992.

\bibitem[LCRY93]{lcry:sigmod93}
Ming-Ling Lo, Ming-Syan Chen, C.V. Ravishankar, and Philip~S. Yu.
\newblock {``On Optimal Processor Allocation to Support Pipelined Hash
  Joins''}.
\newblock In {\em Proceedings of the 1993 {ACM SIGMOD} International Conference
  on Management of Data}, pages 69--78, Washington, D.C., June 1993.

\bibitem[LVZ93]{lvz:vldb93}
Rosana~S.G. Lanzelotte, Patrick Valduriez, and Mohamed Za\"{i}t.
\newblock {``On the Effectiveness of Optimization Search Strategies for
  Parallel Execution Spaces''}.
\newblock In {\em Proceedings of the Nineteenth International Conference on
  Very Large Data Bases}, pages 493--504, Dublin, Ireland, August 1993.

\bibitem[MD95]{md:vldb95}
Manish Mehta and David~J. DeWitt.
\newblock {``Managing Intra-operator Parallelism in Parallel Database
  Systems''}.
\newblock In {\em Proceedings of the 21st International Conference on Very
  Large Data Bases}, pages 382--394, Zurich, Switzerland, September 1995.

\bibitem[MD97]{md:vldbj97}
Manish Mehta and David~J. DeWitt.
\newblock {``Data Placement in Shared-Nothing Parallel Database Systems''}.
\newblock {\em The {VLDB} Journal}, 6(1):53--72, January 1997.

\bibitem[Meh94]{mehta:phdthesis94}
Manish Mehta.
\newblock {\em {``Resource Allocation for Parallel Shared-Nothing Database
  Systems''}}.
\newblock PhD thesis, University of Wisconsin-Madison, 1994.

\bibitem[ND98]{nd:cikm98}
Biswadeep Nag and David~J. DeWitt.
\newblock {``Memory Allocation Strategies for Complex Decision Support
  Queries''}.
\newblock In {\em Proceedings of the Seventh International Conference on
  Information and Knowledge Management}, Bethesda, Maryland, November 1998.

\bibitem[NM98]{nm:vldb98}
Clara Nippl and Bernhard Mitschang.
\newblock {``TOPAZ: a Cost-Based, Rule-Driven, Multi-Phase Parallelizer''}.
\newblock In {\em Proceedings of the 24th International Conference on Very
  Large Data Bases}, New York, USA, August 1998.

\bibitem[NSHL95]{nshl:dimacs95}
Thomas~M. Niccum, Jaideep Srivastava, Bhaskar Himatsingka, and Jianzhong Li.
\newblock {``Query Optimization and Processing in Parallel Databases''}.
\newblock {\em {DIMACS} Series in Discrete Mathematics and Theoretical Computer
  Science}, 22:259--287, 1995.

\bibitem[NZT96]{nzt:sigmodrec96}
Michael~G. Norman, Thomas Zurek, and Peter Thanisch.
\newblock {``Much Ado About Shared-Nothing''}.
\newblock {\em {ACM SIGMOD} Record}, 25(3):16--21, September 1996.

\bibitem[{\"O}RS96]{ors:icmcs96}
Banu {\"O}zden, Rajeev Rastogi, and Avi Silberschatz.
\newblock {``Disk Striping in Video Server Environments''}.
\newblock In {\em Proceedings of the 1996 International Conference on
  Multimedia Computing and Systems}, Hiroshima, Japan, June 1996.

\bibitem[{\"O}RS97]{ors:pods97}
Banu {\"O}zden, Rajeev Rastogi, and Avi Silberschatz.
\newblock {``Multimedia Support for Databases''}.
\newblock In {\em Proceedings of the Sixteenth {ACM SIGACT-SIGMOD-SIGART}
  Symposium on Principles of Database Systems}, Tucson, Arizona, May 1997.

\bibitem[PI96]{pi:vldb96}
Viswanath Poosala and Yannis~E. Ioannidis.
\newblock {``Estimation of Query-Result Distribution and its Application in
  Parallel-Join Load Balancing''}.
\newblock In {\em Proceedings of the 22nd International Conference on Very
  Large Data Bases}, pages 448--459, Mumbai(Bombay), India, September 1996.

\bibitem[PMC{\etalchar{+}}90]{pmcls:dbpds90}
Hamid Pirahesh, C.~Mohan, Josephine Cheng, T.S. Liu, and Pat Selinger.
\newblock {``Parallelism in Relational Data Base Systems: Architectural Issues
  and Design Approaches''}.
\newblock In {\em Proceedings of the Second International Symposium on
  Databases in Parallel and Distributed Systems}, pages 4--29, Dublin, Ireland,
  July 1990.

\bibitem[RM95]{rm:vldb95}
Erhard Rahm and Robert Marek.
\newblock {``Dynamic Multi-Resource Load Balancing in Parallel Database
  Systems''}.
\newblock In {\em Proceedings of the 21st International Conference on Very
  Large Data Bases}, pages 395--406, Zurich, Switzerland, September 1995.

\bibitem[SAC{\etalchar{+}}79]{saclp:sigmod79}
P.~Selinger, M.M. Astrahan, D.D. Chamberlin, R.A. Lorie, and T.G. Price.
\newblock {``Access Path Selection in a Relational Database Management
  System''}.
\newblock In {\em Proceedings of the 1979 {ACM SIGMOD} International Conference
  on Management of Data}, pages 23--34, Boston, Massachusetts, June 1979.

\bibitem[Sch90a]{schneider:phdthesis90}
Donovan~A. Schneider.
\newblock {\em {``Complex Query Processing in Multiprocessor Database
  Machines''}}.
\newblock PhD thesis, University of Wisconsin-Madison, September 1990.

\bibitem[Sch90b]{schwetman:mcc90}
Herb Schwetman.
\newblock {``CSIM User's Guide''}.
\newblock Technical Report ACT-126-90, MCC, Austin, Texas, March 1990.

\bibitem[SE93]{se:pdis93}
Jaideep Srivastava and Gary Elsesser.
\newblock {``Optimizing Multi-Join Queries in Parallel Relational Databases''}.
\newblock In {\em Proceedings of the Second International Conference on
  Parallel and Distributed Information Systems}, pages 84--92, San Diego,
  California, January 1993.

\bibitem[{Sea}]{cheetah:spec00}
{Seagate Technology}.
\newblock {Cheetah 36 (ST136403FC) Technical Specifications}.
\newblock {({\tt http://www.seagate.com/})}.

\bibitem[SG98]{silberschatz-galvin:98}
Abraham Silberschatz and Peter Galvin.
\newblock {\em {``Operating System Concepts''}}.
\newblock Addison-Wesley Publishing Company, 1998.
\newblock (Fifth Edition).

\bibitem[Sha86]{shapiro:tods86}
Leonard~D. Shapiro.
\newblock {``Join Processing in Database Systems with Large Main Memories''}.
\newblock {\em {ACM} Transactions on Database Systems}, 11(3):239--264,
  September 1986.

\bibitem[ST99]{st:ipl99}
Hadas Shachnai and John~J. Turek.
\newblock {``Multiresource malleable task scheduling to minimize response
  time''}.
\newblock {\em Information Processing Letters}, 70(1):47--52, April 1999.

\bibitem[Sto86]{ston:debull86}
Michael Stonebraker.
\newblock {``The Case for Shared Nothing''}.
\newblock {\em {IEEE} Data Engineering Bulletin}, 9(1):4--9, March 1986.

\bibitem[SYT93]{syt:vldb93}
Eugene~J. Shekita, Honesty~C. Young, and Kian-Lee Tan.
\newblock {``Multi-Join Optimization for Symmetric Multiprocessors''}.
\newblock In {\em Proceedings of the Nineteenth International Conference on
  Very Large Data Bases}, pages 479--492, Dublin, Ireland, August 1993.

\bibitem[TL93]{tl:ipl93}
Kian-Lee Tan and Hongjun Lu.
\newblock {``On Resource Scheduling of Multi-join Queries in Parallel Database
  Systems''}.
\newblock {\em Information Processing Letters}, 48:189--195, 1993.

\bibitem[TWPY92]{twpy:sigmetrics92}
John Turek, Joel~L. Wolf, Krishna~R. Pattipati, and Philip~S. Yu.
\newblock {``Scheduling Parallelizable Tasks: Putting it All on the Shelf''}.
\newblock In {\em Proceedings of the 1992 {ACM SIGMETRICS} Conference on
  Measurement \& Modeling of Computer Systems}, pages 225--236, Newport, Rhode
  Island, June 1992.

\bibitem[TWY92]{twy:spaa92}
John Turek, Joel~L. Wolf, and Philip~S. Yu.
\newblock {``Approximate Algorithms for Scheduling Parallelizable Tasks''}.
\newblock In {\em Proceedings of the Fourth Annual {ACM} Symposium on Parallel
  Algorithms and Architectures}, pages 323--332, San Diego, California, June
  1992.

\bibitem[Val93]{valduriez:dpdbs93}
Patrick Valduriez.
\newblock {``Parallel Database Systems: Open Problems and New Issues''}.
\newblock {\em Distributed and Parallel Databases}, 1:137--165, 1993.

\bibitem[WC92]{wc:siamjc92}
Qingzhou Wang and Kam~Hoi Cheng.
\newblock {``A Heuristic of Scheduling Parallel Tasks and its Analysis''}.
\newblock {\em {SIAM} Journal on Computing}, 21(2):281--294, April 1992.

\bibitem[WFA92]{wfa:vldb92}
Annita~N. Wilschut, Jan Flokstra, and Peter~M.G. Apers.
\newblock {``Parallelism in a Main-Memory DBMS: The Performance of
  PRISMA/DB''}.
\newblock In {\em Proceedings of the Eighteenth International Conference on
  Very Large Data Bases}, pages 521--532, Vancouver, Canada, August 1992.

\bibitem[WFA95]{wfa:sigmod95}
Annita~N. Wilschut, Jan Flokstra, and Peter~M.G. Apers.
\newblock {``Parallel Evaluation of Multi-join Queries''}.
\newblock In {\em Proceedings of the 1995 {ACM SIGMOD} International Conference
  on Management of Data}, pages 115--126, San Jose, California, May 1995.

\bibitem[WTCY94]{wtcy:sigmetrics94}
Joel~L. Wolf, John Turek, Ming-Syan Chen, and Philip~S. Yu.
\newblock {``Scheduling Multiple Queries on a Parallel Machine''}.
\newblock In {\em Proceedings of the 1994 {ACM SIGMETRICS} Conference on
  Measurement \& Modeling of Computer Systems}, pages 45--55, Nashville,
  Tennessee, May 1994.

\bibitem[WTCY95]{wtcy:tpds95}
Joel~L. Wolf, John Turek, Ming-Syan Chen, and Philip~S. Yu.
\newblock {``A Hierarchical Approach to Parallel Multiquery Scheduling''}.
\newblock {\em {IEEE} Transactions on Parallel and Distributed Systems},
  6(6):578--590, June 1995.

\bibitem[YC93]{yc:vldbj93}
Philip~S. Yu and Douglas~W. Cornell.
\newblock {``Buffer Management Based on Return on Consumption in a Multi-Query
  Environment''}.
\newblock {\em The {VLDB} Journal}, 2(1):1--37, January 1993.

\end{thebibliography}
